%% file: main.tex
\documentclass[11pt,a4paper]{article} 	
\usepackage[utf8]{inputenc}
\usepackage{geometry}  		
\usepackage{graphicx,subcaption}					
\usepackage{amssymb}
\usepackage{indentfirst}
\usepackage{amsmath}
\usepackage{amsthm}
\usepackage{bm}
\usepackage{lineno}
\usepackage{setspace}
\usepackage{booktabs,multirow}
\usepackage{authblk}
\usepackage{graphicx}
\usepackage{float}
\usepackage[flushleft]{threeparttable} 
\usepackage{titlesec}
\usepackage{tabularx}
\usepackage{rotating}
\usepackage{booktabs} 
\usepackage{amssymb}

\usepackage[colorlinks=true, citecolor=blue, urlcolor=blue]{hyperref} 
\usepackage[backend=biber, style=numeric, sorting=none]{biblatex} 
\DeclareNameAlias{sortname}{family-given}
\input{symbols_def}

\title{\normalfont\bfseries\boldmath\Huge Low-energy Muon-Nucleon scattering experiment: LUNE\\ {\it (White Paper)}}

\author[1]{Chenlei An}
\author[2]{Dong Bai}
\author[1]{Ziyu Bai}
\author[1]{Kai Chen}
\author[3]{Liangwen Chen}
\author[4]{Xiang Chen}
\author[1]{Jianqiao Deng}
\author[3]{Yanxin Dou}
\author[1]{Yicheng Feng}
\author[1]{Zekai Feng}
\author[1]{Lu Gao}
\author[1]{Chang Gong}
\author[3]{Aiqiang Guo}
\author[5]{Liang Han}
\author[6]{Qundong Han}
\author[1]{Defu Hou}
\author[1]{Ruiwen Hou}
\author[1]{Huigang Hu}
\author[1]{Chen Ji}
\author[4]{Xiangdong Ji}
\author[1]{Vijay Kumar}
\author[7]{Dikai Li}
\author[1]{Jiuzhao Li}
\author[4]{Liang Li}
\author[8]{Qite Li}
\author[1]{Xin-Qiang Li}
\author[9]{Yuan Li}
\author[1]{Qiming Liang}
\author[3]{Yutie Liang}
\author[9]{Dong Liu}
\author[1]{Duanqing Liu}
\author[10]{Langtian Liu}
\author[1]{Weijie Liu}
\author[4]{Zejia Lu}
\author[9]{Maowu Nie}
\author[5]{Ziwen Pan}
\author[1]{Hua Pei}
\author[1]{Jinkang Peng}
\author[1]{Shusu Shi}
\author[11]{Qintao Song}
\author[1]{Xiaocheng Song}
\author[1]{Xiangming Sun}
\author[1]{Yichen Sun}
\author[3]{Zhiyu Sun}
\author[12]{Enke Wang}
\author[11]{Fei Wang}
\author[1]{Hulin Wang}
\author[13]{Jike Wang}
\author[1]{Xiang-Peng Wang}
\author[1]{Xiao Wang}
\author[1]{Xin-Nian Wang}
\author[1]{Yangxi Wang}
\author[1]{Yaping Wang}
\author[14]{Zeren Simon Wang}
\author[1]{Yanbing Wei}
\author[1]{Yuehong Xie}
\author[9]{Weizhi Xiong}
\author[1,3]{Nu Xu}
\author[3]{Yu Xu}
\author[5]{Siqi Yang}
\author[15]{Yadong Yang}
\author[1]{Hang Yin\thanks{Corresponding author: yinh@ccnu.edu.cn}}
\author[1]{Xing-Bo Yuan}
\author[1]{Dongliang Zhang}
\author[1]{Ranyu Zhang}
\author[3]{Ruitian Zhang}
\author[3]{Xueheng Zhang}
\author[14]{Yu Zhang}
\author[3]{Yuxiang Zhao}
\author[5]{Zihan Zhao}
\author[1]{Shuai Zhou}
\author[1]{XiaoKang Zhou}
\author[1]{Xiaoyu Zhu}

\affil[1]{{\it \small Key Laboratory of Quark and Lepton Physics (MOE) and Institute of Particle Physics, Central China Normal University, Wuhan 430079, China}}
\affil[2]{{\it \small College of Mechanics and Engineering Science, Hohai University, Nanjing 211100, China}}
\affil[3]{{\it \small State Key Laboratory of Heavy Ion Science and Technology, Institute of Modern Physics, Chinese Academy of Sciences, Lanzhou 730000, China}}
\affil[4]{{\it \small Key Laboratory for Particle Astrophysics and Cosmology (MOE) \& Shanghai Key Laboratory for Particle Physics and Cosmology, Shanghai Jiao Tong University, Shanghai 200240, China}}
\affil[5]{{\it \small Department of Modern Physics, University of Science and Technology of China, Hefei 230026, China}}
\affil[6]{{\it \small Istituto Nazionale di Fisica Nucleare Sezione di Padova INFN, Padova 35131, Italy}}
\affil[7]{{\it \small College of Engineering Physics, Shenzhen Technology University, Shenzhen 518118, China}}
\affil[8]{{\it \small School of Physics, Beijing University, Beijing 100871, China}}
\affil[9]{{\it \small Institute of Frontier and Interdisciplinary Science and Key Laboratory of Particle Physics and Particle Irradiation (MOE), Shandong University, Qingdao 266237, China}}
\affil[10]{{\it \small Department of Physics, School of Science, Shantou University. Shantou 515063, China}}
\affil[11]{{\it \small School of Physics, Zhengzhou University, Zhengzhou 450001, China}}
\affil[12]{{\it \small Key Laboratory of Atomic and Subatomic Structure and Quantum Control, South China Normal University, Guangzhou 510006, China}}
\affil[13]{{\it \small The Institute for Advanced Studies, Wuhan University, Wuhan 430072, China}}
\affil[14]{{\it \small School of Physics, Hefei University of Technology, Hefei 230601, China}}
\affil[15]{{\it \small Institute of Particle and Nuclear Physics, Henan Normal University, Xin xiang 453007, China}}

\date{\today}							

\begin{document}
\renewcommand{\thefootnote}{\arabic{footnote}}
\setcounter{footnote}{0}

\maketitle

\begin{abstract}
The HIAF will provide high-intensity, high-quality muon beams with momenta from 0.5 to 7.5 GeV/$c$. This energy range is uniquely suited for precision muon scattering, bridging the gap between low-energy electron facilities and future high-energy lepton-ion colliders. In particular, HIAF will enable precision measurements with both positive and negative muon beams over a broad kinematic range, complementing existing electron-scattering facilities such as JLab, EicC and EIC.

Based on HIAF muon source, the LUNE Collaboration has been established to address several fundamental questions in nuclear and particle physics, including the proton charge radius puzzle, nucleon electromagnetic structure, and the dynamics of quantum electrodynamics and hadronic interactions. The program proceeds in two phases, from elastic scattering to nucleon structure and beyond-Standard-Model searches.

The experiment is expected to determine the proton charge radius with a precision of approximately 1.0\% using elastic muon-proton scattering. It will also perform systematic measurements of the proton electromagnetic form factors with both $\mu^+$ and $\mu^-$ beams, enabling precise studies of two-photon exchange effects and stringent tests of quantum electrodynamics. Beyond elastic scattering, LUNE will investigate TMD, gravitational form factors, and nuclear charge radii, providing new insights into the 3D structure of nucleons and nuclei. The experiment will further address important topics including Coulomb-distortion corrections, nuclear medium effects, and possible signatures of physics beyond the Standard Model.

This white paper presents the scientific motivation, detector concept, expected performance, and long-term strategy of LUNE.

\end{abstract}

\setcounter{page}{2}
\mbox{~}
\cleardoublepage

\tableofcontents
\cleardoublepage

\input{ExecutiveSummary}
\input{introduction}
\input{muonbeam}
\input{detector}
\input{physicsprojection}
\input{Summary}

\clearpage

\appendix
\input{appendix}

\printbibliography[heading=bibintoc,title=References] 

\end{document}

%% file: symbols_def.tex
\usepackage{xspace} 
\usepackage{upgreek}

 \def\PK      {\ensuremath{K}\xspace}

 \def\Pi      {\ensuremath{i}\xspace}

 \def\Pp      {\ensuremath{p}\xspace}

 \def\Pmu         {\ensuremath{\mu}\xspace}

 \def\Ppi         {\ensuremath{\pi}\xspace}

  \mathchardef\PDelta="7101
 \mathchardef\PXi="7104
 \mathchardef\PLambda="7103
 \mathchardef\PSigma="7106
 \mathchardef\POmega="710A
 \mathchardef\PUpsilon="7107
 
\def\hiaf   {\mbox{HIAF}\xspace}
\def\lune   {\mbox{LUNE}\xspace}
\def\muse   {\mbox{MUSE}\xspace}

\def\psi    {\mbox{PSI}\xspace}

\def\mup        {{\ensuremath{\Pmu^+}}\xspace}
\def\mun        {{\ensuremath{\Pmu^-}}\xspace} 

\def\pion   {{\ensuremath{\Ppi}}\xspace}
\def\piz    {{\ensuremath{\pion^0}}\xspace}

\def\pim    {{\ensuremath{\pion^-}}\xspace}

\def\kaon    {{\ensuremath{\PK}}\xspace}
\def\Kbar    {{\kern 0.2em\overline{\kern -0.2em \PK}{}}\xspace}

\def\KorKbar    {\kern 0.18em\optbar{\kern -0.18em K}{}\xspace}

\def\Km      {{\ensuremath{\kaon^-}}\xspace}
\def\Kpm     {{\ensuremath{\kaon^\pm}}\xspace}

\def\proton      {{\ensuremath{\Pp}}\xspace}

\def\qsq       {{\ensuremath{Q^2}}\xspace}

\newcommand{\aunit}[1]{\ensuremath{\text{\,#1}}}

\newcommand{\tev}{\aunit{Te\kern -0.1em V}\xspace}
\newcommand{\gev}{\aunit{Ge\kern -0.1em V}\xspace}
\newcommand{\mev}{\aunit{Me\kern -0.1em V}\xspace}
\newcommand{\kev}{\aunit{ke\kern -0.1em V}\xspace}
\newcommand{\ev}{\aunit{e\kern -0.1em V}\xspace}

\newcommand{\mevc}{\ensuremath{\aunit{Me\kern -0.1em V\!/}c}\xspace}
\newcommand{\gevc}{\ensuremath{\aunit{Ge\kern -0.1em V\!/}c}\xspace}
\newcommand{\mevcc}{\ensuremath{\aunit{Me\kern -0.1em V\!/}c^2}\xspace}
\newcommand{\gevcc}{\ensuremath{\aunit{Ge\kern -0.1em V\!/}c^2}\xspace}

\def\cm   {\ensuremath{\mathrm{ \,cm}}\xspace}

\def\mm   {\ensuremath{\mathrm{ \,mm}}\xspace}

\def\mum  {\ensuremath{{\,\upmu\mathrm{m}}}\xspace}

\def\fm   {\ensuremath{\mathrm{ \,fm}}\xspace}

\def\pb {\ensuremath{\mathrm{ \,pb}}\xspace}
\def\invpb {\ensuremath{\mbox{\,pb}^{-1}}\xspace}

\def\ps   {\ensuremath{{\mathrm{ \,ps}}}\xspace}

\newcommand{\etc}{\mbox{\itshape etc.}\xspace}


%% file: ExecutiveSummary.tex
\section*{Executive Summary}

The High-Intensity Heavy-ion Accelerator Facility (\hiaf) will provide one of the world's most advanced medium-energy muon ($\mu$) beams, delivering high-intensity positive and negative muon beams ($\mu^{\pm}$) with momenta between 0.5 and 7.5\gevc. These beam characteristics create a unique opportunity for a new generation of precision muon-scattering experiments.

The Low-energy Muon-Nucleon Scattering Experiment (\lune) is proposed as one of nuclear physics programs at the \hiaf muon facility. Its primary objective is to establish a comprehensive experimental program based on precision muon scattering, addressing fundamental questions in hadron structure, nuclear physics, and searches for physics beyond the Standard Model (SM).
Unlike electron scattering, muons provide a substantially heavier lepton probe, significantly reducing radiative effects and enhancing sensitivity to several fundamental observables. The availability of both $\mu^+$ and $\mu^-$ beams further enables direct measurements of charge-dependent processes, particularly two-photon exchange (TPE) contributions, which cannot be isolated using electron beams alone. These unique features make muon scattering an essential complement to precision electron-scattering experiments.

The scientific role of \lune is complementary to existing and future international facilities, as summarized in Table~\ref{tab:facility_capability}:
\begin{itemize}
    \item Jefferson Lab (JLab) provides extremely high-precision electron scattering below approximately 12\gev but cannot perform precision muon scattering.
    \item The Electron-ion collider in China (EicC) and Electron-Ion Collider (EIC) will explore the quark and gluon structure of matter at much higher center-of-mass energies, emphasizing partonic imaging and quantum chromodynamics (QCD) dynamics rather than low four-momentum transfer (\qsq) precision measurements.
    \item \hiaf occupies a unique intermediate region, combining medium-energy muon beams with high intensity, making it ideally suited for precision studies of elastic scattering, nucleon electromagnetic structure, nuclear charge distributions, and precision tests of quantum electrodynamics (QED).
\end{itemize}
Rather than competing with these facilities, \lune fills an important missing component in the worldwide lepton-scattering program by providing precision measurements that cannot be performed elsewhere. The \lune physics program is organized into two stages.

\begin{table*}[!htbp]
\centering
\caption{Comparison of the capabilities of major lepton-scattering facilities. A check mark ($\checkmark$) indicates that the corresponding capability is a primary strength of the facility.}
\label{tab:facility_capability}
\renewcommand{\arraystretch}{1.25}
\begin{tabular}{lcccc}
\hline
\textbf{Capability} &
\textbf{JLab} &
\textbf{MUSE} &
\textbf{EIC/EicC} &
\textbf{LUNE} \\
\hline
High-intensity $\mu^\pm$ beam &-- &$\checkmark$ &-- &$\checkmark$ \\
Precision elastic $\mu p$ scattering &-- &$\checkmark$ &-- &$\checkmark$ \\
Momentum coverage ($0.5-7.5~\gevc$)&-- &-- &-- &$\checkmark$ \\
Proton charge radius &$\checkmark$ &$\checkmark$ &-- &$\checkmark$ \\
TPE with $\mu^\pm$ &-- &Limited &-- &$\checkmark$ \\
Nuclear charge radii &$\checkmark$ &Limited &-- &$\checkmark$ \\
Low-\qsq precision QED tests &$\checkmark$ &$\checkmark$ &-- &$\checkmark$ \\
High-\qsq nucleon tomography &$\checkmark$ &-- &$\checkmark$ &
 Phase-II \\

\hline
\end{tabular}
\end{table*}

\paragraph{Phase-I}
Phase-I focuses on precision elastic scattering using hydrogen and nuclear targets. Its principal goals include:
\begin{itemize}
    \item determination of the proton charge radius with approximately 1.0\% precision;
    \item systematic measurements of proton electromagnetic form factors using both positive and negative muons;
    \item precision determination of TPE effects;
    \item measurements of nuclear charge radii for light and medium-mass nuclei;
    \item studies of Coulomb-distortion effects and other higher-order electromagnetic corrections.
    \item precision searches for dark-sector particles and other phenomena beyond the SM.
\end{itemize}
These measurements will provide stringent tests of QED while substantially improving our understanding of nucleon and nuclear structure.

\paragraph{Phase-II}
Phase II will extend the experimental program toward polarized beams and advanced target systems. This stage will investigate:
\begin{itemize}
    \item transverse-momentum-dependent parton distributions (TMD);
    \item generalized parton distributions (GPD) and nucleon tomography;
    \item gravitational form factors;
    \item spin-dependent observables;
    \item nuclear modifications of partonic structure;
    \item further searches for dark-sector particles and other phenomena beyond the SM.
\end{itemize}

A conceptual detector has been developed specifically for precision muon scattering over the full momentum range of the \hiaf muon beam. The detector combines high-resolution silicon tracking, electromagnetic and hadronic calorimetry, and flexible target systems. The design emphasizes excellent momentum and angular resolutions while maintaining a low material budget to minimize multiple scattering.

Extensive simulation studies demonstrate that the proposed detector can achieve the performance required for the Phase-I physics goals. In particular, the projected precision on the proton charge radius is approximately 1.0\%, while systematic measurements of elastic scattering over a broad kinematic range will provide unprecedented information on nucleon electromagnetic structure.

The \lune collaboration envisions a long-term physics program that will evolve together with future upgrades of the \hiaf muon source, including increased beam intensity, polarized muon beams, and expanded experimental capabilities. Together, these developments will establish \hiaf as a major international center for precision muon scattering and provide a unique experimental platform that complements both JLab and the EicC/EIC.

This white paper summarizes the scientific motivation, detector concept, simulation studies, projected physics performance, and future roadmap of the \lune experiment, demonstrating its potential to become a dedicated facility for precision muon-based nuclear physics.

\clearpage

%% file: introduction.tex
\section{Introduction}
\label{sec:introduction}
Protons and neutrons, which constitute nearly all visible matter in the Universe, are emergent bound states of quarks and gluons governed by QCD~\cite{Gross:1973id,Politzer:1973fx}. While QCD has been established as the fundamental theory of the strong interaction, achieving a quantitative understanding of how nucleon properties emerge from its underlying degrees of freedom remains one of the central challenges in modern nuclear and particle physics. In particular, the transition region between hadronic and partonic descriptions, where neither purely hadronic models nor fully perturbative QCD techniques are sufficient, represents a uniquely rich and yet incompletely explored regime. Precision measurements in this regime can address a series of fundamental questions: How do the mass, charge radius~\cite{Pohl:2010zza}, spin, and mechanical structure~\cite{Burkert:2018bqq} of nucleons arise from quarks and gluons? What is the three-dimensional structure of the nucleon in momentum and coordinate space~\cite{Diehl:2003ny}? Can precision tests of lepton–nucleon interactions reveal deviations from the SM or shed light on unresolved puzzles in hadronic physics?

Lepton–nucleon scattering~\cite{EuropeanMuon:1983wih,BCDMS:1989qop,NewMuon:1991hlj,Breidenbach:1969kd} has long served as one of the most powerful and clean experimental probes of nucleon and nuclear structure. Over the past several decades, electron-scattering experiments~\cite{Breidenbach:1969kd,H1:2015ubc,JeffersonLabHallA:1999epl} have established our current knowledge of nucleon electromagnetic structure and partonic dynamics. Precision measurements of elastic scattering, deep-inelastic scattering (DIS), semi-inclusive DIS (SIDIS), and exclusive processes have provided critical tests of QCD and the SM. Nevertheless, several important challenges remain, including the `proton radius puzzle'~\cite{Goharipour:2025yxm}, the quantitative understanding of TPE effects~\cite{Arrington:2011dn}, the three-dimensional tomography of nucleons, and the role of short-range correlations (SRC)~\cite{Hen:2016kwk} in nuclear systems. Addressing these questions requires complementary measurements with different lepton probes and across a broad kinematic range.

Muon scattering~\cite{MUSE:2013uhu,COMPASS:2007rjf,Adams:2018pwt} provides a unique and highly complementary approach to studies of nucleon and nuclear structure. Owing to its mass, which is approximately 207 times larger than that of the electron~\cite{ParticleDataGroup:2024cfk}, radiative effects~\cite{Carlson:2015jba} associated with bremsstrahlung are strongly suppressed for muons, leading to improved control of systematic uncertainties in precision measurements. This intrinsic feature makes muon-induced reactions particularly well suited for precision studies of nucleon and nuclear structure, especially in the low and intermediate \qsq regime where nonperturbative QCD dynamics dominate.
Furthermore, direct comparisons between electron- and muon-induced processes provide a powerful tool for testing lepton universality~\cite{LHCb:2021trn} and for disentangling higher-order QED contributions, including TPE effects. Such measurements are of particular interest in light of the `proton radius puzzle', which has revealed a persistent discrepancy between determinations of the proton charge radius obtained from electronic and muonic systems. Precision muon-scattering experiments therefore offer a unique opportunity to clarify the origin of this discrepancy and to search for possible deviations from the SM.

\subsection{The \lune experiment}
The \hiaf~\cite{Yang:2013yeb}, constructed in Huizhou, China, by the Institute of Modern Physics (IMP), Chinese Academy of Sciences, is a major next-generation accelerator complex for nuclear and hadron physics. Its high-power heavy-ion accelerator chain will provide intense primary beams that enable the production of high-intensity secondary particle beams, including a dedicated muon beam~\cite{Xu:2025spd}. Together with a broad research program in nuclear structure and strongly interacting matter, this capability opens new opportunities for precision studies using muon probes.

A distinctive capability of \hiaf is its dedicated muon beam facility, which is expected to deliver secondary muon beams with momenta ranging from approximately 0.5 to 7.5~\gevc~\cite{Xu:2025spd}, depending on the primary beam configuration. This largely unexplored kinematic regime bridges the gap between low-energy muon experiments and high-energy DIS facilities, providing direct access to the transition region between hadronic and partonic descriptions of matter. The combination of high beam intensity, broad momentum coverage, and the reduced radiative effects characteristic of muons probes create a unique experimental environment of precision studies of muon-nucleons and muon-nuclei interactions. 

\subsubsection{Scientific Objectives}
To fully exploit these unique capabilities, the \lune has been proposed as a comprehensive physics program at the \hiaf muon facility, complementing the physics covered by Electron-ion collider in EicC~\cite{Anderle:2021wcy}, the Hyperon Nucleon Spectrometer (H-NS)~\cite{Bai:2026syt}, and other proposed experiments~\cite{Chen:2025ppt,An:2025lws,Chen:2026str,Liu:2025vyq}. The scientific objectives of \lune include:
\begin{itemize}
    \item Precision determination of the proton charge radius and systematic measurements of proton electromagnetic form factors, providing input to the `proton-radius puzzle' and TPE studies in elastic muon-proton scattering;
    \item Comparative measurements of electron- and muon-induced reactions to test lepton universality and improve the understanding of higher-order QED radiative corrections;
    \item Investigations of the transition from hadronic to partonic dynamics through measurements of quark-hadron duality over a broad kinematic range, providing new insight into the onset of perturbative QCD;
    \item Studies of the three-dimensional nucleon structure through semi-inclusive and exclusive scattering, including TMDs, GPDs, and the extraction of nucleon gravitational form factors to probe the internal mechanical structure of the nucleon~\cite{Polyakov:2018zvc};
    \item Precision measurements of the charge radii and electromagnetic structure of the deuteron and selected light nuclei, together with investigations of nuclear-medium effects, including the EMC effect, SRCs, and Coulomb distortion in muon-nucleus scattering;
    \item Searches for dark photons, light dark-sector particles, and other manifestations of physics beyond the SM~\cite{Alexander:2016aln}.
\end{itemize}
Together, these measurements will establish a broad scientific program spanning nucleon structure, nuclear structure, precision tests of QCD and QED, and searches for new physics. By combining high-intensity muon beam with a largely unexplored momentum range of 0.5 to 7.5\gevc, \lune will provide a unique experimental platform for addressing some of the most important open questions in hadronic and nuclear physics in the coming decades.

\subsubsection{Staged Physics Program}
The \lune physics program is designed to be implemented in stages, allowing a progressive expansion of its scientific reach as the experimental infrastructure and beam capabilities mature.

In Phase-I, solid or composite nuclear targets, including polyethylene (CH$_2$), deuterated polyethylene (CD$_2$), carbon ($C$), and selected high-$Z$ materials, will be employed, as discussed in Sec.~\ref{sec:targets}. These targets provide high luminosity and excellent statistical precision, enabling an broad physics program of muon–proton and muon–nucleus scattering measurements. The primary objectives of this stage include precision determinations of the the proton charge radius~\cite{Xiong:2019umf}, systematic studies of proton electromagnetic form factors~\cite{Pacetti:2014jai}, and investigations of TPE effects~\cite{Arrington:2011dn} through elastic muon scattering. Measurements on nuclear targets will further allow studies of nuclear electromagnetic structure, Coulomb-distortion effects, and EMC effect. In addition, high-$Z$ targets provide enhanced sensitivity to rare processes and searches for dark photons and other light weakly coupled particles beyond the SM (BSM).

In Phase-II, the experimental program will be extended through the deployment of cryogenic liquid targets, particularly liquid hydrogen (LH$_2$) and liquid deuterium (LD$_2$), together with polarized beam and target technologies. The use of liquid hydrogen targets will substantially reduce nuclear and material-related systematic uncertainties, enabling precision measurements of proton structure at an unprecedented level. Liquid deuterium targets will provide access to neutron structure and proton-neutron dynamics~\cite{Perdrisat:2006hj}, allowing determinations of neutron electromagnetic form factors, neutron charge distributions~\cite{JeffersonLabE93-026:2003tty}, and nuclear binding effects within a unified lepton-nucleus scattering framework~\cite{EuropeanMuon:1983wih}.

The introduction of polarized beams and polarized targets will significantly broaden the scientific scope of \lune. Spin-dependent observables in semi-inclusive and exclusive muon-scattering processes will provide access to TMDs and GPDs, enabling multidimensional imaging of the nucleon in both momentum and coordinate space. These measurements will further offer insights into the spin decomposition and mechanical structure of the nucleon through the extraction of gravitational form factors.

\subsection{The Proton Radius Puzzle}
The proton charge radius ($r_p^E$)~\cite{Gao:2021sml} is one of the most fundamental observables characterizing the spatial distribution of electric charge within the proton. It is defined through the slope of the proton electric form factor ($G_E^p$) at vanishing momentum transfer,
\begin{equation}
\label{eq:proton_radius}
r_p^2 = -6 \left. \frac{dG_E^p(Q^2)}{dQ^2} \right|_{Q^2=0},
\end{equation}
and thus provides direct access to the long-range structure of the proton in the non-perturbative regime of QCD. Its precise determination is also essential for tests of bound-state QED and for high-precision atomic spectroscopy, where finite-size effects contribute to the energy levels of hydrogenic systems.

\begin{figure}[!htbp]
    \centering
    \includegraphics[width=0.95\linewidth]{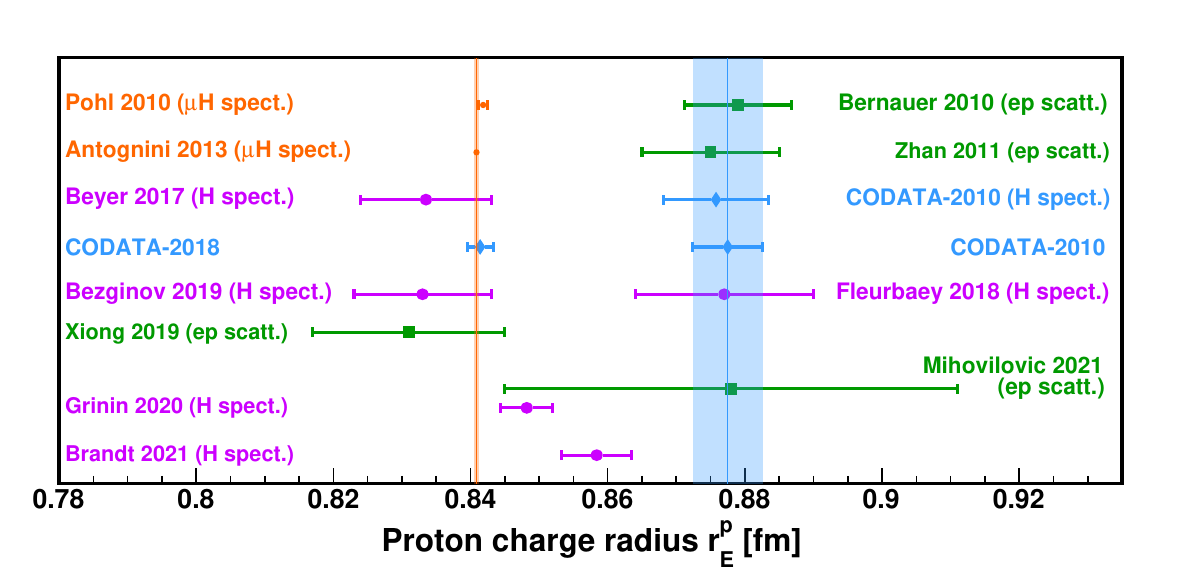}
    \caption{Compilation of proton charge radius measurements obtained from electronic and muonic spectroscopy as well as elastic lepton-proton scattering experiments. The comparison highlights the discrepancy between different determinations, commonly referred to as the `proton charge-radius puzzle'. Figure adapted from Ref.~\cite{Xiong:2023zih}.}
    \label{fig:rp_measurement}
\end{figure}

Experimentally, the proton charge radius is determined via two complementary approaches. The first is precision spectroscopy of electronic~\cite{Beyer:2017gug,Bezginov:2019mdi,Grinin:2020txk,Maisenbacher:2026nau} and muonic hydrogen~\cite{Pohl:2010zza,Antognini:2013txn}, where the finite proton size induces measurable shifts in atomic energy levels, with particularly high sensitivity in the Lamb shift of muonic hydrogen. The second approach is elastic lepton-proton scattering~\cite{Xiong:2019umf,Zhan:2011ji,A1:2010nsl,Mihovilovic:2019jiz,A1:2010nsl}, where the charge radius is extracted from the slope of the proton electric from factor at low \qsq. Figure~\ref{fig:rp_measurement} presents a compilation of proton charge radius measurements obtained from both spectroscopy and scattering experiments, highlighting the long-standing discrepancy known as the `proton radius puzzle'.

\subsubsection{The Proton Charge Radius}
In the one-photon-exchange approximation, the elastic scattering cross section is given by:
\begin{equation}
\frac{d\sigma}{d\Omega} = \left( \frac{d\sigma}{d\Omega} \right)_{\rm Mott} \frac{1}{1+\tau}
\left[(G_E^p(Q^2))^2 + \frac{\tau}{\epsilon}
(G_M^p(Q^2))^2 \right],
\label{eq:proton_radius_mott}
\end{equation}
where $G_E^p$ and $G_M^p$ denote the electric and magnetic form factors of the proton, respectively, and $\tau = Q^2/(4M_p^2)$, $\epsilon = \left[1+2(1+\tau)\tan^2(\theta/2)\right]^{-1}$
is the virtual-photon polarization parameter. The extraction of the charge radius requires a controlled extrapolation of $G_E^p(\qsq)$ to $\qsq \rightarrow 0$, making the treatment of radiative corrections and normalization uncertainties critical in precision measurements~\cite{Arrington:2011dn}.

For several decades, results from electronic hydrogen spectroscopy and electron-proton scattering were mutually consistent~\cite{Mohr:2012tt}, yielding a proton charge radius of approximately 0.88\fm. This situation changed with the 2010 measurement of the Lamb shift in muonic hydrogen~\cite{Pohl:2010zza}, which reported a significantly smaller value of approximately 0.84\fm, with a discrepancy exceeding five standard deviations. This observation, commonly referred to as the `proton radius puzzle', has since become one of the most prominent open problems in precision hadron physics~\cite{Carlson:2015jba,Antognini:2013txn}.
Since then, a wide range of new measurements has been performed in both spectroscopy and scattering channels. While recent results have reduced the tension, a fully consistent picture has not yet emerged. In particular, high-precision electron-proton scattering measurements such as PRad at JLab~\cite{Xiong:2019umf} and MAMI A1 data analyses~\cite{A1:2010nsl} have demonstrated that the extraction of the radius is highly sensitive to systematic effects in the low \qsq region. Consequently, an independent and systematically controlled determination of the proton charge radius remains a high-priority objective in modern nuclear and particle physics.

Currently, a global experimental effort is addressing this problem using complementary techniques. Electron-scattering experiments such as PRad-II~\cite{PRad:2020oor} at JLab, MAGIX at Mainz~\cite{A1:2021njh}, and ULQ2~\cite{Suda:2022hsm} at Tohoku University, aim to improve control of systematics in the extremely low-\qsq regime. In parallel, muon-based experiments (more discussions can be found in Appendix~\ref{app:history}) including MUSE experiment~\cite{MUSE:2013uhu} at Paul Scherrer Institute (PSI) and AMBER experiment~\cite{Adams:2018pwt} at European Organization for Nuclear Research (CERN), provide an independent determination of the proton charge radius through muon–proton scattering, enabling stringent tests of lepton universality and radiative corrections.

The \hiaf muon facility provides a complementary and potentially advantageous environment for precision studies of the proton charge radius. Compared with very low-energy muon beams employed by \muse, the higher beam energies at \hiaf significantly reduce multiple Coulomb scattering and improve tracking performance, leading to reduced experimental systematics. Compared with high-energy muon beams such as those employed in AMBER, the \hiaf energy range is well matched to the low-\qsq region relevant for charge-radius extraction, while avoiding extreme forward-angle kinematics that complicate acceptance and radiative correction control. Furthermore, the intermediate-energy regime provides favorable kinematic conditions for investigating TPE effects with well-controlled systematic uncertainties.
These features place \hiaf in a competitive position for precision muon–proton scattering measurements. By combining high beam intensity, favorable kinematics, and improved control of systematic uncertainties, the \lune experiment will enable a high-precision, independent determination of the proton charge radius and provide critical input toward resolving the `proton radius puzzle'.

\subsubsection{Two-Photon Exchange Effects}

\begin{figure}[!htbp]
    \centering
    \includegraphics[width=0.9\linewidth]{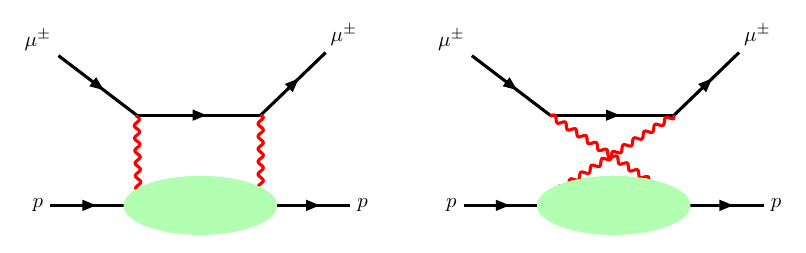}
    \caption{Leading TPE diagrams contributing to elastic muon-proton scattering.}
    \label{fig:tpe}
\end{figure}

TPE constitutes one of the most important higher-order QED corrections in elastic lepton-nucleon scattering~\cite{Arrington:2011dn,Afanasev:2017gsk}. While the dominant contribution to the scattering process arises from single-photon exchange, TPE originates from amplitudes involving the exchange of two virtual photons between the incident lepton and the nucleon, corresponding to the box and crossed-box diagrams shown in Fig.~\ref{fig:tpe}. Although formally suppressed by an additional factor of the electromagnetic coupling constant, TPE effects can generate corrections at the percent level and therefore become increasingly important in modern high-precision scattering experiments.

The importance of TPE extends beyond a conventional radiative correction. Precision determinations of nucleon electromagnetic form factors and charge radii rely on an accurate description of the elastic scattering cross section at the sub-percent level. In particular, the extraction of the proton charge radius from low-\qsq scattering measurements is highly sensitive to small distortions of the cross-section shape. Consequently, an incomplete treatment of TPE effects can lead to systematic biases in the determination of both the proton electric form factor and its slope at $\qsq \rightarrow 0$.

\begin{figure}[!htbp]
    \centering
    \includegraphics[width=0.7\linewidth]{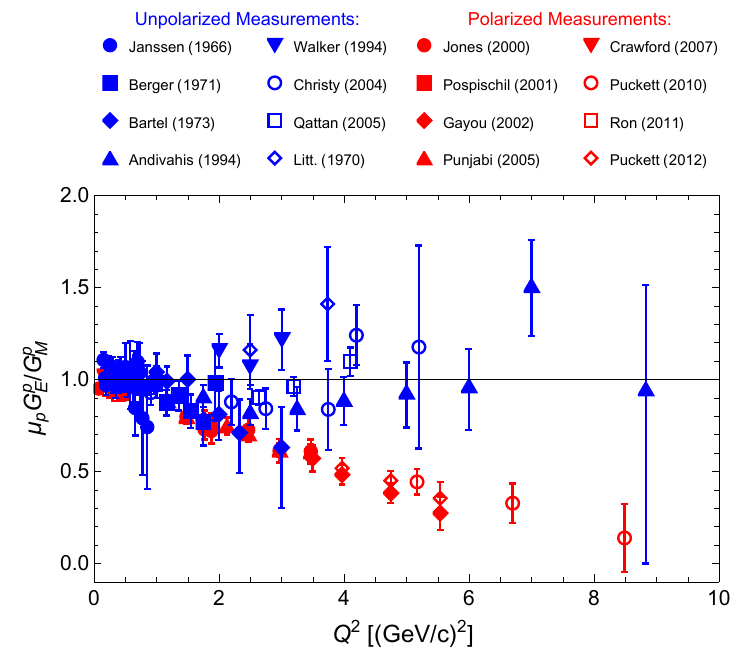}
    \caption{Comparison of proton form-factor ratio measurements extracted from unpolarized cross sections (Rosenbluth method) and polarization observables. The observed disagreement is now understood to arise largely from previously neglected TPE contributions. Figure adapted from Ref.~\cite{Alarcon:2023fcc}.}
    \label{fig:tpe_exp}
\end{figure}

The significance of TPE first became evident through the long-standing discrepancy~\cite{Qattan:2004ht} between proton form-factor measurements obtained using the Rosenbluth separation technique~\cite{Rosenbluth:1950yq} and those based on polarization-transfer observables~\cite{JeffersonLabHallA:2001qqe}, as shown in Fig.~\ref{fig:tpe_exp}. While Rosenbluth measurements suggested approximate scaling of the ratio $G_E^p/G_M^p$, polarization experiments revealed a pronounced decrease of this ratio with increasing \qsq. Subsequent theoretical and experimental studies established TPE effects as a major contributor to the observed discrepancy, with the inclusion of TPE corrections leading to improved consistency between the two approaches~\cite{Blunden:2003sp,McRae:2023zgu}. As a result, TPE is now recognized as an essential ingredient in precision analyses of elastic lepton-proton scattering.

Beyond their practical role in reducing systematic uncertainties, TPE observables provide direct access to hadronic structure~\cite{Arrington:2011dn,Tomalak:2014dja}. The TPE amplitude involves intermediate hadronic states and depends on the full electromagnetic structure of the nucleon over a broad kinematic range. Contributions from nucleon resonances, continuum excitations, and partonic degrees of freedom enter naturally into the TPE process. Consequently, precision measurements of TPE offer a unique probe of non-perturbative QCD dynamics that complements conventional form-factor studies.
$\frac{\sigma(l^+p)}{\sigma(l^-p)}$
provides one of the most direct and model-independent experimental observables for isolating TPE contributions~\cite{OLYMPUS:2016gso}.

The availability of both $\mup$ and $\mun$ beams at \hiaf offers a unique opportunity for such measurements. The broad momentum range of the \hiaf muon source, spanning approximately 0.5-7.5\gevc, covers a kinematic region that bridges the gap between MUSE and AMBER. This energy range enables systematic studies of TPE effects over a wide range of momentum transfer while maintaining favorable detector acceptance and manageable radiative corrections. By comparing $\mup p$ and $\mun p$ scattering under identical experimental conditions, \lune can provide high-precision measurements of TPE observables with unprecedented sensitivity.

These measurements will serve a dual purpose. On one hand, they will reduce a major source of systematic uncertainty in the extraction of the proton charge radius and electromagnetic form factors. On the other hand, they will provide valuable constraints on theoretical descriptions of hadronic structure, nucleon excitations, and higher-order QED effects. Together with precision radius measurements, TPE studies constitute one of the central components of the \lune physics program.

\subsection{Hadron Structure and QCD Dynamics}
One of the central goals of modern nuclear physics is to understand how the quark and gluon degrees of freedom of QCD generate the observed properties of hadrons. This challenge spans several interconnected questions: how the resonance description of the nucleon evolves into a partonic one; how quark momentum, transverse position, and spin are correlated; how the nucleon spin is built from intrinsic and orbital degrees of freedom; and how the energy, momentum, and internal forces of QCD are distributed inside hadronic matter.

The \lune program is designed to address these questions with precision muon scattering in the intermediate-energy regime. The available muon momentum range (0.5-7.5\gevc) bridges the resonance and scaling domains and enables a continuous exploration of the transition from hadronic to partonic dynamics. Compared with electron scattering, the larger muon mass suppresses lepton-line collinear radiation and substantially reduces bremsstrahlung-induced radiative tails. In addition, the availability of both positive and negative muon beams provides a distinctive handle on charge-odd interference effects in hard exclusive reactions. Together with hydrogen, deuterium, and other light-nuclear targets, these features make \lune complementary to existing electron-scattering facilities and high-energy muon programs.

\subsubsection{From Hadronic to Partonic Dynamics}
At large momentum transfer, QCD provides a successful description of hard scattering in terms of quarks and gluons. At lower energies, however, confinement, dynamical chiral-symmetry breaking, nucleon resonances~\cite{Aznauryan:2011qj}, meson-baryon dynamics~\cite{Bernard:1995dp}, and multi-parton correlations dominate the observed response. Understanding how these two descriptions are connected remains a central open problem in strong-interaction physics.

The transition is conventionally characterized by the invariant mass of the hadronic final state ($W$),
\begin{equation}
    W^2 = M^2 + Q^2\left(\frac{1}{x_B}-1\right),
\end{equation}
where $M$ is the nucleon mass, and $x_B$ (hereafter denoted simply by $x$) is the Bjorken scaling variable. At low $W$, the inclusive response is dominated by resonance excitation and meson-baryon final states. At sufficiently large $W$ and \qsq, the same response is described in terms of parton distribution functions (PDFs) and perturbative QCD evolution~\cite{Gribov:1972ri,Dokshitzer:1977sg,Altarelli:1977zs}. The intermediate region, where both descriptions are relevant, provides a particularly sensitive laboratory for non-perturbative QCD.

A remarkable manifestation of this connection is quark-hadron duality~\cite{Bloom:1970xb}. In inclusive lepton-nucleon scattering, resonance-region structure functions, when averaged over an appropriate interval in $x$ or the Nachtmann scaling variable $\xi$ ($\xi=\frac{2x}{1+\sqrt{1+4M^2x^2/Q^2}}$), approximately follow the scaling curves measured in the deep-inelastic regime. The phenomenon indicates that suitably averaged hadronic excitations encode the same underlying quark-gluon dynamics that appear in the partonic description. Its dynamical origin, onset scale, flavor and spin dependence, and extension to semi-inclusive or exclusive channels remain incompletely understood.

For unpolarized inclusive scattering, the measured cross section can be expressed in terms of the transverse and longitudinal structure functions,
\begin{equation}
    \frac{d^2\sigma}{dxdQ^2} =
    \frac{4\pi\alpha^2}{xQ^4}
    \left[
        \left(1-y-\frac{Mxy}{2E}\right)F_2(x,Q^2)
        +y^2xF_1(x,Q^2)
    \right],
    \label{eq:inclusive_DIS_cross_section}
\end{equation}
up to the usual convention-dependent target-mass and lepton-mass terms. Here $E$ is the incident-lepton energy in the target-rest frame and $y$ is the inelasticity. It is customary to separate the cross section into transverse ($\sigma_T$) and longitudinal ($\sigma_L$) components. These are related to the structure functions by $\sigma_T\propto F_1$ and 
$\sigma_L\propto F_L$, where the longitudinal structure function is defined as $F_L=\left(1+4M^2x^2/Q^2\right)F_2-2xF_1$. The principal observables are therefore,
\begin{equation}
    R(x,Q^2)=\frac{\sigma_L}{\sigma_T}=\frac{F_L}{2xF_1},
\end{equation}
together with their moments and their evolution with \qsq.

A practical test of global duality compares an averaged resonance response with a scaling reference evaluated over the same kinematic interval:
\begin{equation}
    \mathcal{R}_{\rm dual}(Q^2) =
    \frac{
        \displaystyle\int_{x_{\rm min}}^{x_{\rm max}}
        dx\,F_2^{\rm res}(x,Q^2)
    }{
        \displaystyle\int_{x_{\rm min}}^{x_{\rm max}}
        dx\,F_2^{\rm DIS}(x,Q^2)
    }.
    \label{eq:duality_ratio}
\end{equation}
Duality corresponds to $\mathcal{R}_{\rm dual}\simeq 1$ within the experimental and theoretical uncertainties. Local duality can be tested by applying the same comparison separately in individual resonance intervals. Measurements of the \qsq dependence of these ratios provide direct sensitivity to target-mass effects, higher-twist contributions, and multi-parton correlations.

\begin{figure}[!htbp]
    \centering
    \includegraphics[width=0.8\linewidth]{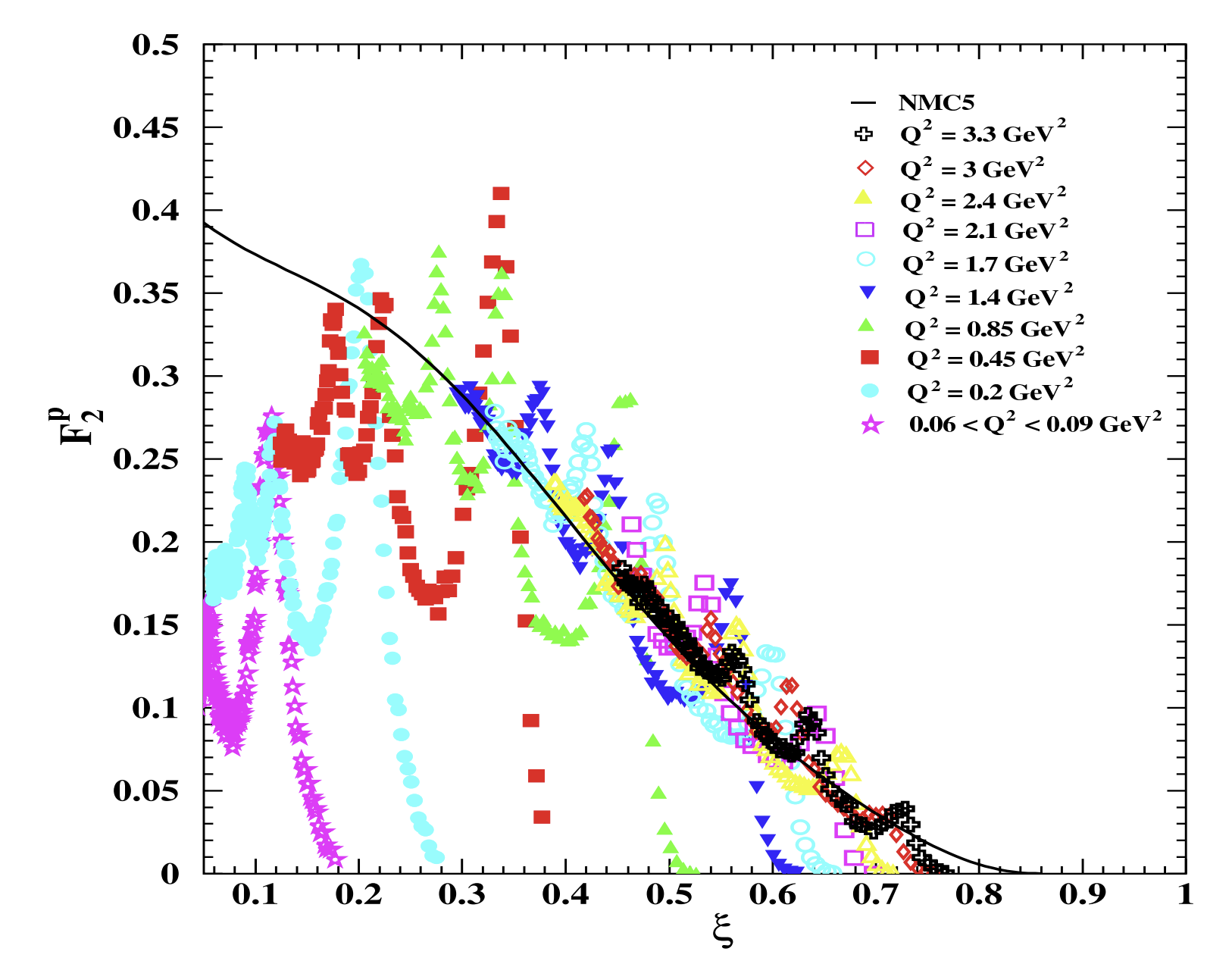}
    \caption{Illustration of quark-hadron duality, demonstrating that the structure function in the nucleon resonance region follows the scaling behavior observed in DIS when averaged over resonances. Adapted from Ref.~\cite{Melnitchouk:2005zr}.}
    \label{fig:duality}
\end{figure}

Inclusive electron-scattering experiments at SLAC and JLab, together with high-energy DIS measurements, have established the principal experimental evidence for duality and for the transition toward partonic scaling~\cite{Niculescu:2000tj,Dudek:2012vr}. These experiments demonstrated the validity of the partonic description over a broad kinematic range and provided extensive evidence for quark-hadron duality, as shown in Fig.~\ref{fig:duality}. 
Precision measurements in the resonance region are, however, particularly sensitive to QED radiation. Because collinear photon emission is enhanced for light leptons, electron scattering exhibits substantial bremsstrahlung, radiative tails, and kinematic bin migration~\cite{Mo:1968cg}. These effects can couple neighboring resonance and continuum regions and introduce model dependence in the extraction of resonance-averaged structure functions.

Muon scattering provides a complementary probe because the larger lepton mass suppresses the collinear enhancement of lepton-line radiation. The resulting reduction in radiative smearing is especially valuable in the resonance-DIS transition region, where rapidly varying cross sections and overlapping resonances make unfolding intrinsically challenging. \lune can therefore measure the evolution of inclusive structure functions across the resonance boundary with a different systematic profile from existing electron-scattering data. Comparisons between electron- and muon-induced reactions will additionally test the robustness of nucleon-structure extractions and provide sensitivity to possible lepton-mass-dependent electromagnetic effects.

With its broad momentum scan, \lune can map the evolution of $F_2$, $R$, and duality ratios over a continuous region in $(x,\qsq,W)$. The program will address the onset of scaling, the size and kinematic dependence of higher-twist contributions~\cite{Melnitchouk:2005zr,Deur:2018roz}, the validity of global and local duality, and the modification of these phenomena in deuterium and other light nuclei. These measurements will establish the inclusive foundation for the multidimensional and exclusive studies discussed below.

\subsubsection{Three-Dimensional Nucleon Structure}
Conventional collinear PDFs~\cite{Cridge:2021qjj} describe the longitudinal momentum distributions of quarks and gluons, but they do not provide a complete image of the nucleon. In particular, they contain limited information on transverse partonic motion, transverse spatial distributions, and correlations between momentum and spin.
The development TMDs~\cite{Aidala:2012mv,Bacchetta:2019sam} and GPDs~\cite{Diehl:2003ny,Belitsky:2005qn} has extended hadron-structure studies from one-dimensional momentum distributions to a multidimensional imaging program.

TMDs describe the dependence of parton distributions on both the longitudinal momentum fraction $x$ and the intrinsic transverse momentum $\bm{k}_T$. They encode transverse motion, spin-momentum correlations, and the role of initial- and final-state QCD interactions~\cite{Anselmino:2007fs}. Their primary experimental access is provided by SIDIS,
\begin{equation}
    \mu N \rightarrow \mu' h X,
    \label{eq:SIDIS_process}
\end{equation}
where an identified final-state hadron $h$ supplies sensitivity to flavor and transverse momentum. Differential SIDIS multiplicities and azimuthal modulations measured as functions of $(x,Q^2,z,P_{hT},\phi_h)$ provide access to convolutions of TMDs~\cite{Bacchetta:2006tn} and fragmentation functions. Here $z$ is the fractional energy carried by the detected hadron, $P_{hT}$ is its transverse momentum relative to the virtual photon, and $\phi_h$ is its azimuthal angle between the lepton plane and hadron plane.

\begin{figure}[!htbp]
    \centering
    \includegraphics[width=0.3\linewidth]{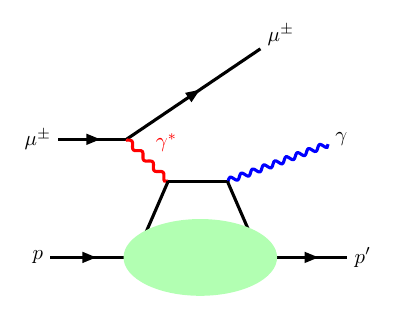}
    \includegraphics[width=0.6\linewidth]{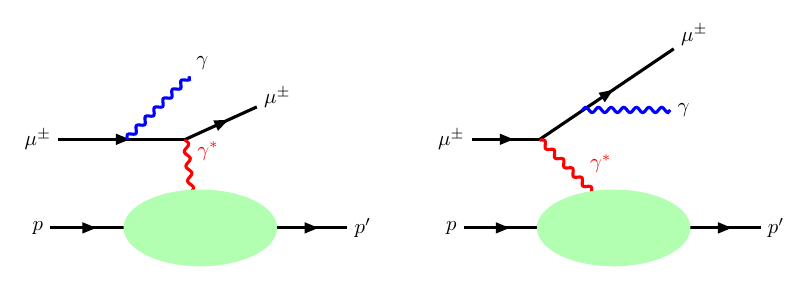}
    \caption{Leading-order diagrams for the DVCS process (left) and the Bethe-Heitler process (right).}
    \label{fig:dvcs}
\end{figure}

GPDs provide complementary information by correlating longitudinal parton momentum with transverse momentum transfer. They depend on $x$, skewness $\xi$, and invariant momentum transfer $t$. In the appropriate kinematic limit, Fourier transformation in transverse momentum transfer relates their $t$ dependence to transverse spatial distributions of partons. GPDs thus connect form factors, PDFs, and spatial imaging within a common QCD framework. Their first and second moments are related to the electromagnetic form factors and the QCD energy-momentum tensor (EMT) form factors, respectively. The latter provide access to the total angular momentum carried by quarks and gluons through Ji's sum rule~\cite{Ji:1996ek}.
The cleanest experimentally established channel for accessing GPDs is DVCS,
\begin{equation}
    \mu N \rightarrow \mu' N' \gamma.
    \label{eq:DVCS_process}
\end{equation}
As shown in Fig.~\ref{fig:dvcs}, the observed final state receives contributions from both the DVCS amplitude and the Bethe-Heitler (BH) process, in which the real photon is emitted from the lepton line. The cross section can be written schematically as
\begin{equation}
    d\sigma =
    d\sigma_{\rm BH}
    +d\sigma_{\rm DVCS}
    +d\sigma_{\rm INT},
    \label{eq:DVCS_decomposition}
\end{equation}
where $d\sigma_{\rm INT}$ denotes the interference term. The azimuthal and beam-charge dependence, and, where available, beam- and target-spin dependence provide complementary constraints on the real and imaginary parts of Compton form factors, which are convolutions of GPDs with perturbatively calculable hard kernels.

Existing measurements from HERMES~\cite{HERMES:2004mhh,HERMES:2009lmz}, COMPASS~\cite{COMPASS:2010hbb,COMPASS:2017jbv} and JLab~\cite{Dudek:2012vr} have demonstrated the power of SIDIS and DVCS for nucleon imaging, but the intermediate region in which factorization emerges from non-perturbative dynamics remains comparatively underexplored. \lune will investigate this region with a muon probe and can test the kinematic domain in which TMD and collinear-factorization descriptions become quantitatively reliable. The program will measure semi-inclusive hadron production and exclusive photon-production observables over a broad kinematic range of $Q^2$, $x$, $P_{hT}$, and $t$.

A particularly distinctive opportunity is the comparison of $\mu^+$ and $\mu^-$ induced DVCS. Under lepton-charge reversal, the pure BH and DVCS contributions are charge even in the cross section, whereas the
interference term is charge odd. The beam-charge asymmetry~\cite{HERMES:2008abz,CLAS:2007clm},
\begin{equation}
    A_C(\phi) =
    \frac{
        d\sigma^{\mu^+}(\phi)-d\sigma^{\mu^-}(\phi)
    }{
        d\sigma^{\mu^+}(\phi)+d\sigma^{\mu^-}(\phi)
    },
    \label{eq:beam_charge_asymmetry}
\end{equation}
therefore isolates the interference contribution and provides enhanced sensitivity to the real parts of the relevant Compton form factors. This capability complements electron-only measurements and will provide valuable input to global TMD and GPD analyses.

\subsubsection{Nucleon Spin and Orbital Dynamics}
The origin of the nucleon spin remains one of the central questions of hadron physics. The observation by the EMC collaboration that quark helicities account for only a limited fraction of the proton spin~\cite{EuropeanMuon:1989yki,Bass:2004xa} led to an extensive experimental and theoretical program aimed at resolving the contributions from quark spin, gluon spin, and orbital angular momentum. In a schematic decomposition,
\begin{equation}
 \frac{1}{2}=\frac{1}{2}\Delta\Sigma + \Delta G + L_q +L_g,
\end{equation}
where $\Delta\Sigma$ and $\Delta G$ denote the quark and gluon spin contributions, while $L_q$ and $L_g$ represent the corresponding orbital angular momenta. The individual terms are scheme- and scale-dependent, but the decomposition provides a useful framework for organizing the experimental program.

Spin-dependent TMDs provide direct access to spin-orbit correlations in the nucleon. For example, the Sivers function~\cite{Sivers:1989cc,Sivers:1990fh} describes a correlation between the transverse momentum of an unpolarized parton and the transverse spin of the parent nucleon, while transversity and the Collins fragmentation function~\cite{Collins:1992kk} generate characteristic azimuthal modulations in the distribution of produced hadrons. Measurements with transversely-polarized targets can access observables of the form
\begin{equation}
    A_{UT}^{\sin(\phi_h-\phi_S)}
    \propto
    f_{1T}^{\perp}\otimes D_1,
    \qquad
    A_{UT}^{\sin(\phi_h+\phi_S)}
    \propto
    h_1\otimes H_1^{\perp},
    \label{eq:SIDIS_spin_asymmetries}
\end{equation}
which are commonly referred to as Sivers and Collins asymmetries,
respectively. Their multidimensional dependence on $x$, $Q^2$,
$z$, and $P_{hT}$ is essential for disentangling non-perturbative
transverse motion from QCD evolution.

Longitudinal beam polarization, if available with sufficient magnitude and control, provides complementary sensitivity to beam-spin observables. For an unpolarized target, the beam single-spin asymmetry is defined as
\begin{equation}
    A_{LU} =
    \frac{
        d\sigma^{\rightarrow}-d\sigma^{\leftarrow}
    }{
        d\sigma^{\rightarrow}+d\sigma^{\leftarrow}
    }.
    \label{eq:beam_SSA}
\end{equation}
In SIDIS, this observable probes subleading-twist structure and quark-gluon correlations; it should not be interpreted as a direct measurement of the leading-twist Sivers or Collins functions. In DVCS, beam-spin asymmetries provide sensitivity primarily to the imaginary parts of Compton form factors. A future polarized-target program would substantially broaden the reach by enabling single- and double-spin asymmetries~\cite{CLAS:2006krx} and by providing direct access to transverse-spin phenomena.

The \lune spin program is scientifically most compelling when framed as a study of how spin-momentum and spin-orbit correlations emerge in the transition from hadronic to partonic dynamics. The moderate energy range is well suited to investigating the interplay of leading-twist dynamics, higher-twist effects, and quark-gluon correlations~\cite{Aidala:2012mv,Ji:2006ub}. Measurements on proton and deuteron targets will provide complementary flavor sensitivity, while a future polarized target upgrade could enable a comprehensive mapping of TMD asymmetries and tests of their evolution.

\subsubsection{From Nucleon Tomography to Mechanical Structure}

Beyond spatial imaging, GPDs provide a pathway to the mechanical properties of the nucleon. The matrix elements of the QCD EMT can be parametrized in terms of generalized form factors, commonly referred to as EMT or gravitational form factors. For the quark contribution, it can be written schematically
\begin{equation}
    \langle p'|T_q^{\mu\nu}|p\rangle
    \longrightarrow
    A_q(t),\; B_q(t),\; D_q(t),
    \label{eq:EMT_form_factors}
\end{equation}
with analogous gluon contributions. The second Mellin moments of GPDs yield  these gravitational  form factors~\cite{Ji:1996ek,Polyakov:2002yz}. The form factors $A(t)$ and $B(t)$ are related to momentum and angular momentum distributions, whereas the $D$-term governs the spatial distribution of internal stress.

\begin{figure}[!htbp]
    \centering
    \includegraphics[width=0.6\linewidth]{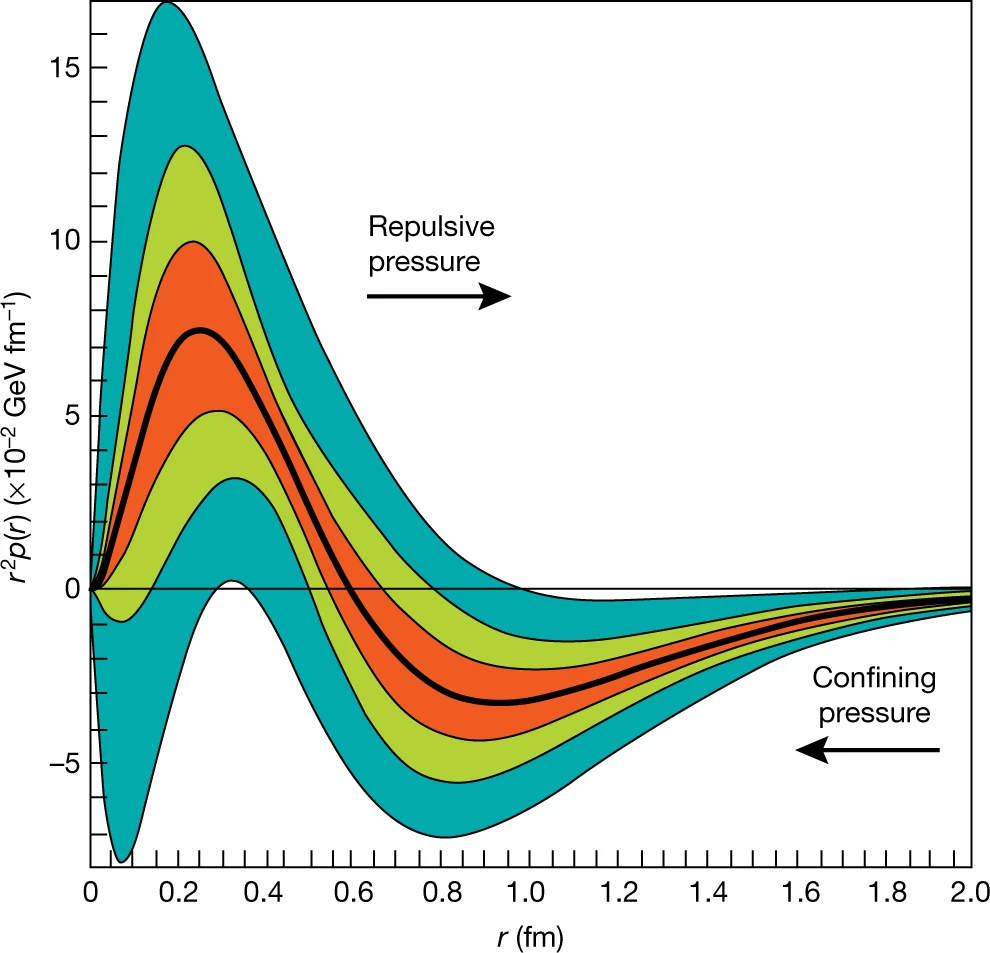}
    \caption{Pressure distribution inside the proton extracted from DVCS measurements and generalized parton distribution analyses~\cite{Burkert:2018bqq}.
    }
    \label{fig:proton_pressure}
\end{figure}

In the Breit-frame interpretation, the Fourier transform of the $D$-term determines the pressure $p(r)$ and shear-force distribution $s(r)$ inside the nucleon. These quantities characterize the balance between repulsive and attractive forces required for mechanical stability. The stability condition may be expressed through the von Laue relation,
\begin{equation}
    \int_0^\infty dr\,r^2 p(r)=0.
    \label{eq:von_Laue_condition}
\end{equation}
The first phenomenological extractions based on DVCS data suggested a strong repulsive pressure in the central region balanced by an attractive outer region~\cite{Burkert:2018bqq}, as shown in Fig.~\ref{fig:proton_pressure}. These results are important demonstrations of the potential of hard exclusive scattering, but the inferred pressure profiles remain dependent on the assumed GPD parameterization, the extrapolation in $t$, and the treatment of the $D$-term. Future measurements must therefore aim not only at improved statistical precision but also at expanded kinematic coverage and reduced model dependence.

DVCS is the principal experimental channel through which present constraints on the quark gravitational form factors are obtained. Differential cross sections, beam-spin asymmetries, target-spin asymmetries, and beam-charge asymmetries constrain complementary linear combinations of Compton form factors. In particular, the charge-odd interference term accessible through $\mup/\mun$ comparisons provides information that is difficult to isolate with a single lepton charge. Such measurements can improve the separation of real and imaginary components of the DVCS amplitude and thereby strengthen global constraints on GPD parameterizations relevant to the extraction of $A(t)$, $B(t)$, and $D(t)$.

\lune will contribute to this program by measuring exclusive photon-production observables in a kinematic region complementary to both high-luminosity electron facilities and high-energy muon experiments. Its most important role is not a standalone, model-independent determination of the proton pressure distribution, which requires broad $t$ coverage and global analysis, but rather the provision of high-value constraints on charge-sensitive DVCS observables in the transition regime. Combined with measurements from other facilities and with lattice-QCD calculations, these data can reduce the uncertainty on the GPDs and gravitational form factors that underlie quantitative studies of the nucleon's mechanical structure~\cite{Guidal:2013rya,Kumericki:2016ehc}.

In this way, the \lune hadron-structure program forms a coherent physics chain:
\begin{equation}
\begin{aligned}
    \text{resonance dynamics} 
     &\rightarrow \text{partonic structure} \\
    & \rightarrow \text{multidimensional tomography} \\
    & \rightarrow \text{spin-orbit correlations, spin decomposition, internal forces}.
\end{aligned}
\end{equation}
Precision muon scattering in the intermediate-energy domain will therefore provide a distinctive bridge between confinement-scale QCD dynamics and the quark-gluon description of nucleon structure.

\subsection{Nuclear Structure Physics}

Following its precision proton-scattering program, \lune can naturally extend its physics reach to a broad range of light nuclear systems through measurements on deuterium, helium, and heavier nuclear targets. 

\subsubsection{Light-Nuclei Charge Radii}
Precision determinations of electromagnetic form factors and charge radii of light nuclei~\cite{Sick:2008zza,Sick:2014yha} provide critical benchmarks for modern nuclear theory, effective field theories~\cite{Epelbaum:2008ga}, and {\it ab initio} calculations of nuclear structure~\cite{Carlson:2014vla}.
As the only stable two-nucleon bound system, consisting of a proton and a neutron, deuteron provides the most fundamental bridge between single-nucleon structure and the complex many-body dynamics of heavier nuclei. Because its properties can be calculated using well-established two-body dynamical frameworks, the deuteron serves as a benchmark system for testing nucleon-nucleon interaction models, including modern chiral effective field theory (ChEFT) descriptions. Precise knowledge of the deuteron charge radius is therefore essential for constraining nuclear-force models, validating few-body calculations, and understanding the evolution of nuclear structure from individual nucleons to light nuclei.

\begin{figure}[!htbp]
    \centering
    \includegraphics[width=0.8\linewidth]{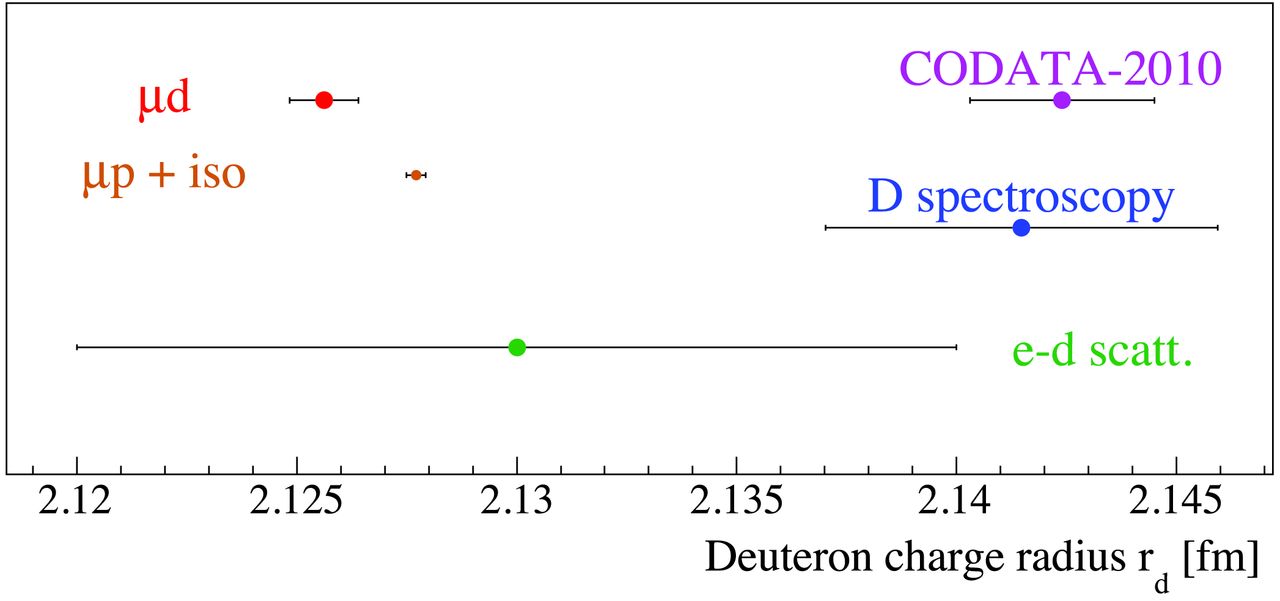}
    \caption{Compilation of existing determinations of the deuteron charge radius from electron scattering, electronic deuterium spectroscopy, and muonic-deuterium spectroscopy~\cite{CREMA:2016idx}.}
    \label{fig:deuteron}
\end{figure}

For many years, measurements from electron-deuteron scattering~\cite{Sick:1998cvq} and ordinary deuterium spectroscopy~\cite{Pohl:2016glp} were found to be mutually consistent within experimental uncertainties, leading to a relatively stable determination of the deuteron charge radius. However, the situation changed dramatically following the high-precision measurement of muonic deuterium by the CREMA collaboration~\cite{CREMA:2016idx} in 2016, as shown in Fig.~\ref{fig:deuteron}. The extracted charge radius was found to be significantly smaller than the value recommended by CODATA-2010~\cite{Mohr:2012tt}, with a discrepancy approaching seven standard deviations. This result mirrored the trend previously observed in the `proton radius puzzle' and extended the issue from a single nucleon to the simplest nuclear bound state.

The observed discrepancy has stimulated considerable theoretical interest. Possible explanations range from unresolved higher-order QED corrections in atomic spectroscopy and uncertainties in TPE contributions to more profound questions concerning lepton universality or limitations in current descriptions of lepton-nucleus bound states. Determining whether the discrepancy originates from nucleon structure, nuclear binding effects, or experimental systematics remains an important task for contemporary nuclear physics.

\lune aims to provide an independent determination of the deuteron charge radius through precision muon-deuteron scattering. Building upon the experience gained from the proton radius measurement program, the experiment is expected to improve upon the precision achieved in previous electron-scattering measurements and deliver a complementary muon-scattering determination of the deuteron radius. 

Beyond deuterium, measurements of elastic muon scattering from helium isotopes and other light nuclei will provide direct access to nuclear charge distributions and electromagnetic form factors over a broad kinematic range~\cite{CREMA:2025zpo}. Such studies will offer stringent tests of few-body nuclear calculations and establish important links between nucleon structure, nuclear forces, and emergent nuclear phenomena. 

\subsubsection{Coulomb Distortion Effects}

A unique opportunity offered by \lune is the systematic study of Coulomb distortion effects~\cite{Aste:2005wc} in muon-nucleus scattering. As a charged lepton traverses the strong electromagnetic field of a nucleus, its trajectory and wave function are modified, leading to corrections in measured cross sections and kinematic distributions. These effects become increasingly important for heavy nuclei and low-to-intermediate beam energies, where precision extractions of nuclear observables require accurate treatment of higher-order electromagnetic corrections.

Coulomb distortion represents one of the major theoretical challenges in precision electron-nucleus scattering. Although substantial progress has been achieved through distorted-wave Born approximation (DWBA) calculations~\cite{Yennie:1954zz} and related theoretical approaches, uncertainties associated with Coulomb corrections remain an important limitation in a variety of precision measurements. Reliable experimental benchmarks are therefore highly desirable.

Muon scattering provides a particularly clean environment for such studies~\cite{MUSE:2013uhu}. Owing to the much larger muon mass, Coulomb deflection and wave-function distortion are significantly reduced compared with electron scattering at comparable beam energies. Measurements of muon scattering from nuclei ranging from light to heavy targets can therefore provide valuable tests of theoretical treatments of Coulomb corrections and help quantify their impact on extracted nuclear observables.

The availability of both positive and negative muon beams at \hiaf further enhances these capabilities. Comparisons between \mup and \mun scattering under identical experimental conditions provide additional sensitivity to charge-dependent electromagnetic effects and offer important constraints on theoretical descriptions of higher-order corrections~\cite{Udias:1993zs}. Such measurements would establish a unique experimental program that complements ongoing efforts in electron scattering and contributes directly to the broader goal of precision nuclear structure studies. Measurements of the ratio
\begin{equation}
R = \sigma(\mun A)/\sigma(\mup A),
\end{equation}
for a series of nuclei from carbon to lead would provide direct experimental benchmarks for theoretical descriptions of Coulomb distortion effects.

\subsubsection{Nuclear Medium Modifications and the EMC Effect}

The EMC effect~\cite{EuropeanMuon:1983wih}, first observed in deep-inelastic muon scattering, revealed that the partonic structure of bound nucleons differs from that of free nucleons, demonstrating that nuclei cannot be described simply as collections of independent protons and neutrons.
Despite extensive experimental and theoretical efforts, the microscopic origin of the EMC effect remains incompletely understood~\cite{Norton:2003cb,Hen:2013oha}. Proposed explanations involve nuclear binding, nucleon off-shell effects, short-range nucleon correlations~\cite{Weinstein:2010rt}, and modifications of nucleon structure induced by the nuclear medium. Disentangling these mechanisms requires precise measurements over a wide range of nuclear targets and kinematic conditions.

The intermediate-energy muon beams available at \lune provide a complementary opportunity to investigate the transition region between traditional nuclear descriptions and partonic degrees of freedom. Measurements on light and medium-mass nuclei can probe the onset of nuclear modifications in structure functions and provide independent constraints on models of nuclear parton distributions~\cite{Eskola:2016oht}. Such studies are particularly valuable because they employ the same lepton species that historically led to the discovery of the EMC effect while operating in a kinematic regime that remains comparatively less explored.

\subsubsection{Future Opportunities: Short-Range Correlations}

Short-range correlations (SRCs)~\cite{Hen:2016kwk} arise from strongly interacting nucleon pairs at distances below approximately 1\fm and are responsible for the high-momentum components of nuclear wave functions. Over the past two decades, electron-scattering experiments~\cite{CLAS:2005ola} have demonstrated the dominance of neutron-proton correlated pairs~\cite{Subedi:2008zz} and established important connections between SRCs, nuclear structure, and the EMC effect.

Although the primary focus of the \lune nuclear-physics program lies in precision studies of electromagnetic structure and Coulomb effects, future developments of the facility may enable complementary investigations of SRC phenomena. In particular, semi-inclusive measurements involving recoil nucleons could provide additional information on high-momentum nuclear configurations and their relation to modifications of nucleon structure inside nuclei.
Realization of such a program would require dedicated detector capabilities, including efficient detection of recoil protons and neutrons, as well as detailed feasibility studies of achievable kinematic coverage and statistical precision. Nevertheless, SRC measurements remain an attractive long-term opportunity that could further expand the scientific scope of the \lune experiment.

\subsection{Searches for Physics Beyond the Standard Model}
\label{sec:newphysics}
Despite the remarkable successes of the SM, a number of fundamental questions remain unresolved, including the nature of dark matter, the origin of neutrino masses, the baryon asymmetry of the Universe, and the possible existence of new fundamental interactions.
These open questions provide strong motivation for searches for new physics beyond the SM~\cite{Essig:2013lka}.

Muon beams offer a powerful and complementary probe of such new physics. A growing class of theoretical models predicts new bosonic states that couple preferentially to muons or exhibit enhanced couplings in the lepton sector.
These include dark photons~\cite{Okun:1982xi,Galison:1983pa,Holdom:1985ag,Boehm:2003hm,Pospelov:2008zw,Curtin:2014cca}, light $Z'$ bosons~\cite{Foot:1990mn,He:1990pn,He:1991qd,
Harnik:2012ni,Altmannshofer:2014pba}, axion-like particles (ALPs)~\cite{Peccei:1977hh,Peccei:1977ur,Witten:1984dg,Conlon:2006tq,Arkani-Hamed:2006emk,Arvanitaki:2009fg,Cicoli:2012sz}, light scalar mediators~\cite{OConnell:2006rsp,Wells:2008xg,Bird:2004ts,Pospelov:2007mp,Krnjaic:2015mbs,Boiarska:2019jym}, and other light BSM states~\cite{Alimena:2025kjv}, often motivated by the above-mentioned SM issues.

\begin{figure}[!htbp]
    \centering
    \includegraphics[width=0.45\linewidth]{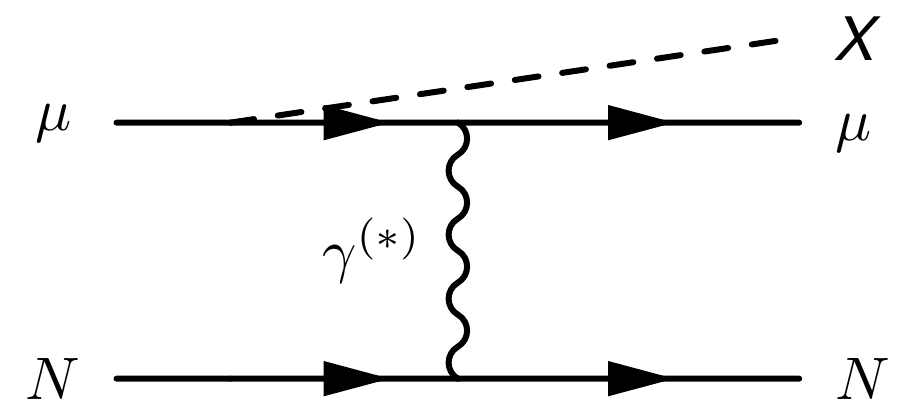}
    \includegraphics[width=0.45\linewidth]{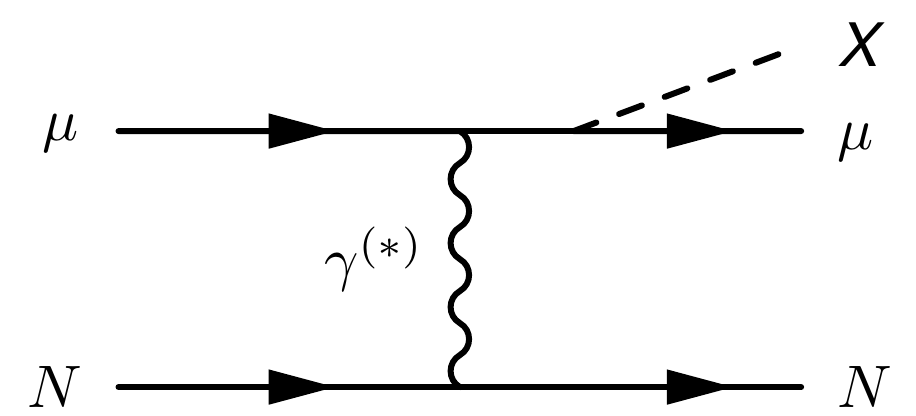}
    \caption{Feynman diagrams for the radiative production of a new boson $X$ beyond the SM.}
    \label{fig:npfd}
\end{figure}

At \lune, high-intensity muon beams incident on a heavy nuclear target, such as lead, can produce new particles through bremsstrahlung-like radiation processes~\cite{Gninenko:2014pea,Kahn:2018cqs},
$\mu N \rightarrow \mu N + X$, where $X$ denotes a hypothetical BSM boson.
Representative production diagrams are shown in Fig.~\ref{fig:npfd}.
The $X$ state can correspond to any of the above-mentioned possibilities.

A key feature of this framework is that the observable signatures depend primarily on the lifetime and decay modes of $X$, leading to a rich set of experimentally accessible final states. These signatures can be classified as follows:
\begin{enumerate}
\item Invisible final states characterized by missing transverse momentum, \\
$\mu N \rightarrow \mu N + {\rm MET}$,
\item Lepton-flavor-violating signatures, \\
$\mu N \rightarrow e N + {\rm MET}$,
\item Visible decays into charged leptons, \\
$\mu N \rightarrow \mu N + X,\quad X \rightarrow \ell^+\ell^-$, \\
leading to multi-lepton final states,
\item Visible decays into photons, \\
$\mu N \rightarrow \mu N + X,\quad X \rightarrow \gamma\gamma$.
\end{enumerate}

The phenomenology is strongly governed by the lifetime of $X$, which determines whether the signal appears as prompt final states, displaced objects, or missing energy.
This unified production mechanism therefore allows \lune to probe a wide range of BSM scenarios within a single experimental framework. 
Combined with the broad momentum coverage of the \hiaf muon beam and the excellent lepton, photon, and hadron identification capabilities of the \lune detector, this setup provides a versatile platform for dark-sector searches and lepton-sector precision tests.
By simultaneously exploring missing-energy signatures, lepton-flavor-violating processes, multi-lepton final states, and di-photon resonances, \lune is expected to provide sensitivities to parameter regions complementary to collider, fixed-target, and beam-dump experiments~\cite{Beacham:2019nyx}. 

\subsection{Summary of the Physics Program}
Overall, the \lune experiment is designed to fully exploit the high-quality muon beams provided by the \hiaf facility and to address a broad range of fundamental questions in nuclear and particle physics. The primary physics topics, together with their priorities, are summarized in Table~\ref{tab:phase1} (Phase-I) and Table~\ref{tab:phase2} (Phase-II).
The physics program is organized into two complementary stages, distinguished not only by beam polarization capabilities but also by detector configuration. 

\textbf{Phase-I (1-2\gevc muon beam): Precision Muon Scattering and New Physics Searches.}
The first stage of the \lune program focuses on precision measurements in the low-energy regime, where elastic and quasi-elastic scattering dominate. The primary objectives include precision determinations of the proton and light-nuclei charge radii, measurements of electromagnetic form factors, studies of TPE effects, investigations of Coulomb distortion in lepton-nucleus scattering, and searches for BSM physics using high-$Z$ targets. Owing to the large cross sections and modest detector requirements, this phase is expected to deliver the earliest high-impact physics results and establish the experimental foundations of the \lune program.

\textbf{Phase-II (3-7\gevc muon beam): Nucleon Tomography and QCD Structure.}
The second stage exploits higher-energy and polarized muon beams together with (polarized) cryogenic targets to access the partonic structure of nucleons and nuclei. This phase enables comprehensive studies of DIS, SIDIS, DVCS, GPDs, TMDs, and spin-dependent observables. The ultimate goal is to achieve a three-dimensional tomographic picture of the nucleon and to investigate the decomposition of nucleon spin and mechanical structure in terms of quark and gluon degrees of freedom.

\begin{table}[!htbp]
\centering
\caption{Phase-I physics program of \lune using low-energy muon beams (1-2\gevc). The emphasis is on precision muon scattering, nucleon and nuclear electromagnetic structure, and searches for BSM physics.}
\label{tab:phase1}
\renewcommand{\arraystretch}{1.2}
\begin{tabular}{lll}
\toprule
\textbf{Physics Topic} & \textbf{Target} & \textbf{Priority} \\
\midrule
Proton charge radius                     & CH$_2$          & Core  \\
Light-nuclei charge radii                & CD$_2$, $^3$He, $C$ & Core  \\
TPE                & CH$_2$          & Core  \\
Searches for BSM physics                 & Pb                    & Core  \\
Proton electromagnetic form factors      & CH$_2$         & Major  \\
Coulomb distortion effects               & $C$, Fe, Pb         & Major  \\
Baryon resonance region                  & CH$_2$          & Exploratory  \\
EMC effect and nuclear medium effects    & $C$, Fe, Pb         & Exploratory  \\
\bottomrule
\end{tabular}
\end{table}

\begin{table}[!htbp]
\centering
\caption{Phase-II physics program of \lune using higher-energy ($\gtrsim 3\gevc$) polarized muon beams and (polarized) cryogenic targets. The focus shifts toward nucleon tomography and QCD structure studies.}
\label{tab:phase2}
\renewcommand{\arraystretch}{1.2}
\begin{tabular}{lll}
\toprule
\textbf{Physics Topic} & \textbf{Target} & \textbf{Priority} \\
\midrule
DIS structure functions                  & LH$_2$                 & Core \\
SIDIS and TMDs                           & LH$_2$       & Core \\
DVCS and GPDs                            & LH$_2$       & Core \\
Nucleon spin decomposition               & (polarized) LH$_2$       & Core \\
Neutron structure studies                & $^3$He       & Major \\
Generalized form factors (GFFs)          & LH$_2$       & Major \\
Mechanical structure of the nucleon      & (polarized) LH$_2$       & Major \\
EMC effect at higher $Q^2$               & $C$, Fe, Pb        & Exploratory \\
Short-range correlations (SRCs)          & $C$, Fe, Pb        & Exploratory \\
\bottomrule
\end{tabular}
\end{table}

\clearpage

%% file: muonbeam.tex
\section{Experimental Conditions at \hiaf}
\label{sec:muonbeam}

\subsection{Overview of the \hiaf Facility}
\hiaf is designed to provide high-intensity beams ranging from protons to uranium ions for a broad physics program including nuclear structure, nuclear astrophysics, hadron physics, atomic physics, and applied sciences~\cite{Yang:2013yeb}.

\begin{figure}[!htbp]
    \centering
    \includegraphics[width=0.9\linewidth]{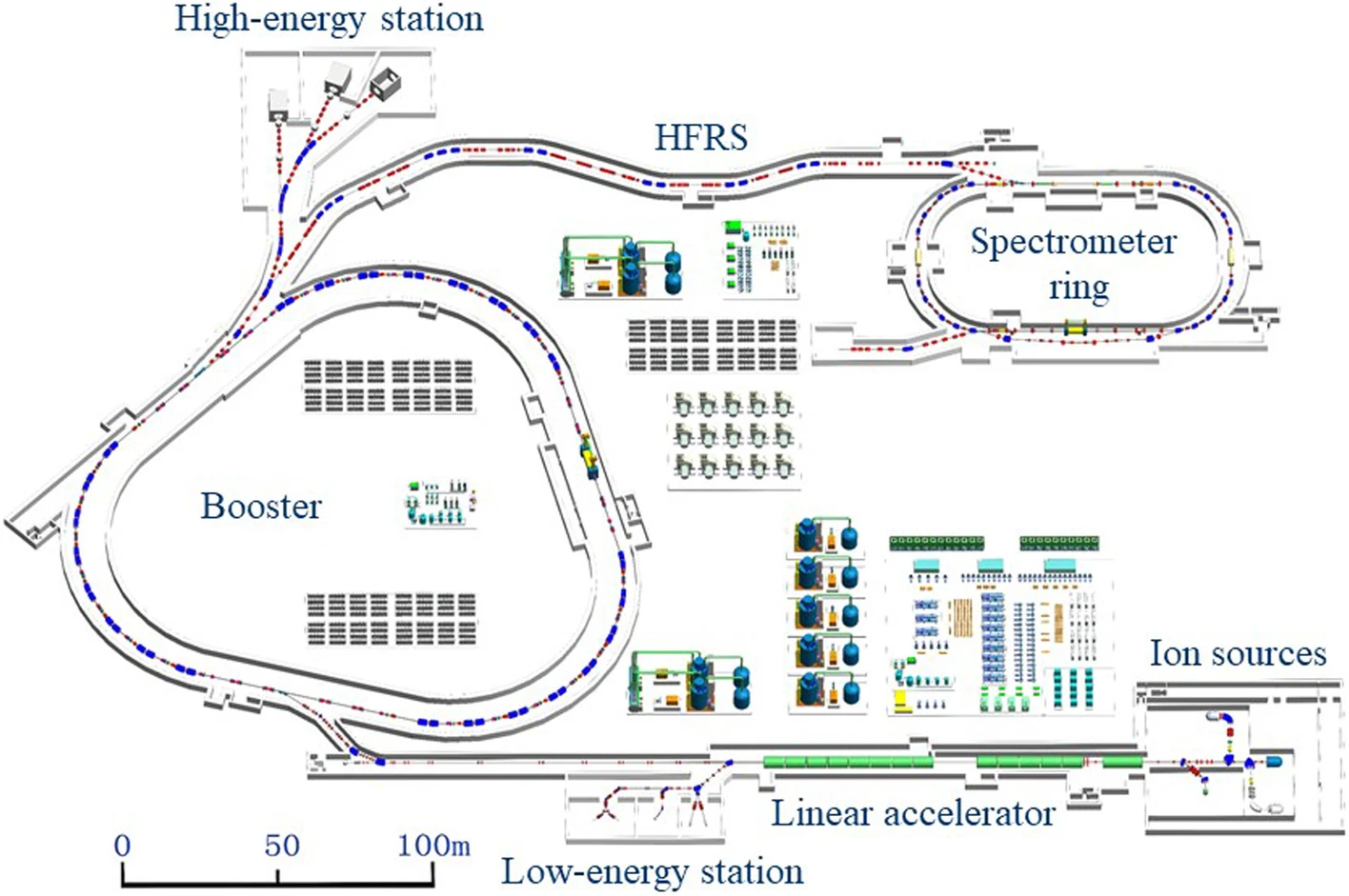}
    \caption{Schematic layout of the HIAF accelerator complex~\cite{Liu:2018pgv}, showing the iLinac, BRing, HIRIBL and SRing subsystems.}
    \label{fig:hiaf}
\end{figure}

As shown in Fig.~\ref{fig:hiaf}, The facility consists of a superconducting ion linear accelerator (iLinac), the Booster Ring (BRing), the Spectrometer Ring (SRing), and the High Intensity Radioactive Ion Beam Line (HIRIBL), formerly known as the High Energy Fragment Separator (HFRS). The BRing can accelerate proton beams up to approximately 9.3\gev and deliver beam intensities approaching $10^{12}$ particles per pulse. Both fast and slow extraction modes are available. Fast extraction provides high-intensity, short-duration spills that maximize secondary-particle yields, whereas slow extraction offers a continuous and nearly constant beam intensity over spill durations of several seconds, significantly reducing instantaneous detector occupancy and enabling precision fixed-target measurements. The slow-extracted proton beam is particularly well suited for the production of the high-quality secondary muon beam required by the \lune experiment.

\begin{figure}[!htbp]
    \centering
    \includegraphics[width=0.9\linewidth]{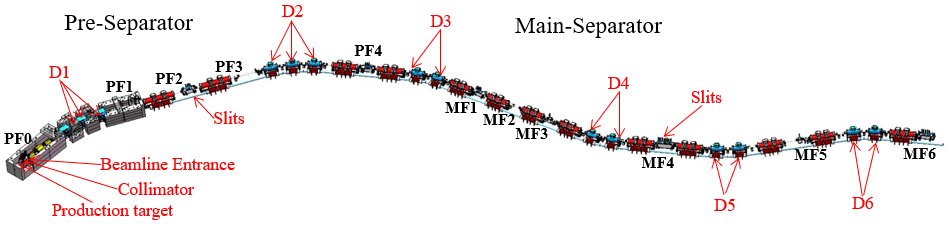}
    \caption{Layout of the HIRIBL~\cite{Xu:2025spd}, which serves as the transport and decay channel for GeV-energy muon production.}
    \label{fig:HFRS}
\end{figure}

A key component for the \lune project is HIRIBL, a 192~m long in-flight separator originally designed for rare-isotope beam production and transport, as shown in Fig.~\ref{fig:HFRS}. The HIRIBL provides a maximum magnetic rigidity of 25~Tm, corresponding to the transport of charged particles with momenta up to approximately 7.5\gevc. Its large length is particularly advantageous for pion decay in flight, enabling the generation of GeV-scale muon beams suitable for precision particle and nuclear physics experiments.

\subsubsection{HIAF Upgrade Program}

The planned HIAF Upgrade (HIAF-U)~\cite{Zhou:2022pxl} represents an important opportunity for expanding the scientific scope of the \lune program. By increasing the maximum proton beam energy from 9.3\gev to approximately 25\gev, the upgraded accelerator complex is expected to significantly extend the accessible secondary muon momentum range while providing higher beam intensities and greater operational flexibility. These improvements will enable precision muon-scattering measurements over a substantially broader kinematic region, thereby strengthening the physics reach in nucleon structure, electromagnetic form factors, and future spin-dependent studies.

In parallel with the accelerator upgrade, HIAF-U will provide new experimental halls with substantially larger floor space and supporting infrastructure. 
The \lune project is therefore envisioned as a staged experimental program that naturally follows the long-term development roadmap of HIAF. The Phase-I experiment employs a relatively compact detector system and can be implemented either in the downstream area of the existing HIRIBL beamline or in the dedicated experimental hall planned as part of the HIAF Phase-I Upgrade, depending on the construction schedule and facility availability. This flexibility allows the early physics program to begin while taking advantage of the evolving HIAF infrastructure.

Looking beyond the initial stage, the Phase-II detector is expected to incorporate a large-acceptance magnetic spectrometer, extended tracking systems and high-performance electromagnetic and hadronic calorimeters. Owing to its significantly increased dimensions and infrastructure requirements, the Phase-II experiment is naturally foreseen to be installed in the new experimental hall provided by the HIAF Upgrade. The expanded beam momentum coverage and improved experimental environment will enable a broad research program extending from precision measurements of proton and nuclear electromagnetic structure to spin physics, three-dimensional nucleon tomography, and other frontier studies with high-intensity muon beams.

This staged deployment strategy minimizes technical risk, allows the detector and physics program to evolve together with the HIAF accelerator complex, and establishes \lune as a long-term flagship experiment within the future HIAF muon science program. In this paper, we will focus on the current constructed HIAF complex and do not consider the planned HIAF-U.

\subsection{Production of GeV-Energy Muon Beams}
The \hiaf muon beam is produced through the decay of secondary pions generated when the primary proton beam impinges on a production target:
\begin{align*}
\pi^+ \rightarrow \mu^+ \nu_\mu, \\ 
\pi^- \rightarrow \mu^- \bar{\nu}_\mu.
\end{align*}

For a pion momentum of several GeV, the decay length is of the order of 100-200~m, comparable to the total length of the HIRIBL beamline. Consequently, a substantial fraction of the produced pions decay while propagating through the separator, naturally generating a high-energy muon beam without requiring a dedicated decay tunnel.
Detailed simulations based on the \hiaf beam parameters indicate that proton beams are the most effective source for producing high-momentum muons above approximately 1.5\gevc, while heavier ion beams may provide advantages at lower momenta~\cite{Xu:2025spd}.

The HIRIBL momentum acceptance of approximately $\pm2\%$ allows the transport of selected pion and muon momenta over a broad range extending up to about 7.5\gevc, making \hiaf one of the few facilities capable of providing tunable GeV-energy muon beams for nuclear and hadron physics. 
The comparison in Table~\ref{tab:muon_facilities} demonstrates that the HIAF-HIRIBL muon beam occupies a unique region of parameter space. Existing facilities are primarily optimized either for low-energy muon science (PSI and J-PARC) or for very-high-energy muon beams (CERN M2), while Fermilab focuses on storage-ring precision measurements. In contrast, \hiaf provides high-intensity muon beams in the 0.5-7.5\gevc momentum range, ideally suited for fixed-target studies of nucleon structure. More discussions can be found in Appendix~\ref{app:international}.

\begin{table}[!htbp]
\centering
\caption{Representative muon beam facilities worldwide. HIAF-HIRIBL covers the intermediate GeV-energy region between low-energy muon facilities and very-high-energy muon beamlines, providing a dedicated platform for precision muon scattering studies of nucleon structure.}
\label{tab:muon_facilities}
\begin{tabular}{lcc}
\hline
Facility &
Muon Momentum &
Typical Intensity \\
\hline

PSI PiM1 &
$0.1-0.5\gevc$ &
$\sim10^{8}~\mu/\mathrm{s}$ \\

J-PARC MUSE &
$\sim29\mevc$ &
$10^{6}$--$10^{8}~\mu/\mathrm{s}$ \\

Fermilab Muon Campus &
$3.1\gevc$ &
$\sim10^{5}~\mu/\mathrm{s}$ ($g$-$2$ beam)\\

CERN M2 &
$20-250\gevc$ &
$\sim10^{7}~\mu/\mathrm{s}$ \\

HIAF-HIRIBL (\lune) &
$0.5-7.5\gevc$ &
$\sim10^{6}~\mu/\mathrm{s}$ \\
\hline
\end{tabular}
\end{table}

\subsection{Expected Muon Beam Performance}

Simulation studies of the HIAF-HIRIBL system have demonstrated the feasibility of producing intense muon beams with momenta between approximately 0.5 and 7.5\gevc~\cite{Xu:2025spd}. Depending on the chosen beam momentum and production configuration, muon intensities can reach the order of $10^{6}~\mu/\mathrm{s}$, before particle-identification selections.

Typical beam parameters relevant for the \lune program are summarized in Table~\ref{tab:HIAFmuon}.
The beam parameters summarized in Table~\ref{tab:HIAFmuon} should be regarded as conservative estimates based on the current HIRIBL design and preliminary simulation studies. The existing HIRIBL optics has been optimized primarily for radioactive-ion beam transport rather than for muon production and collection. Consequently, several key parameters, including the beam spot size, beam divergence, momentum acceptance, and muon transmission efficiency, are expected to benefit from dedicated optimization studies in the future. In addition, ongoing upgrade plans for the \hiaf accelerator complex~\cite{Zhou:2022pxl} are expected to substantially increase the intensity of the primary proton beam. 
Together with improvements in the production target, collection optics, and muon transport system, the available muon flux could increase by up to two orders of magnitude ($10^2$) compared with the current baseline estimates. 

\begin{table}[!htbp]
\centering
\caption{Representative parameters of the \hiaf muon beam relevant for the \lune physics program. Values are based on current HIRIBL simulations and design studies.}
\label{tab:HIAFmuon}
\begin{tabular}{lc}
\hline
Parameter & Value \\
\hline
Momentum range & $0.5-7.5\gevc$ \\
Typical intensity & $\sim10^{6}~\mu/\mathrm{s}$ \\
Momentum acceptance & $\pm2\%$ \\
Beamline length & $192~\mathrm{m}$ \\
Beam spot size (90\% containment) &
$2$-$10~\mathrm{cm}$ \\
Beam divergence& 10 mrad \\
Charge selection & $\mu^{+}$ or $\mu^{-}$ \\
Beam polarization & Naturally polarized (under study) \\
Extraction mode & Fast or slow extraction \\
\hline
\end{tabular}
\end{table}

\begin{figure}[!htbp]
    \centering
    \includegraphics[width=0.45\linewidth]{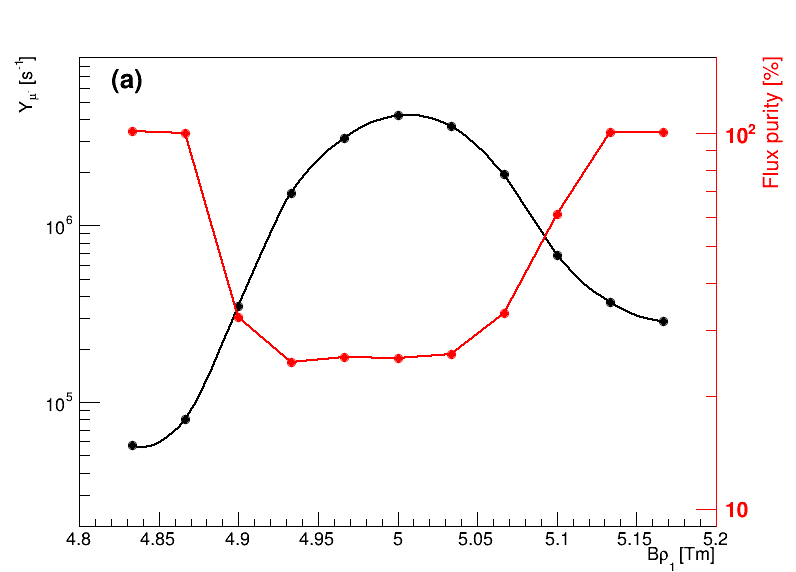}
    \includegraphics[width=0.45\linewidth]{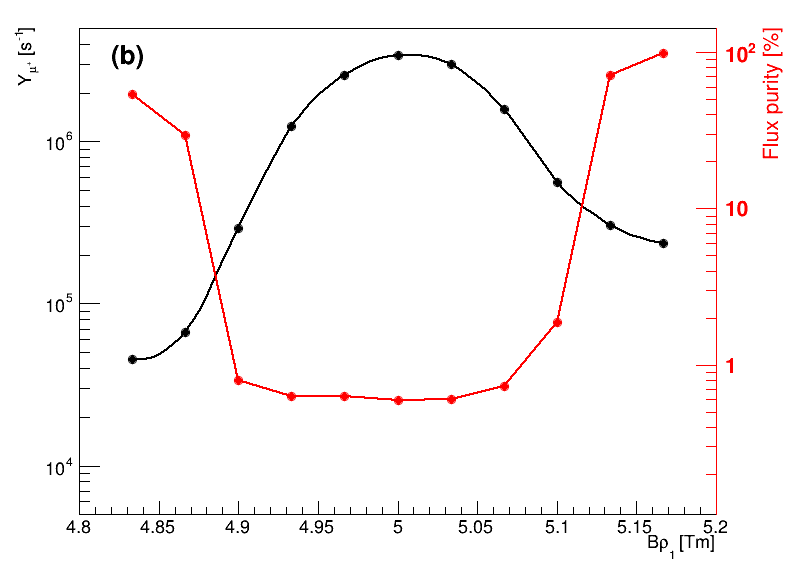}
    \caption{Expected muon flux after beam purification for (left) \mun and (right) \mup beam~\cite{Xu:2025spd}.}
    \label{fig:muon_flux}
\end{figure}

For current facility, with suitable beam purification and particle-identification systems, high-purity muon beams can be obtained while retaining intensities at the level of $10^{4}-10^{5}$~$\mu$/s, as shown in Fig.~\ref{fig:muon_flux}. In addition, since the muons originate from the weak decay of pions, a substantial longitudinal polarization is naturally expected. Such polarized muon beams would provide unique opportunities for precision studies of nucleon spin structure and polarization observables. 

\subsection{Targets}
\label{sec:targets}
The target system plays a central role in determining both the achievable luminosity and the physics reach of the \lune experiment. 

The baseline Phase-I configuration adopts a $5.0~\mathrm{mm}$ thick CH$_2$ target together with a $2.24~\mathrm{mm}$ thick graphite target providing an equivalent carbon areal density for background subtraction. A $5.0~\mathrm{mm}$ thick CD$_2$ target is also foreseen for deuteron measurements, following the approach adopted in previous fixed-target scattering experiments such as ULQ2~\cite{Legris:2025lxo}. These thicknesses represent the reference configurations used in the present design study rather than fixed engineering parameters; targets with different thicknesses can be prepared and optimized for individual physics channels by balancing luminosity, multiple scattering, and systematic uncertainties. Solid targets are selected as the baseline option for Phase-I due to their mechanical robustness, operational simplicity, and cost effectiveness compared with cryogenic hydrogen and deuterium targets. A modular target wheel design enables rapid exchange among CH$_2$, CD$_2$, carbon, and other nuclear targets, including Al, Cu, Fe, and Pb, providing additional opportunities for studies of nuclear structure, nuclear-medium effects, and muon-nucleus scattering processes.

To fully exploit the available muon intensity, Phase-II foresees the installation of cryogenic LH$_2$ and LD$_2$ targets. Compared with solid targets, cryogenic targets eliminate the need for carbon subtraction and substantially increase the achievable luminosity owing to their much larger target length. Two representative target lengths are currently considered. A short target of approximately $5~\mathrm{cm}$ is optimized for precision measurements of nucleon electromagnetic form factors and charge radii by reducing vertex uncertainty and multiple scattering. A long target of approximately $50~\mathrm{cm}$ maximizes luminosity for DIS and other measurements requiring very large event samples. Kapton and aluminum will be selected as the target cell materials for the short and long cells, respectively.

Representative target configurations adopted in the present white paper are summarized in Table~\ref{tab:targets}. 

\begin{table*}[!htbp]
\centering
\caption{Representative target configurations considered for the \lune physics program. The equivalent carbon target provides the same carbon areal density as the $5.0~\mathrm{mm}$ CH$_2$ target for background subtraction.}
\label{tab:targets}
\begin{tabular}{lccccc}
\hline
Target &Density &Thickness &Areal density &Phase &Primary physics \\
&(g/cm$^{3}$) &&(g/cm$^{2}$) && \\ \hline
CH$_2$        & 0.94& 5 mm & 0.47 & I & Proton measurements \\
Equivalent C  & 1.80& 2.24 mm & 0.40 & I & Carbon subtraction \\
CD$_2$        & 1.06& 5 mm& 0.53 & I & Deuteron measurements \\ \hline
LH$_2$        & 0.0708 & 5 cm & 0.354 & II & Radius and form factors \\
LH$_2$        & 0.0708 & 50 cm & 3.54& II & High-luminosity DIS \\
LD$_2$        & 0.169 & 5 cm & 0.845 & II & Deuteron structure \\
LD$_2$        & 0.169 & 50 cm & 8.45& II & High-luminosity DIS \\
\hline
\end{tabular}
\end{table*}

\subsection{Expected Luminosity}
The instantaneous luminosity of a fixed-target experiment is determined by the incident muon flux and the target areal density,
\begin{equation}
\mathcal{L} = \Phi_{\mu} n_t,
\end{equation}
where $\Phi_{\mu}$ denotes the incident muon rate and
\begin{equation}
n_t = \rho L \frac{N_A}{A},
\end{equation}
is the areal density of scattering centers (number of target nuclei per unit area). Here $\rho$ is the target density, $L$ is the target thickness, $A$ is the molar mass, and $N_A$ is Avogadro's constant ($6.022214076\times 10^{23}~\text{mol}^{-1}$)~\cite{ParticleDataGroup:2024cfk}.

Assuming the baseline \hiaf muon intensity of $10^{6}~\mu/\mathrm{s}$, the representative instantaneous luminosities for the target configurations considered in this white paper are summarized in Table~\ref{tab:lumi}.  
As discussed in the previous subsection, future upgrades of the primary proton beam together with dedicated optimization of the muon production target, collection optics, and transport system are expected to increase the available muon flux by up to two orders of magnitude. Consequently, the achievable luminosities are expected to scale proportionally with the beam intensity.

\begin{table*}[!htbp]
\centering
\caption{Representative instantaneous luminosities for the target configurations considered in this white paper. Baseline values assume a muon intensity of $10^{6}~\mu/\mathrm{s}$. The target areal density denotes the number of target nuclei (or scattering centers) per unit area used in the luminosity calculation.}
\label{tab:lumi}
\begin{tabular}{lccc}
\hline
Target & Target Areal density &Baseline &Integrated luminosity \\
&(cm$^{-2}$) &(cm$^{-2}$\,s$^{-1}$) &(\invpb/year) \\ \hline
CH$_2$       & $4.04\times10^{22}$ & $4.04\times10^{28}$ & 1.27 \\
Equivalent C & $2.02\times10^{22}$ & $2.02\times10^{28}$ & 0.64 \\
CD$_2$       & $3.18\times10^{22}$ & $3.18\times10^{28}$ & 1.00 \\
LH$_2$ (5 cm) & $2.13\times10^{23}$ & $2.13\times10^{29}$ & 6.72 \\
LH$_2$ (50 cm) & $2.13\times10^{24}$ & $2.13\times10^{30}$ & 67.2 \\
LD$_2$ (5 cm) & $2.54\times10^{23}$ & $2.54\times10^{29}$ & 8.01 \\
LD$_2$ (50 cm) & $2.54\times10^{24}$ & $2.54\times10^{30}$ & 80.1 \\
\hline
\end{tabular}
\end{table*}

The target configurations span nearly two orders of magnitude in instantaneous luminosity, providing considerable flexibility for different physics programs. 
The luminosity values summarized in Table~\ref{tab:lumi} provide the baseline assumptions adopted throughout the following physics sections for estimating event yields, statistical precision, and the integrated luminosity per year.

\subsection{Representative Event Yields}
The expected event yield for a given physics process can be estimated from
\begin{equation}
N = \mathcal{L}\,\sigma\,\varepsilon\,T,
\end{equation}
where $\mathcal{L}$ is the instantaneous luminosity, $\sigma$ is the corresponding production cross section within the selected kinematic region, $\varepsilon$ denotes the overall detection and reconstruction efficiency, and $T$ is the data-taking time.

In this white paper, representative event yields are evaluated using the baseline luminosities summarized in Table~\ref{tab:lumi}. The production cross sections are estimated using dedicated Monte Carlo (MC) event generators appropriate for each physics process, including \textsc{ESEPP}~\cite{Gramolin:2014pva} for elastic scattering, \textsc{DJANGOH}~\cite{Charchula:1994kf} for DIS and SIDIS, and \textsc{EpIC}~\cite{Aschenauer:2022aeb} for DVCS. Unless otherwise specified, the numbers quoted below correspond to generator-level estimates before detector acceptance and event-selection requirements. A full detector simulation incorporating tracking efficiency, particle identification (PID), trigger response, and reconstruction performance will be presented in future technical studies.

Table~\ref{tab:eventyield} summarizes representative generator-level cross sections and the corresponding annual event yields for several benchmark physics channels in the \lune physics program. The quoted event yields are estimated using the baseline beam intensity and target configurations described in Table~\ref{tab:muon_facilities} and Table~\ref{tab:lumi}, assuming an effective beam time of one operational year.
For elastic scattering, the cross sections are evaluated within the fiducial scattering-angle range of $1^\circ-10^\circ$. As expected, the elastic cross section decreases rapidly with increasing beam momentum. 
The inclusive DIS cross section is obtained with the Phase-II kinematic requirements listed in Sec.~\ref{sec:phaseII_det}. The SIDIS ($K^\pm$) cross section is not calculated independently. Instead, it is estimated from the inclusive DIS sample by multiplying the inclusive DIS cross section by the fraction of generated events containing at least one charged kaon ($K^+$ or $K^-$) in the final state. 
The projected event yields demonstrate that \lune can accumulate nearly $10^7$ inclusive DIS events and approximately $6\times10^5$ charged-kaon SIDIS events per year with the baseline Phase-II configuration, providing sufficient statistics for comprehensive studies of nucleon structure. In addition, more than $6\times10^4$ DVCS events are expected annually, enabling the first exploration of GPDs using a muon beam in this energy regime.
Nevertheless, these numbers provide a useful reference for evaluating the statistical reach of the \lune physics program and serve as common baseline assumptions throughout the following physics sections.

\begin{table*}[!htbp]
\centering
\caption{Representative generator-level event yields for benchmark physics processes considered in the \lune physics program. }
\label{tab:eventyield}
\begin{tabular}{lccc}
\hline
Process &Target &Cross section &Expected events\\
&&(\pb)&(per year) \\
\hline
elastic scattering ($1^\circ-10^\circ$, 1.0\gevc) &CH$_2$ (5 mm) &$8.32\times 10^8$ &$1.06\times10^9$ \\
elastic scattering ($1^\circ-10^\circ$, 1.5\gevc) &CH$_2$ (5 mm) &$3.75\times 10^8$ &$0.47\times10^9$ \\
elastic scattering ($1^\circ-10^\circ$, 2.0\gevc) &CH$_2$ (5 mm) &$2.09\times 10^8$ &$0.27\times10^9$ \\ \hline
Inclusive DIS (5.0\gevc) &LH$_2$ (50 cm) & $144\times 10^3$ &$9.68\times10^6$ \\
SIDIS ($\Kpm$, 5.0\gevc) &LH$_2$ (50 cm) &$8.8\times 10^3$ &$5.91\times10^5$ \\
DVCS (5.0\gevc) &LH$_2$ (50 cm) &$0.99\times 10^3$ & $ 6.65\times10^4$\\
\hline
\end{tabular}
\end{table*}

\clearpage

%% file: detector.tex
\section{Detector}
\label{sec:detector}

\subsection{Phase-I Physics requirements}
The primary objective of the \lune experiment Phase-I is a precision determination of the proton charge radius through elastic muon-proton scattering at low momentum transfer. Achieving a relative precision of approximately (1.0\%) on the extracted radius requires stringent control of both statistical and systematic uncertainties. The resulting detector performance requirements, summarized in Table~\ref{tab:phys_req}, have been derived from detailed MC simulations of elastic scattering and the relevant background processes.

\begin{table}[!htbp]
\centering
\caption{Key physics requirements for the \lune Phase-I detector.}
\label{tab:phys_req} 
\begin{tabular}{l c c} 
\toprule
Subsystem & Parameter & Requirement \\
\midrule
Tracking& Momentum resolution& $<2.0\%$ ($0.25-2\gevc$) \\
& Angular resolution& $<0.8$ mrad \\
& Reconstruction efficiency& $>97\%$ \\
\midrule
PID& $\mu/\pi$ separation efficiency& $>90\%$ ($>95\%$ desirable) \\
& ECal energy resolution& $<8\%/\sqrt{E}\oplus C$ \\
\midrule
Magnet& Field strength& $\sim0.5$ T \\
& Field uniformity& $<\pm5\times10^{-3}$ \\
\midrule
Target& Positioning precision& $<\pm0.1$ mm \\
& Number of stations& $\ge4$ \\
\midrule
DAQ& Beam rate capability& $>10^6~\mu/s$ \\
& Dead time& Negligible \\
& Clock synchronization& $<10\ps$ \\
\bottomrule
\end{tabular}
\end{table}

The proton charge radius is related to the slope of the electric form factor, $G_E^p(Q^2)$, at vanishing momentum transfer ($\qsq \to 0$),
The experimental challenge is therefore to determine the differential cross section with sufficient precision at the lowest accessible values of \qsq. For elastic scattering,
\begin{equation}
 \qsq = 4EE^\prime\sin^2\frac{\theta}{2} \approx EE^\prime \theta^2
\end{equation}
and in the forward-scattering region relevant for the radius extraction one has approximately $Q^2 \approx E^2\theta^2$. Consequently, the angular resolution of the tracking system is one of the most critical detector parameters. 
Dedicated simulation studies indicate that, to limit the corresponding systematic uncertainty to below 0.002\fm, the detector must achieve an angular resolution better than $0.8\,\text{mrad}$.

The momentum resolution directly affects the determination of \qsq and the separation of elastic events from semi-elastic and inelastic backgrounds. Insufficient momentum resolution broadens the reconstructed kinematic distributions and can bias the background subtraction procedure. Simulation studies show that a momentum resolution better than 2.0\% over the momentum range $0.25-2\gevc$ is sufficient for the proton-radius measurement while also satisfying the requirements of future DIS measurements.
In addition, the overall track reconstruction efficiency must exceed 97\% to preserve the statistical precision of the measurement.

Although the \hiaf muon beam is expected to have a pion contamination ($f(\pi/\mu)$) below $10^{-6}$, the $\pi p$ elastic scattering cross section is approximately several orders of magnitude larger than the corresponding $\mu p$ cross section. The residual pion background scales approximately as
\begin{equation}
\frac{N_{\pi\text{-bg}}}{N_{\mu p}}
\sim f\left(\frac{\pi}{\mu}\right)\left(\frac{\sigma_{\pi p}}{\sigma_{\mu p}}\right)(1-\varepsilon_{\rm PID}),
\end{equation}
where $\varepsilon_{\rm PID}$ denotes the pion rejection efficiency. For a rejection efficiency of 90\%, the residual contamination remains at the level of $10^{-3}$ relative to the signal. Therefore, a $\mu/\pi$ separation efficiency exceeding 90\% is required, while a performance above 95\% is highly desirable to further suppress hadronic backgrounds.
An electromagnetic calorimeter is not essential for the primary proton-radius measurement but significantly improves electron identification and background rejection. Furthermore, it provides the capability for future studies of exclusive processes such as DVCS, where energetic photons must be detected with high efficiency. A Shashlyk-type calorimeter with an energy resolution better than
\begin{equation}
\frac{\sigma_E}{E} < \frac{8\%}{\sqrt{E(\mathrm{GeV})}} \oplus C
\end{equation}
is sufficient for these physics goals.

Momentum reconstruction is provided by a dipole magnetic field. 
Detector simulations indicate that a dipole field of approximately $0.5~\text{T}$, with a field uniformity better than ($\pm 5\times10^{-3}$), is sufficient to satisfy the required momentum resolution.
The target system must provide a positioning precision better than $\pm0.1\mm$ in order to control systematic uncertainties associated with the interaction vertex. Multiple target stations are foreseen to accommodate different physics targets and dedicated background measurements.

Finally, the experiment will operate in a triggerless data-taking mode at beam intensities approaching $10^6$ muons per second. The data acquisition system must therefore sustain continuous high-rate readout with negligible dead time while maintaining sub-10\ps clock synchronization across the detector subsystems.

\subsection{Detector Design Concept}
The \lune detector is designed around the unique characteristics of the \hiaf muon beam, which provides high-intensity muons with momenta between 0.5 and 7.5\gevc. The detector concept is guided by several key principles: the use of mature and proven technologies, optimization for small-angle measurements, broad momentum coverage, precise tracking capabilities, and long-term upgradeability. The overall design aims to meet the immediate requirements of precision nucleon-radius measurements while providing a clear pathway toward an expanded DIS program in the future.
To maximize technical readiness and minimize development risks, the baseline detector relies primarily on detector technologies that have already been developed and validated. 
This strategy reduces construction risks, shortens development time, and improves the reliability and maintainability of the experimental apparatus.

\begin{figure}[!htbp]
    \centering
    \includegraphics[width=0.95\linewidth]{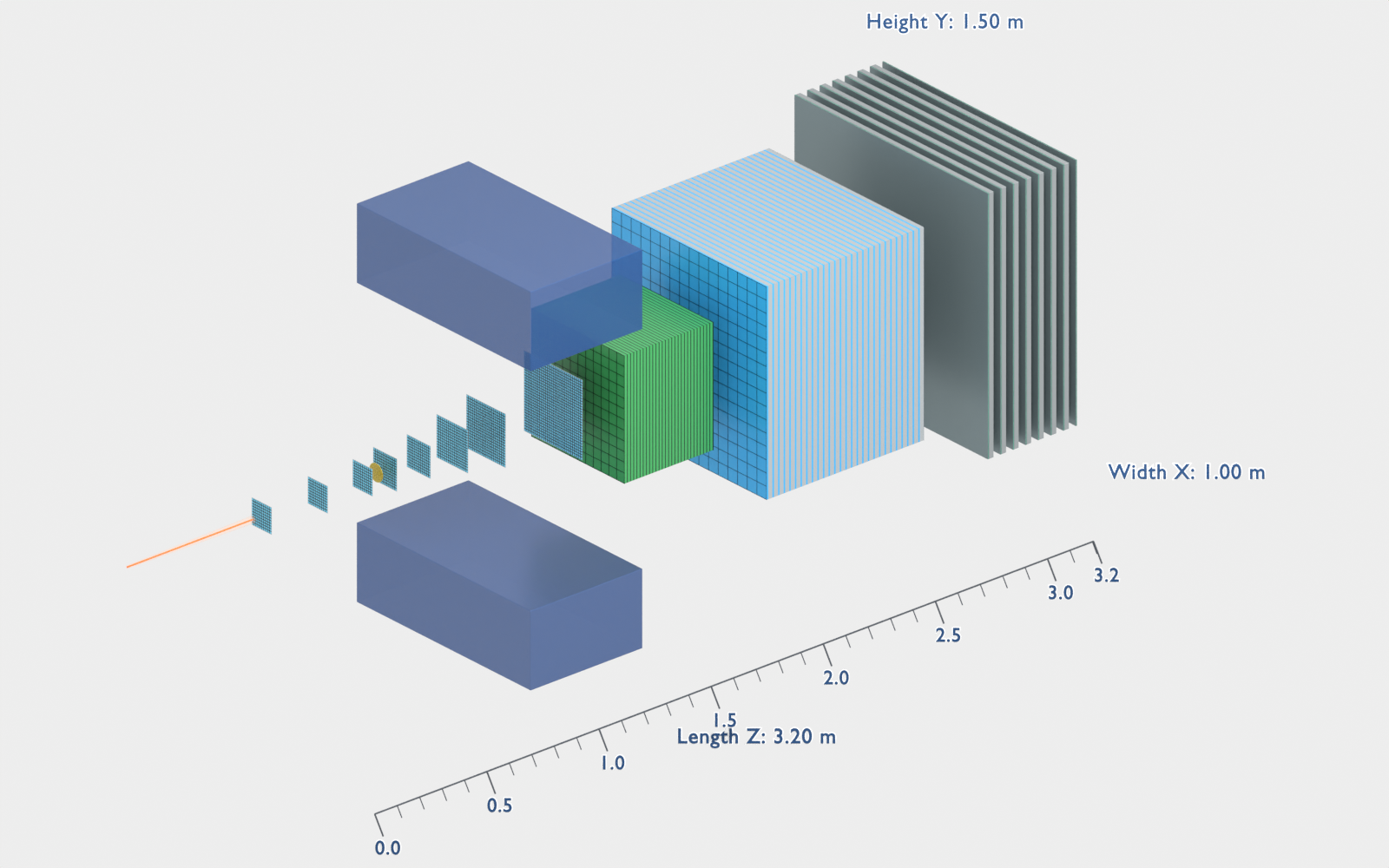}
    \caption{Schematic layout of the Phase-I \lune detector optimized for precision measurements of elastic muon-proton scattering and the extraction of the proton charge radius. }
    \label{fig:phaseI}
\end{figure}

\begin{figure}[!htbp]
    \centering
    \includegraphics[width=0.95\linewidth]{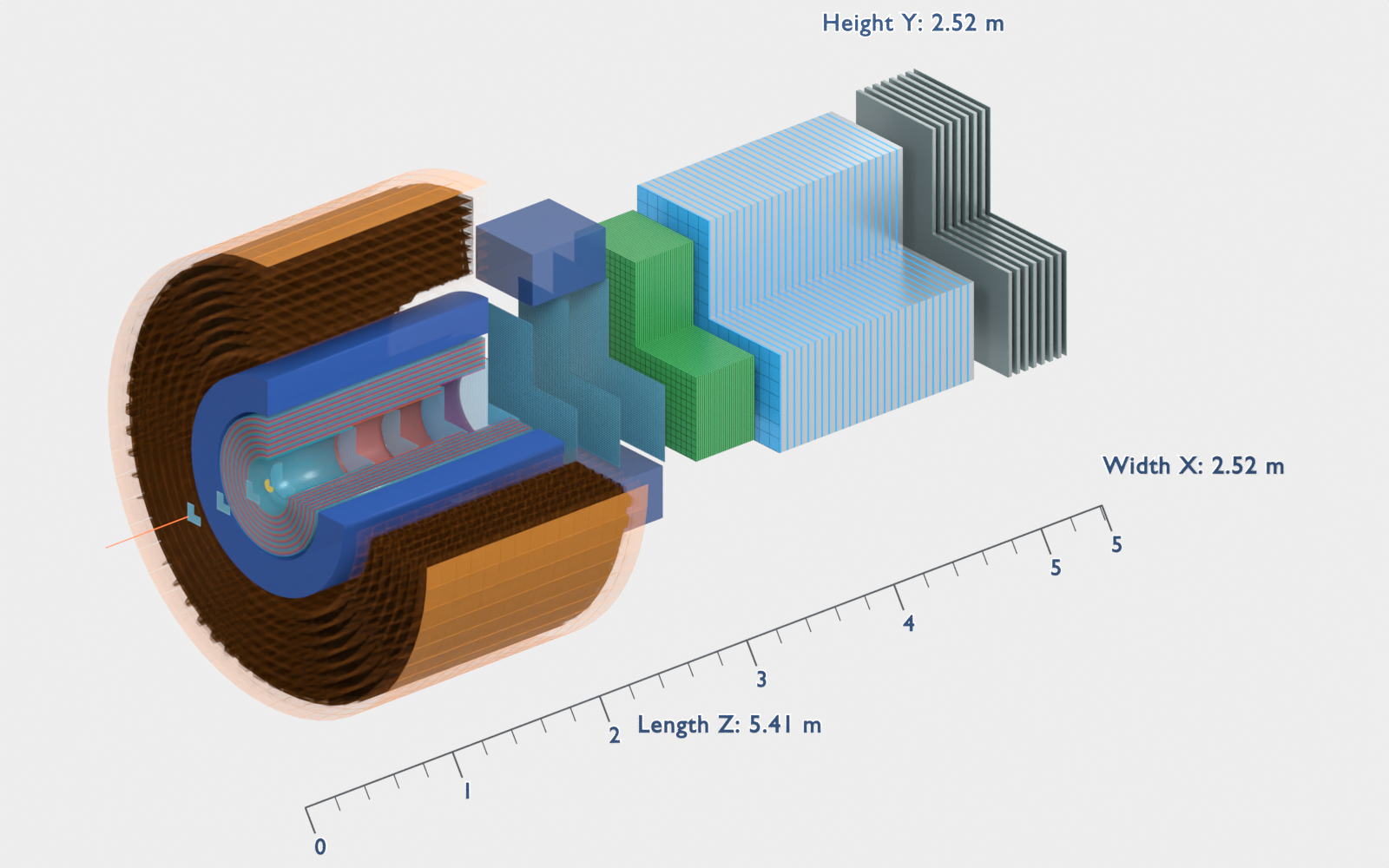}
    \caption{Conceptual layout of the Phase-II \lune detector. Building upon the Phase-I spectrometer, the upgraded configuration extends the acceptance and particle-identification capabilities through the addition of barrel detectors and larger tracking coverage.}
    \label{fig:phaseII}
\end{figure}

The detector is planned to be constructed in two phases. Phase-I is optimized for precision measurements of elastic muon-proton scattering in the 1-2\gevc energy range, focusing on nucleon charge-radius and electromagnetic form-factor studies. Phase-II extends the physics program to the 3-7\gevc region and enables comprehensive investigations of nucleon structure through DIS, SIDIS, and exclusive scattering processes. The concept design of detector for Phase-I is shown in Fig.~\ref{fig:phaseI}, and for Phase-II in Fig.~\ref{fig:phaseII}. The detector architecture has therefore been designed from the outset with future upgrades in mind, allowing major subsystems to be reused with minimal modifications.
To address the experimental requirements of both phases, the detector design incorporates the following major subsystems, also summarized in Table~\ref{tab:det}:

\begin{table}[!htbp]
\centering
\caption{Staged implementation and technology choices for the \lune detector.}
\label{tab:det}
\begin{tabular}{lccc}
\toprule
\textbf{Subsystem} & \textbf{Phase-I} & \textbf{Phase-II} & \textbf{Technology} \\
\midrule
Beam monitor & \checkmark & \checkmark & Silicon pixel \\
Forward tracking detector & \checkmark & \checkmark & Silicon pixel \\
Forward dipole magnet & \checkmark & \checkmark & Room-temperature dipole \\
Electromagnetic calorimeter & \checkmark & \checkmark & Shashlyk \\
Hadronic calorimeter & \checkmark & \checkmark & Shashlyk \\
Multi-target system & \checkmark & -- & -- \\
Solenoid magnet & -- & \checkmark & Superconducting \\
Barrel tracking detector & -- & \checkmark & Drift chamber \\
Barrel PID detector & -- & \checkmark & Shashlyk (or others) \\
Polarized target (optional) & -- & (\checkmark) & -- \\
\bottomrule
\end{tabular}
\end{table}

\begin{itemize}
\item \textbf{Low-material tracking system.}
Precise reconstruction of scattering angles is essential for charge-radius measurements, particularly for events scattered at very small angles. To minimize multiple-scattering effects for GeV-scale muons while maintaining excellent spatial resolution under high-rate conditions, \lune adopts silicon pixel tracking detectors instead of conventional gas-based tracking systems in the beam and forward regions. The use of thin sensors and lightweight support structures minimizes the material budget while providing high-precision measurements of both incoming and outgoing particle trajectories.

\item \textbf{High-precision momentum spectrometer.}
Accurate determination of particle momentum is crucial for reconstructing kinematic variables and suppressing backgrounds from inelastic and quasi-elastic processes. In Phase-I, a forward dipole spectrometer combined with high-resolution tracking detectors provides momentum analysis for scattered muons and produced charged particles. The system is designed to achieve momentum resolutions at the percent level across the primary kinematic region of interest.

\item \textbf{Particle-identification and calorimeter system.}
A sampling calorimeter system based on Shashlyk technology provides PID and energy measurements. The system consists of electromagnetic and hadronic calorimeter sections employing alternating absorber and scintillator layers. By exploiting differences in shower development and energy deposition patterns, efficient separation of muons, electrons, photons, and hadrons can be achieved. The use of solid-state calorimetry offers a robust and cost-effective solution with excellent operational stability.
\item \textbf{Versatile target system.}
A modular target system is foreseen to support a broad range of physics measurements involving hydrogen, deuterium, carbon, and heavier nuclear targets. The design allows remote operation, precise positioning, and rapid target exchange while maintaining strict control of target-related systematic uncertainties.

\item \textbf{Large-angle spectrometer upgrade.}
Phase-II introduces a superconducting solenoidal magnet surrounding the target region together with additional barrel tracking and PID detectors. This upgrade substantially increases the geometric acceptance and enables measurements of semi-inclusive and exclusive final states over a much wider kinematic range. The solenoidal spectrometer is particularly important for studies of nucleon spin structure, TMDs, GPDs, and other multidimensional nucleon-structure observables.

\end{itemize}

The detector architecture is designed to maximize reuse of major subsystems throughout the staged implementation of the experiment. Core components, including the forward spectrometer, tracking detectors, calorimeter systems, and data-acquisition infrastructure, will remain integral parts of the experiment after the Phase-II upgrade. This approach will reduces long-term construction costs while preserving continuity between successive physics programs.
The detector parameters presented in this White Paper should be regarded as baseline design targets. Further optimization studies are ongoing, particularly for the beam transport system, detector acceptance, material budget, particle-identification performance, and integration with future \hiaf upgrades. 

\subsubsection{Phase-I detector Configuration}
The \lune Phase-I detector adopts a compact forward-spectrometer configuration with a total length of approximately 3~m downstream of the target, as illustrated in Fig.~\ref{fig:phaseI}.

The incoming muon trajectory is measured by a beam tracking system located upstream of the target. The system consists of three silicon pixel tracking stations with an active area of $10\times10~\cm^2$ each, providing precise event-by-event determination of the beam position and incident angle.

The target system is positioned at the interaction point and serves as the scattering vertex for muon-proton interactions. The target region is designed in a modular configuration that allows rapid exchange among different target materials for physics measurements and dedicated background studies.

Downstream of the target, the scattered particles are reconstructed by a high-precision tracking system composed of five silicon tracking stations distributed between $z=5$ and 80\cm. To match the increasing angular spread of the scattered particles, the active area of the tracking detectors gradually increases along the beam direction from $12\times12\cm^2$ to $30\times30\cm^2$. This geometry provides efficient acceptance for scattering angles in the range of approximately $1^\circ$-$10^\circ$, corresponding to the low-\qsq region most relevant for the proton-radius determination.

Momentum reconstruction is achieved using a warm dipole magnet positioned around the downstream tracking stations. The magnet provides a vertical magnetic field of approximately $0.5~\text{T}$ within an effective field volume of $0.9\times0.9\times0.5~m^3$, enabling precise measurement of the momentum of scattered particles.

At the downstream end of the spectrometer $z\approx90\cm$, a PID system consisting of an electromagnetic calorimeter and a hadronic calorimeter provides discrimination between muons, pions, and electrons. The electromagnetic and hadronic calorimeters cover active areas of approximately $48\times48\cm^2$ and $80\times80\cm^2$, respectively, matching the geometrical acceptance of the tracking system.

\subsubsection{Phase-II detector Concept}
\label{sec:phaseII_det}
Building upon the Phase-I proton-radius program, the Phase-II \lune detector is designed as a general-purpose fixed-target spectrometer for studies of nucleon structure with muon beams in the momentum range of 0.5-7.5\gevc. The detector extends the angular acceptance, momentum coverage, and PID capabilities of the Phase-I apparatus, enabling a broad physics program including elastic scattering, DIS, SIDIS, and exclusive processes.

A schematic view of the detector is shown in Fig.~\ref{fig:phaseII}. The detector adopts a forward-spectrometer configuration centered on a liquid-hydrogen target located inside a superconducting solenoid. From upstream to downstream, the system consists of a beam monitoring system, a target system, a central tracking detector, a barrel PID system, a dipole spectrometer, forward tracking stations, and forward PID detectors.

The target is positioned along the central axis of a superconducting solenoid, starting from the upstream entrance and extending to a predetermined axial location, where a magnetic field of approximately $2~\text{T}$ is provided. The solenoid has an inner diameter of about 0.8~m and a length of approximately 1.5~m, defining the central tracking volume. Charged particles emitted at moderate and large scattering angles are reconstructed within this magnetic field region, allowing precise momentum determination and charge identification.
Downstream of the solenoid, a dipole spectrometer with a magnetic field strength of approximately $0.5~\text{T}$ provides additional momentum analysis for forward-going particles. The dipole magnet covers an effective field volume of approximately and is separated from the solenoid by a short transition region to accommodate detector services and support structures.

The tracking system combines central and forward detectors to provide nearly continuous charged-particle reconstruction over the full acceptance. The central tracker is optimized for particles produced at large angles, while the forward tracking stations are designed for the high-rate forward region where most DIS and elastic scattering products are concentrated.
PID is achieved through a combination of barrel and forward detector systems surrounding the tracking volume and positioned downstream of the spectrometer. Together they provide efficient separation of electrons, muons, pions, kaons, and protons over a broad momentum range.

Compared with the compact Phase-I detector, the Phase-II apparatus significantly increases the geometrical acceptance and extends the accessible kinematic range in \qsq, Bjorken-$x$, and hadron momentum. The design philosophy emphasizes the use of mature detector technologies, modular construction, and maximum reuse of detector components developed during the Phase-I program.

\subsection{Detector Overview}

The individual components of the \lune Phase-I and Phase-II detector systems are described in the following sections. The parameters presented here represent the current conceptual design and serve as a baseline for detector development. Further optimization studies based on detailed MC simulations of elastic scattering, DIS, SIDIS, and exclusive processes will be performed to refine the detector geometry, evaluate alternative technologies, and establish the final detector configuration.

\subsubsection{Target System}

The target system provides a stable and well-controlled interaction environment for the \lune physics program. As shown in Table~\ref{tab:targets}, its design follows the staged development strategy of the experiment, supporting both the precision elastic scattering measurements of Phase-I and the high-luminosity DIS program envisioned for Phase-II.

\paragraph{Phase-I Solid Target Station}

As discussed in Sec.~\ref{sec:targets}, the baseline target configuration for Phase-I employs a solid-target station located at the beginning of the tracking system. The station is optimized for precision measurements of elastic muon scattering, where strict control of target-related systematic uncertainties is essential.

\begin{figure}[!htbp]
    \centering
    \includegraphics[width=0.4\linewidth]{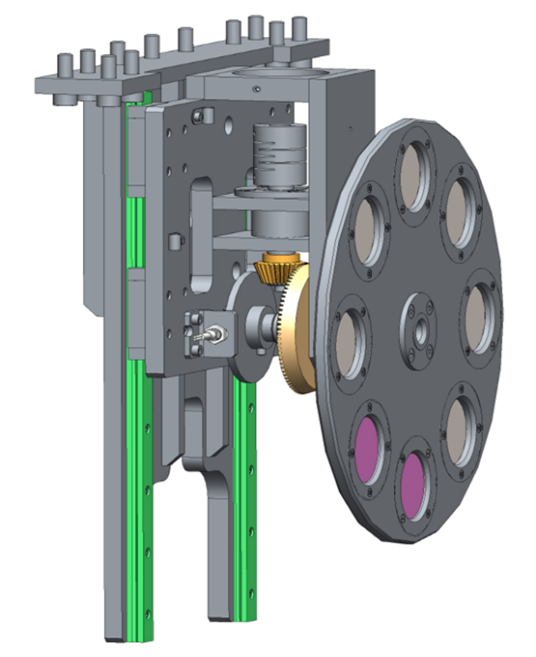}
    \includegraphics[width=0.3\linewidth]{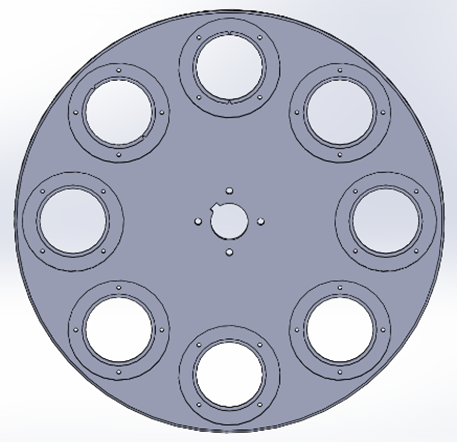}
    \caption{Conceptual design of the modular target station. (left) CAD model of the target station equipped with a motor-driven rotating target wheel. (right) front view of the wheel, which accommodates up to eight interchangeable target positions for rapid switching between different target materials.}
    \label{fig:target_station}
\end{figure}

As shown in Fig.~\ref{fig:target_station}, a modular target wheel allows rapid switching between multiple target materials during data taking. Such capability enables relative cross-section measurements under identical beam and detector conditions, thereby reducing systematic uncertainties associated with detector acceptance, beam intensity variations, and long-term stability.
The target exchange mechanism is based on a mature rotating-target technology developed at IMP. The system provides micrometer-level positioning precision and excellent reproducibility, ensuring accurate alignment with respect to the beam axis. Also, the target thicknesses should be controlled at the level of better than 1\%.

\paragraph{Phase-II Cryogenic Target System}

To fully exploit the luminosity potential of the \hiaf muon beam and enable a comprehensive nucleon-structure program, a cryogenic LH$_2$ and LD$_2$ target system is foreseen for Phase-II operation. Two target configurations are currently under consideration. A short target with a length of approximately 5~cm is optimized for precision measurements of nucleon electromagnetic form factors and radius extractions, where minimizing vertex smearing and multiple-scattering effects is particularly important. A longer target with a length of approximately 50~cm is designed to maximize luminosity for DIS, SIDIS, and exclusive reactions such as DVCS.

The cryogenic target infrastructure will also establish a flexible platform for future upgrades. In particular, the integration of polarized targets is under consideration as a long-term development toward a comprehensive spin-physics program. Polarized solid-state proton and deuteron targets based on dynamic nuclear polarization (DNP), which have been successfully employed in several high-energy scattering experiments~\cite{Tateishi:2025vma}, represent a promising option. More target options are also discussed in the H-NS white paper~\cite{Bai:2026syt}.

\subsubsection{Magnet System}

The magnet system is designed to provide precise momentum measurements for charged particles over a broad momentum range while maintaining excellent angular reconstruction capability at low momentum transfer. The baseline configuration employs a warm dipole spectrometer magnet for Phase-I operation, while a large-aperture solenoidal magnet is foreseen for the Phase-II physics program, as summarized in Table~\ref{tab:magnet}. 

\begin{table}[!htbp]
\centering
\caption{Main parameters of the \lune magnet system.}
\label{tab:magnet}
\begin{tabular}{lcc}
\hline
Parameter & Phase-I/II Dipole & Phase-II Solenoid \\
\hline

Magnet type& Warm dipole& Solenoidal magnet \\
Central field& 0.5 T& 2.0 T \\
Field direction& $B_y$& $B_z$ \\
Magnetic length& 50 cm& 150 cm \\
Inner aperture& $90 \times 90\times 50$ cm$^3$& 80 cm diameter \\
Iron yoke height& 30 cm& 20 cm radial thickness \\
Field uniformity& $<0.5\%$& TBD \\
Position relative to target& Downstream& Near upstream entrance \\
\hline
\end{tabular}
\end{table}

\paragraph{Dipole Magnet}

For the Phase-I program, momentum analysis is performed using a high-uniformity warm dipole magnet located downstream of the tracking detectors. The magnet is optimized for precise reconstruction of scattered-particle momenta in the 0.5-2.0\gevc momentum range while preserving acceptance for small-angle scattering events.

\begin{figure}[!htbp]
    \centering
    \includegraphics[width=0.7\linewidth]{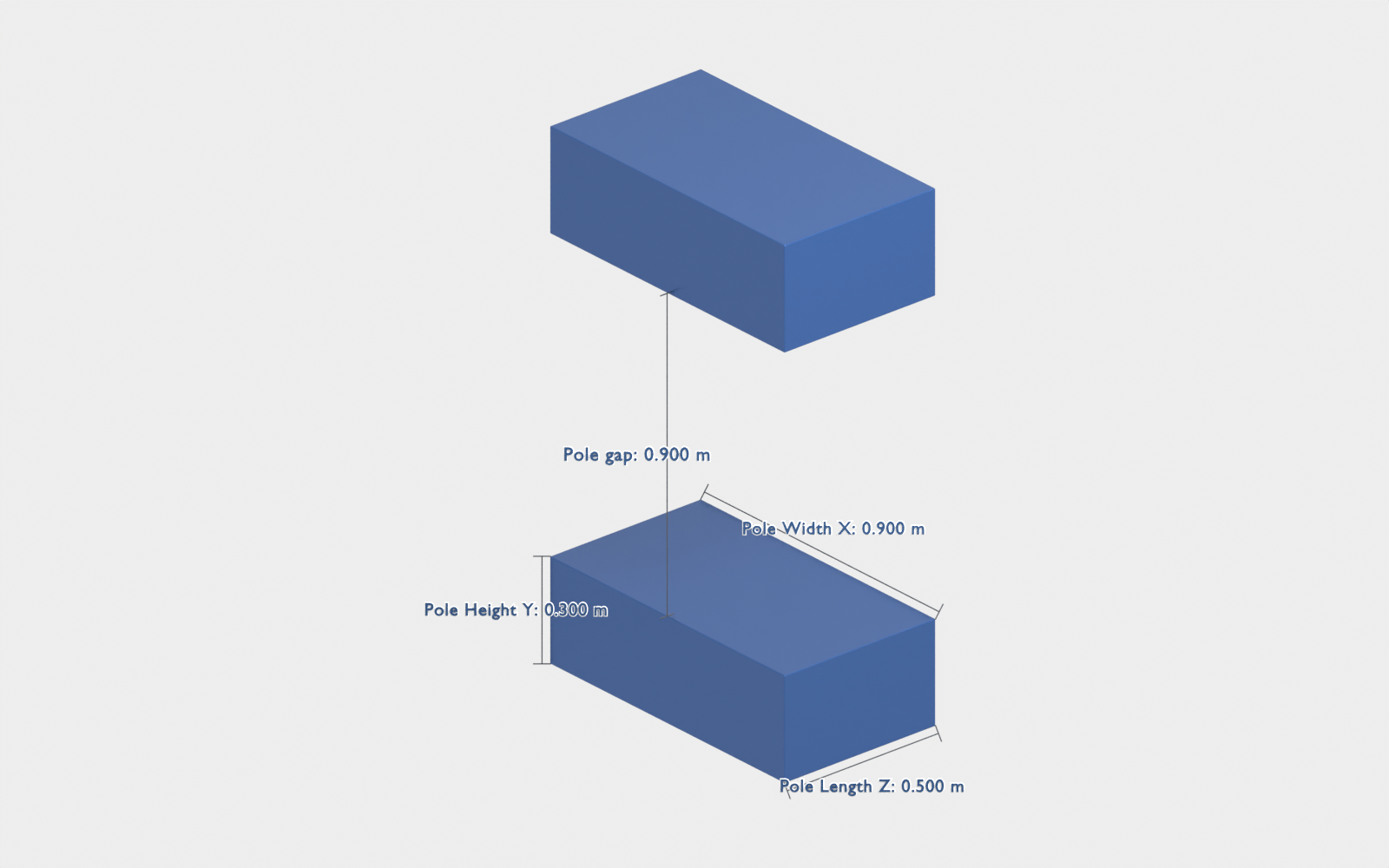}
    \caption{Conceptual layout of the Phase-I/II dipole magnet, showing the magnet geometry.}
    \label{fig:dipole}
\end{figure}

As shown in Fig.~\ref{fig:dipole}, the effective magnetic volume is approximately $0.9 \times 0.9 \times 0.5~\mathrm{m}^3$ (length $\times$ height $\times$ width), with a field uniformity better than $\pm0.5\%$ throughout the tracking region. Such field quality is required to minimize systematic effects in momentum reconstruction and to ensure long-term stability of the spectrometer response.
The magnetic circuit is based on high-permeability soft-magnetic iron and employs a monolithic yoke structure to maximize magnetic-field uniformity and mechanical rigidity. Field-shaping techniques, including optimized pole profiles and correction shims, are incorporated to further improve field homogeneity and suppress edge-field distortions. The operating magnetic flux density is chosen to avoid saturation effects and preserve a highly linear relationship between track curvature and particle momentum.

The dipole magnet works in conjunction with the downstream silicon tracking system. The first tracking layers measure the scattering angle in a field-free region, while the final tracking stations record the trajectory deflection within the magnetic field. This configuration enables simultaneous determination of particle angle and momentum while minimizing systematic uncertainties in low-\qsq measurements.

\begin{figure}[!htbp]
    \centering
    \includegraphics[width=0.95\linewidth]{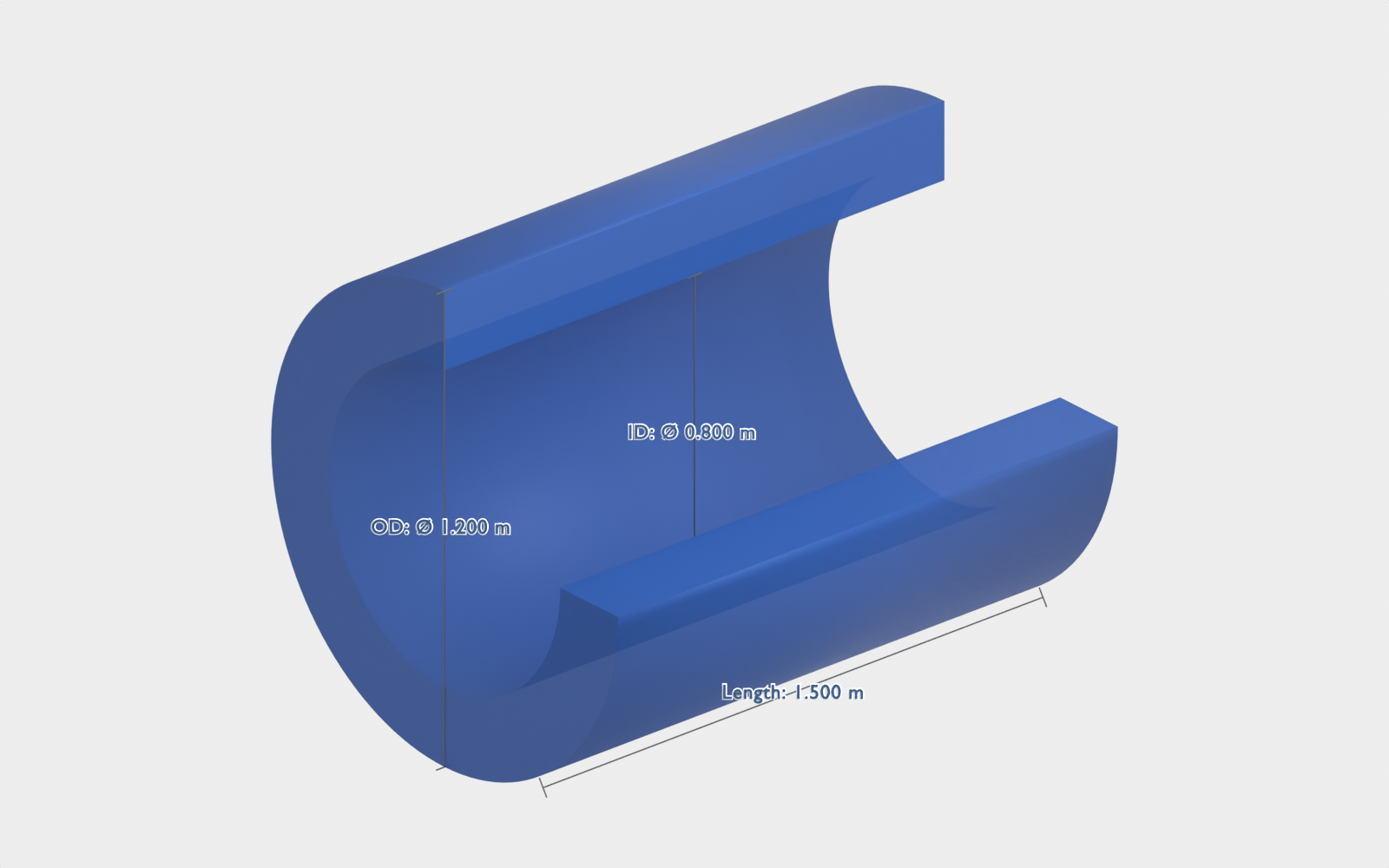}
    \caption{Conceptual layout of the Phase-II solenoid magnet.}
    \label{fig:solenoid}
\end{figure}

\paragraph{Solenoidal Magnet for Phase-II}
To support the expanded Phase-II physics program, \lune foresees the installation of a large-aperture superconducting solenoidal magnet surrounding the target and central tracking detectors. The conceptual design features an inner diameter of approximately 80~cm, a radial thickness of about 20~cm, and an overall length of approximately 1.5~m, as shown in Fig.~\ref{fig:solenoid}. The solenoid technology is based on developments for the H-NS project~\cite{Bai:2026syt} at \hiaf.

The superconducting solenoid is designed to deliver an approximately 2~T axial magnetic field with high uniformity across the full tracking acceptance. 
The solenoidal configuration further enables a compact and integrated detector layout, allowing the installation of barrel tracking detectors within the magnetic volume. 

\subsubsection{Silicon Pixel Tracking System}
The silicon pixel tracking system constitutes the primary charged-particle tracking detector of the \lune experiment, providing precise measurements of particle trajectories throughout the experimental acceptance. The main geometrical parameters of the tracking detectors, including their positions and active dimensions, are summarized in Table~\ref{tab:tracking}.
To minimize multiple-scattering effects and maintain excellent tracking performance at low momentum transfer, \lune adopts a unified tracking concept based on Monolithic Active Pixel Sensor (MAPS) technology for both the upstream beam tracker and the downstream scattering tracker.

\begin{table*}[htbp]
\centering
\caption{Main parameters of the beam monitor and scattering tracking system.}
\label{tab:tracking}
\begin{tabular}{lllll}
\hline
Subsystem &Layer &Position ($z$) &Active size &Purpose \\
\hline
Beam monitor& 1-3& upstream & $10\times10$ cm$^2$& Beam trajectory \\ \hline
Phase-I scattering& 1& 5 cm& $12\times12$ cm$^2$& Angle \\
& 2& 20 cm& $12\times12$ cm$^2$& Angle \\
& 3& 35 cm& $16\times16$ cm$^2$& Angle \\
& 4& 50 cm& $20\times20$ cm$^2$& Momentum \\
& 5& 80 cm& $30\times30$ cm$^2$& Momentum \\\hline
Phase-II tracker& 1& 10 cm& $R=10$ cm& Vertexing \\
& 2& 60 cm& $R=15$ cm& Track seeding \\
& 3& 90 cm& $R=15$ cm& Track seeding \\
& 4& 120 cm& $R=19$ cm& Momentum \\
& 5& 150 cm& $R=25$ cm& Momentum \\
& 6& 180 cm& $70\times70$ cm$^2$& Forward matching \\
& 7& 205 cm& $80\times80$ cm$^2$& Forward matching \\
& 8& 232 cm& $88\times88$ cm$^2$& Forward matching \\\hline
\end{tabular}
\end{table*}

\begin{figure}[!htbp]
    \centering
    \includegraphics[width=0.7\linewidth]{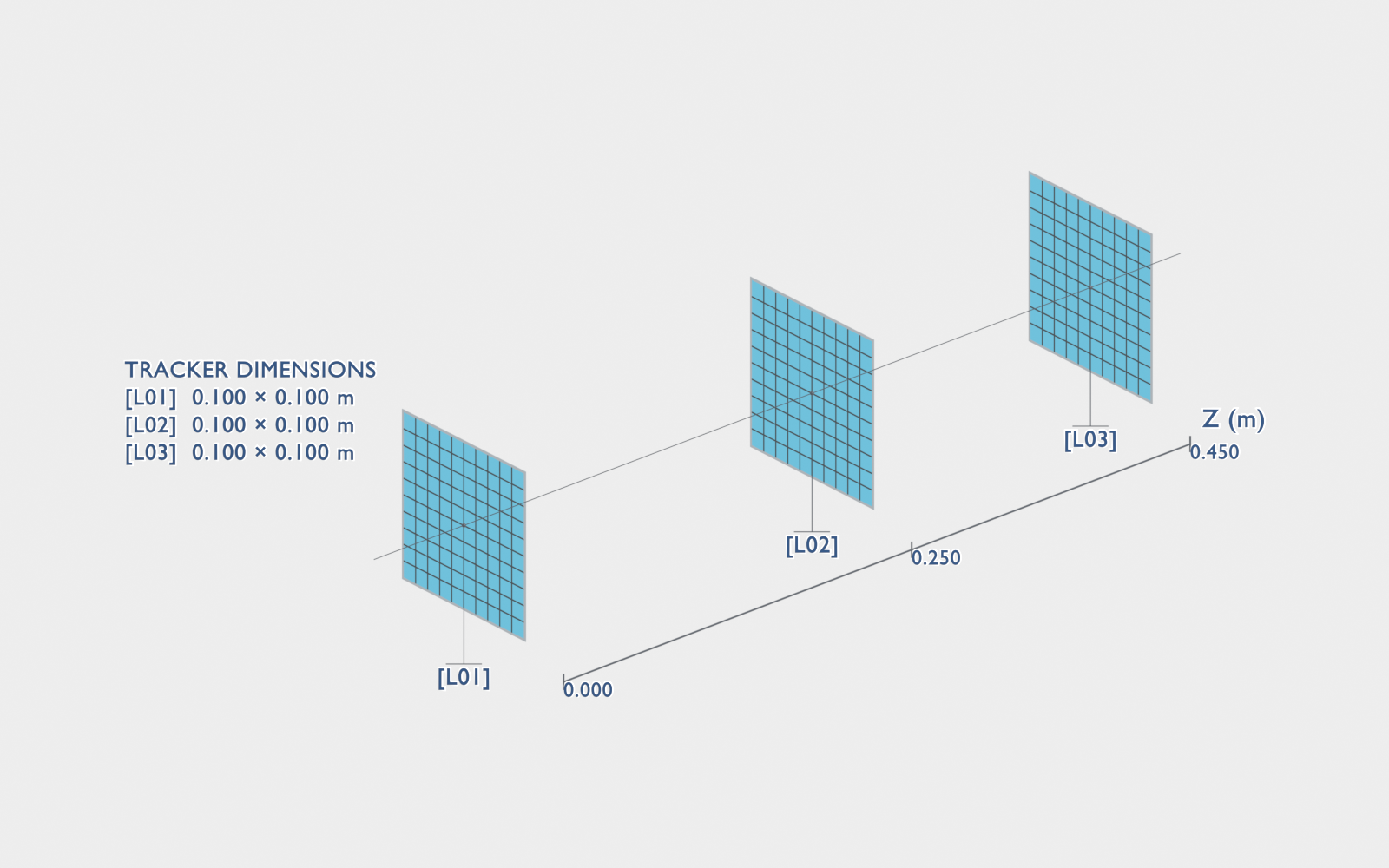}
    \caption{Conceptual layout of the silicon pixel beam-monitoring system upstream of the target.}
    \label{fig:beam_monitor}
\end{figure}

\paragraph{Beam Monitor detector} The beam monitor tracking system is installed upstream of the target and consists of three high-resolution silicon pixel detector planes, as shown in Fig.~\ref{fig:beam_monitor}. Its primary purpose is to provide event-by-event measurements of the position and direction of the incident muon beam before it enters the target volume.

Using multi-plane track fitting, the system reconstructs the incoming muon trajectory with high precision, providing the spatial coordinates and flight direction of each beam particle. These measurements serve as essential inputs to the scattering-vertex reconstruction and significantly improve the determination of the scattering kinematics. The precise track extrapolation capability also allows efficient identification and rejection of beam-halo particles as well as background events originating from interactions in non-target materials, such as detector supports and surrounding structures.

In addition to its physics role, the beam tracker serves as a continuous beam diagnostic tool. It provides real-time monitoring of the beam profile, beam position stability, angular divergence, and intensity fluctuations, thereby supplying valuable feedback for accelerator operation and beam-quality control.

\paragraph{Scattering Tracking System} 
The scattering tracking system provides precision measurements of charged particles emerging from the target and constitutes the primary detector for the reconstruction of scattering kinematics. 
The detector concept follows the staged development strategy of the experiment. In Phase-I, the tracking system is optimized for precision measurements of small-angle elastic scattering events. In Phase-II, the tracking configuration is extended to support the larger acceptance and more complex final states associated with DIS, SIDIS, and exclusive reactions.

\begin{figure}[!htbp]
    \centering
    \includegraphics[width=0.7\linewidth]{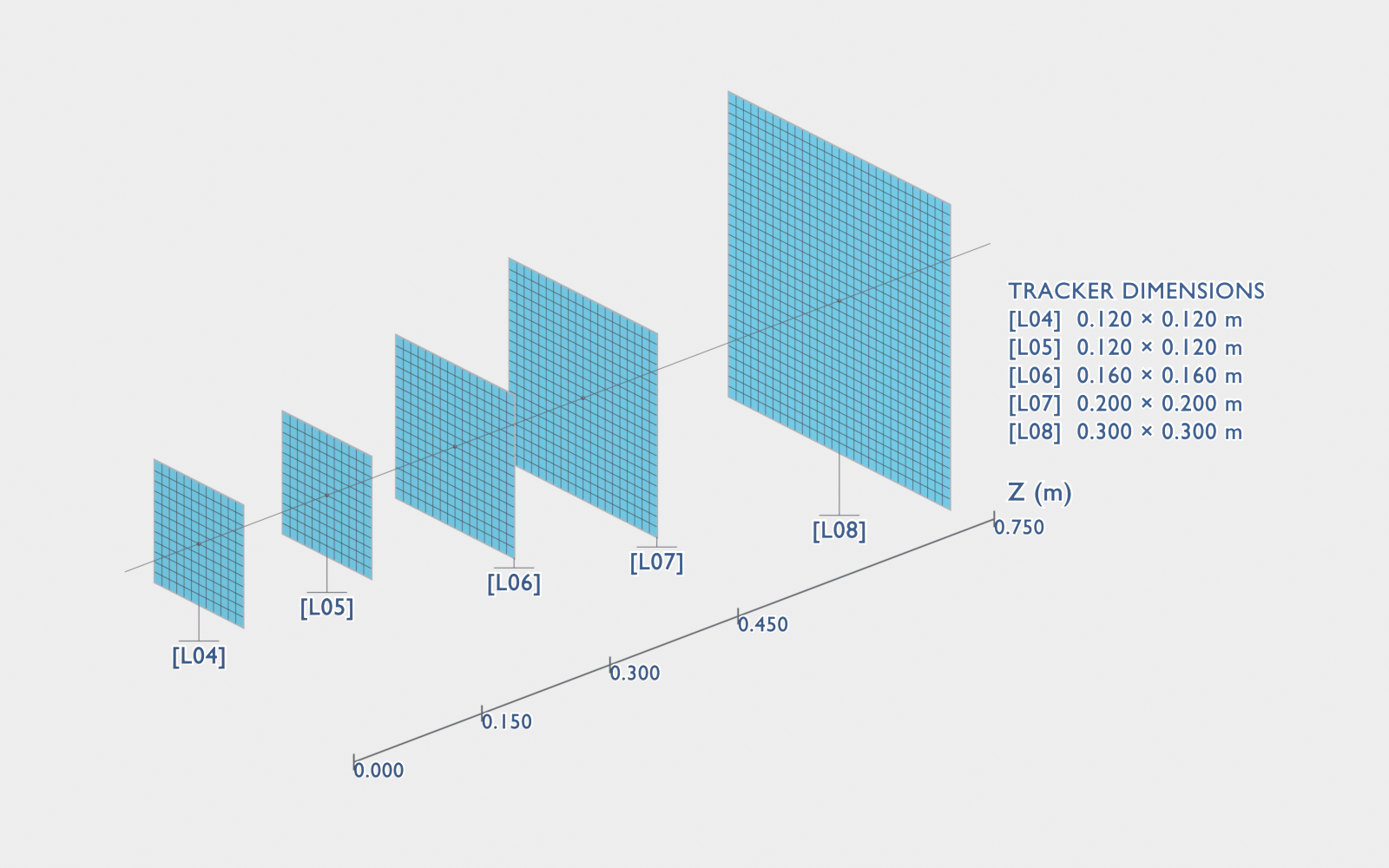}
    \caption{Phase-I silicon tracking system showing the five tracking stations used for scattering-angle and momentum reconstruction.}
    \label{fig:scattering_tracker}
\end{figure}

\subparagraph{Phase-I Configuration} The Phase-I scattering tracker consists of five silicon pixel detector stations positioned between approximately 5 and 80~cm downstream of the target, as shown in Fig.~\ref{fig:scattering_tracker}. To maintain high acceptance for particles scattered at angles between approximately $1^\circ$ and $10^\circ$, the active detector area increases progressively with distance from the target. The five tracking stations have active areas of $12\times12$, $12\times12$, $16\times16$, $20\times20$, and $30\times30~\mathrm{cm}^2$, respectively.

The first three detector layers are located upstream of the dipole spectrometer and provide a precise measurement of the initial scattering angle. The final layers are positioned inside the magnetic field region and measure the trajectory deflection induced by the Lorentz force. By combining hit information from all five tracking stations, the system simultaneously reconstructs the scattering angle and momentum of the outgoing particle.

\subparagraph{Phase-II Upgrade}
For the Phase-II program, the tracking system is expanded to operate in conjunction with the superconducting solenoid and central drift chamber. Several additional silicon pixel tracking layers are installed along the beam direction to provide precise space points for vertex determination, track seeding, and momentum reconstruction.

Within the solenoidal magnetic volume, four circular silicon tracking stations are positioned at approximately 10, 60, 90, and 120~cm downstream of the target, with active radii ranging from 10 to 19~cm. The cylindrical geometry matches the acceptance of the central drift chamber and provides uniform coverage in azimuth.

A larger circular tracking layer with a radius of approximately 25~cm is placed near the downstream end of the solenoid to improve momentum resolution by extending the tracking lever arm. Downstream of the solenoid, three additional planar silicon tracking stations are installed at approximately 180, 205, and 232~cm from the target. These stations provide precise matching between tracks reconstructed in the central detector and particles entering the forward spectrometer.

\paragraph{MAPS Technology Choice} 
Both tracking subsystems are based on MAPS technology, which has reached a high level of maturity through successful deployment in several modern particle-physics experiments, including the ALICE Collaboration Inner Tracking System upgrade~\cite{Reidt:2021tvq} and the STAR Collaboration Heavy Flavor Tracker~\cite{Qiu:2014dha}.
Compared with conventional hybrid pixel detectors and gaseous tracking systems, MAPS sensors integrate the sensing and readout electronics within a single silicon substrate. This architecture enables an exceptionally low material budget while maintaining excellent spatial resolution. Detector layers with thicknesses below 100\mum are routinely achievable, while intrinsic position resolutions better than 5\mum can be obtained with pixel pitches of approximately 30\mum.

The low material budget is particularly important for \lune, where multiple scattering represents one of the dominant sources of systematic uncertainty in low-\qsq measurements. To suppress multiple-scattering effects and achieve the angular precision required for low-\qsq scattering measurements, the material budget of each tracking layer is kept below 0.5\% of a radiation length ($X_0$). At the same time, the fine pixel granularity provides good hit resolution, excellent pattern-recognition capability, and strong pileup tolerance under high-rate beam conditions. The baseline design adopts MAPS technology developed for the H-NS experiment~\cite{Bai:2026syt}, specifically the MIC6-v4, the fourth-generation MAPS chip developed by Central China Normal University (CCNU). 

To maximize detector uniformity and simplify system integration, the beam tracker and scattering tracker share a common MAPS sensor technology, front-end readout architecture and mechanical support concept. This unified approach reduces development risk, simplifies calibration procedures, and ensures consistent tracking performance throughout the experiment.

\begin{figure}[!htbp]
    \centering
    \includegraphics[width=0.95\linewidth]{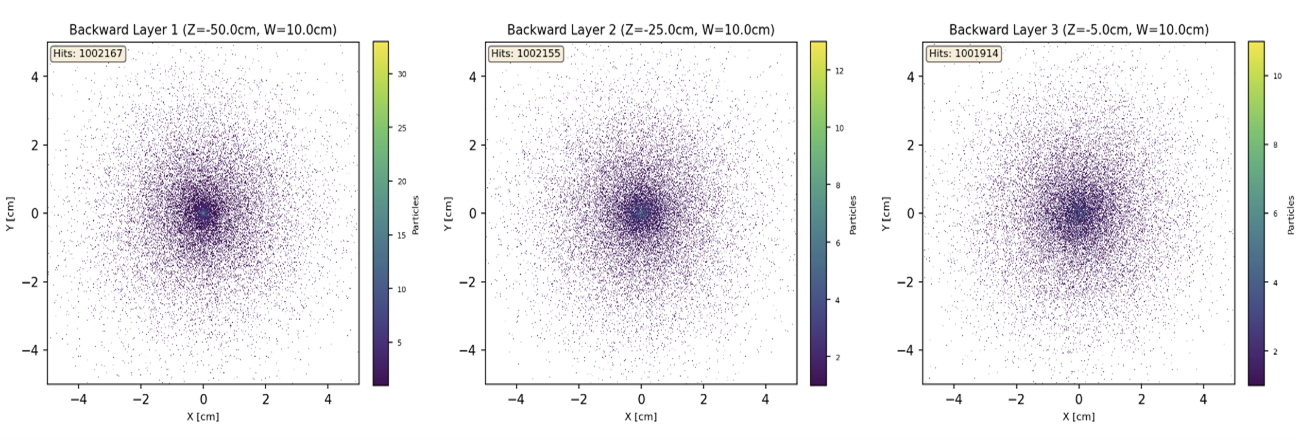}
    \caption{Occupancy of the three-layer silicon pixel detectors under given beam conditions.}
    \label{fig:beam_rate}
\end{figure}

\paragraph{Occupancy and Rate Capability} Detailed simulations have been performed to evaluate detector occupancy under the nominal \hiaf muon-beam conditions. Assuming a beam intensity of $10^6 \mu/s$, a beam spot size of approximately 2~cm, and a beam divergence below 10 mrad, the occupancy of the three-layer beam tracking system was studied using MAPS sensors with a pixel size of 30\mum, as shown in Fig.~\ref{fig:beam_rate}.

The simulation indicates that the maximum hit rate per pixel is approximately 40 Hz. Given a typical MAPS readout integration time on the order of several hundred microseconds ($\mu s$), the resulting occupancy remains far below the saturation limit of the readout system. Stable detector operation is therefore expected under the nominal \lune running conditions.
Furthermore, substantial performance margins remain available in both occupancy and readout bandwidth. The detector concept is expected to remain viable for future \hiaf upgrades, including operation with muon beam intensities approaching $10^8~\mu/\mathrm{s}$, thereby providing a clear path toward higher-luminosity physics programs.

\subsubsection{Draft Chamber}
For the Phase-II physics program, a cylindrical multi-wire drift chamber (CDC) is foreseen as the principal large-volume tracking detector inside the superconducting solenoid. The chamber complements the silicon pixel tracking system by providing a large number of spatial measurements with a minimal material budget, thereby improving momentum resolution and tracking efficiency for charged particles emitted at intermediate and large polar angles.

The CDC is located downstream of the target and surrounds the forward region covered by the silicon tracking detectors. Operating inside the 2~T solenoidal magnetic field, it measures the curvature of charged-particle trajectories and provides additional tracking information for momentum reconstruction. Together with the silicon pixel detectors, the chamber forms a hybrid tracking system capable of reconstructing charged particles over a broad angular and momentum range.

\begin{figure}[!htpb]
    \centering
    \includegraphics[width=0.8\linewidth]{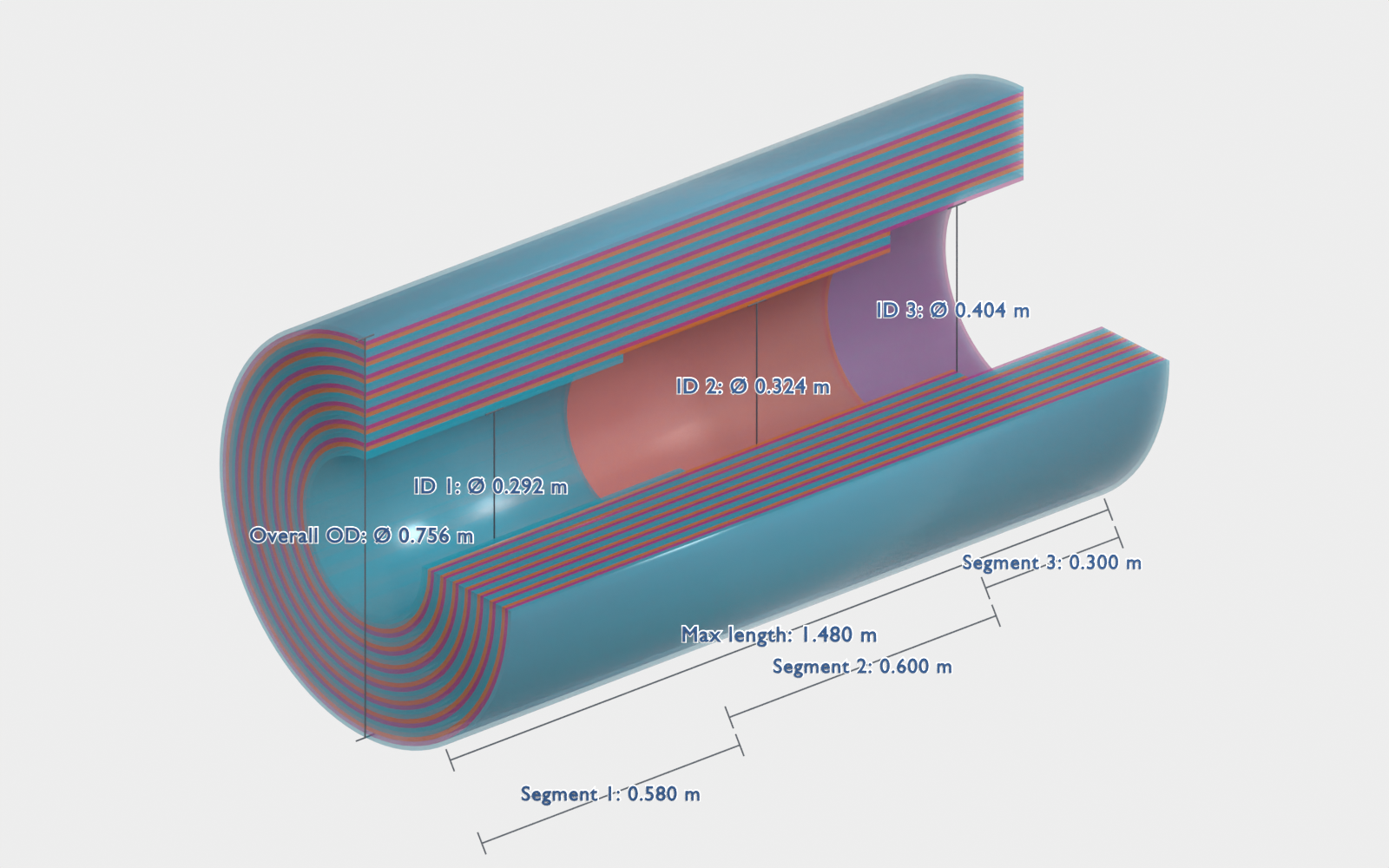}
    \caption{Conceptual layout of the Phase-II central drift chamber (CDC).}
    \label{fig:MDC}
\end{figure}

The baseline design consists of twelve cylindrical drift-chamber layers with radii ranging from approximately 15.0~cm to 37.4~cm. The inner layers employ axial wire configurations to provide precise measurements in the transverse plane, while the outer layers utilize stereo wire arrangements to determine the longitudinal coordinate and enable full three-dimensional track reconstruction.
The chamber length gradually increases with radius, providing efficient acceptance matching to the forward geometry of the experiment. The active volume extends from approximately 1.0~m to 1.5~m in length, corresponding to the expected distribution of charged-particle trajectories emerging from the target region.
The detector is operated with an Ar-CO$_2$ (70:30) gas mixture, providing stable operation, low multiple scattering, and good spatial resolution. Based on experience from modern drift-chamber systems, a transverse position resolution of approximately 100-150~$\mu$m is expected~\cite{BESIII:2009fln,Achasov:2023gey}, while the stereo layers provide longitudinal coordinate measurements at the millimeter level. The main design parameters of the CDC are summarized in Table~\ref{tab:cdc}, while its conceptual layout is illustrated in Fig.~\ref{fig:MDC}.

\begin{table}[!htbp]
\centering
\caption{Main parameters of the Phase-II central drift chamber.}
\label{tab:cdc}
\begin{tabular}{lc}
\hline
Parameter & Value \\
\hline
Detector type & Cylindrical drift chamber \\
Number of layers & 29 \\
Layer configuration &
15 axial + 7 stereo$+$ + 7 stereo$-$ \\
Inner radius & 15.0 cm \\
Outer radius & 37.4 cm \\
Radial spacing & 0.8 cm \\
Active length & 1.0-1.5 m \\
Stereo angle & $\pm5^\circ$ \\
Gas mixture & Ar-CO$_2$ (70:30) \\
Transverse resolution & 100-150 $\mu$m \\
Longitudinal resolution & $\sim1$ mm \\
\hline
\end{tabular}
\end{table}

\subsubsection{Calorimeter System}
\begin{figure}[!htbp]
    \centering
    \includegraphics[width=0.9\linewidth]{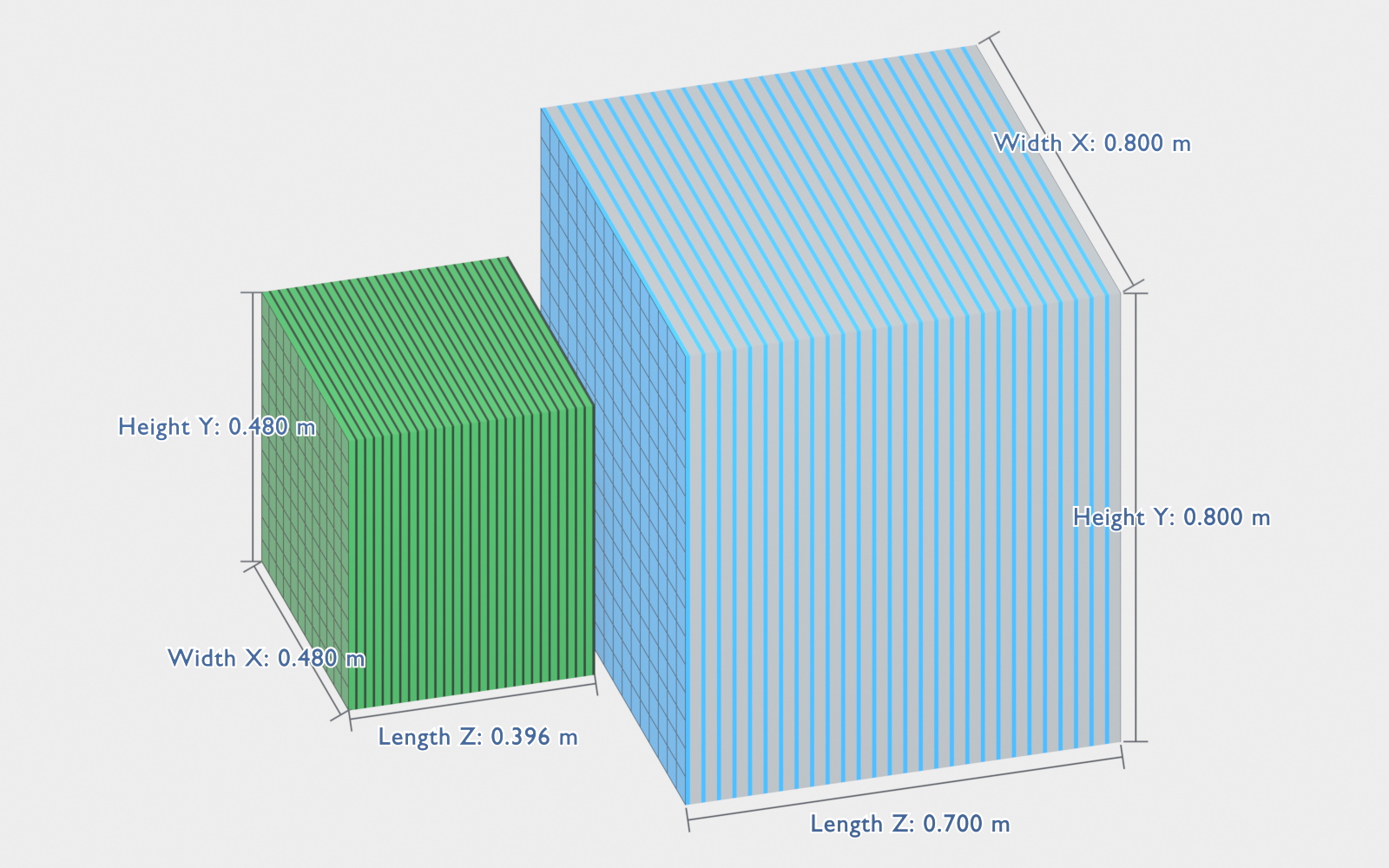}
    \caption{Schematic diagram of the \lune Phase-I calorimeter. The detector consists of an electromagnetic calorimeter followed by a hadronic calorimeter.}
    \label{fig:calo_PhaseI}
\end{figure}

The calorimeter system provides energy measurements, PID, and event-topology reconstruction for the \lune physics program. In combination with the tracking detectors, it enables the identification of electrons, photons, muons, and hadrons, and plays a central role in the reconstruction of exclusive and semi-inclusive final states.
The calorimeter concept follows the staged development strategy of the experiment. During Phase-I, a forward calorimeter system is optimized for precision measurements of elastic and quasi-elastic scattering at small angles, as shown in Fig.~\ref{fig:calo_PhaseI}. In Phase-II, the calorimeter acceptance is significantly expanded to support DIS, SIDIS, and exclusive reactions involving multi-particle final states.

Both phases employ sampling calorimeters based on plastic scintillators, wavelength-shifting fibers, and silicon photomultiplier (SiPM) readout. This technology offers a robust combination of high detection efficiency, compact mechanical design, and scalability for future upgrades. More details on the calorimeter are shown in Table~\ref{tab:calorimeter} and~\ref{tab:calo_segmentation}.

\begin{figure}[!htbp]
    \centering
    \includegraphics[width=0.95\linewidth]{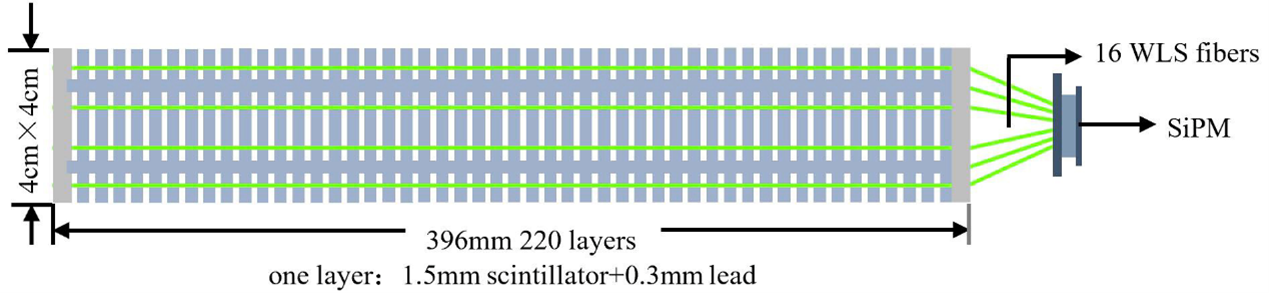}
    \caption{Schematic diagram of the \lune EM tower, consisting of 220 layers. Each layer comprises a 1.5 mm plastic scintillator and a 0.3 mm lead absorber.}
    \label{fig:em_tower}
\end{figure}

\paragraph{Electromagnetic Calorimeter}
The electromagnetic calorimeter is based on the Shashlyk technology developed for the MPD experiment at NICA~\cite{Shen:2019ibk,Li:2020mjj}. As shown in Fig.~\ref{fig:em_tower}, the baseline tower design consists of 220 alternating layers of 1.5-mm plastic scintillator and 0.3-mm lead absorber plates, with scintillation light collected by 16 wavelength-shifting fibers and read out by SiPMs. 
The transverse dimensions of a tower are approximately $4\times4~\mathrm{cm}^2$, and the total depth corresponds to approximately $11.8\,X_0$. The effective Moli\`ere radius is about 6.2~cm.
An energy resolution of approximately 5\% for 1-GeV electromagnetic particles and a timing resolution of about 500~ps are expected. These characteristics provide sufficient performance for electron and photon reconstruction, neutral-cluster identification, and background rejection in exclusive reaction channels.

The same tower design is adopted for both Phase-I and Phase-II, as shown in Table~\ref{tab:calorimeter}. The Phase-II upgrade increases the transverse acceptance by enlarging the number of calorimeter towers while preserving the proven module design and readout architecture, as shown in Table~\ref{tab:calo_segmentation}.

\paragraph{Hadronic Calorimeter}
A sampling hadronic calorimeter (HCAL) will be installed downstream of the ECAL to measure hadronic shower energy and improve particle-identification capabilities. Each tower consists of alternating layers of plastic scintillator and iron absorber plates. Four wavelength-shifting fibers collect scintillation light and transport it to SiPM-based readout electronics.

The total tower length is approximately 80~cm, providing substantial containment of hadronic showers in the momentum range relevant to the \lune experiment. The HCAL tower has transverse dimensions of $5\times 5 \cm^2$ and consists of 56 alternating layers of 0.25 cm plastic scintillator and 1.0 cm iron absorber plats. The total depth corresponds to approximately 3.5 interaction lengths ($\lambda_I$). The modular tower structure allows straightforward scaling of the calorimeter acceptance between the Phase-I and Phase-II detector configurations.
The forward HCAL primarily serves the reconstruction of forward-going hadrons produced in DIS and exclusive reactions.

\paragraph{Barrel Calorimeter}

To support the broad angular acceptance required for the Phase-II physics program, a barrel calorimeter is installed outside the superconducting solenoid and surrounding the central tracking detectors.

The barrel calorimeter employs a lead-scintillator sampling structure arranged in a projective geometry around the beam axis. The detector consists of eight alternating layers of plastic scintillator and lead absorber material. The calorimeter is segmented into 60 azimuthal sectors and 15 longitudinal sections, providing nearly uniform coverage over the full azimuthal acceptance.

Operating in conjunction with the central drift chamber and silicon tracking detectors, the barrel calorimeter serves primarily as a PID device for hadrons and muons emerging at intermediate and large polar angles. Owing to the solenoidal magnetic field, electrons and photons are contained within the inner detector volume and do not reach the barrel calorimeter. This PID capability is particularly crucial for SIDIS and exclusive reactions, where final-state hadrons and muons populate a broad angular range. 

The barrel calorimeter significantly extends the acceptance of the detector beyond the forward spectrometer and represents a key component of the Phase-II upgrade.

\paragraph{Particle Identification}
The combined ECAL and HCAL systems provide complementary information on energy deposition, shower topology, and longitudinal shower development. Together with momentum measurements from the tracking system, these observables enable efficient discrimination among electrons, photons, muons, and hadrons.

Electromagnetic particles produce compact showers largely contained within the ECAL, while hadrons generate broader cascades extending into the HCAL. Muons behave predominantly as minimum-ionizing particles and deposit only limited energy in both calorimeters. These characteristic response patterns provide a robust basis for PID and background suppression throughout the \lune physics program.

\begin{table*}[!htbp]
\centering
\caption{Main parameters of the calorimeter system in Phase-I and Phase-II.}
\label{tab:calorimeter}
\begin{tabular}{lccc}
\hline
Parameter &Phase-I ECAL &Phase-I HCAL &Phase-II Barrel \\ \hline
Technology& Shashlyk& Sampling HCAL& Sampling HCAL \\
Active material& Plastic scintillator& Plastic scintillator& Plastic scintillator \\
Absorber& Pb& Fe& Pb \\
Number of layers& 220& 56& 8 \\
Tower size& $4\times4$ cm$^2$& $5\times5$ cm$^2$& Projective towers \\
Readout (+ SiPM)& 16 WLS fibers & 4 WLS fibers & WLS fibers\\
Longitudinal depth& 396 mm& 700 mm& 560 mm \\
Energy resolution& $\sim8\%$ @ 1 GeV& --& TBD \\
Timing resolution& $\sim500$ ps& --& TBD \\
\hline
\end{tabular}
\end{table*}

\begin{table}[!htbp]
\centering
\caption{Calorimeter segmentation in the staged detector configurations.}
\label{tab:calo_segmentation}
\begin{tabular}{lcc}
\hline
Subsystem &Phase-I &Phase-II \\\hline
Forward ECAL& $12\times12$ towers& $24\times24$ towers \\
Forward HCAL& $16\times16$ towers& $56\times56$ towers \\
Barrel calorimeter& --& $60\times15$ towers \\
\hline
\end{tabular}
\end{table}

\subsubsection{Forward Muon Identification System}
The forward muon identification system is designed to enhance the separation of muons from charged hadrons in the forward spectrometer. Although the hadronic calorimeter provides substantial absorption of strongly interacting particles, a fraction of pions may either pass through the calorimeter without undergoing significant hadronic interactions or interact only in the final layers of the absorber. Such particles can mimic the response of penetrating muons and therefore constitute a potential source of background in physics channels involving muons.

To improve the discrimination between muons and hadrons, a dedicated Resistive Plate Chamber (RPC)-based~\cite{Santonico:1981sc} muon identification system~\cite{Abbrescia:2003bn,Ban:2005zxa} will be installed downstream of the forward hadronic calorimeter. 
The detector exploits the different penetration characteristics of muons and hadrons after passage through the calorimeter system. Muons are expected to pass through the absorber stack and produce correlated hits in multiple RPC layers, while most hadrons are absorbed or undergo secondary interactions before reaching the final detector layers.

As summarized in Table~\ref{tab:rpc}, the baseline design consists of eight RPC tracking layers interleaved with iron absorber plates. Each RPC layer comprises a 2 mm gas gap enclosed between two 2 mm thick insulating electrode plates. The detector is operated with an Ar-CO$_2$ gas mixture and provides fast timing signals together with high detection efficiency for minimum-ionizing particles.

Between adjacent RPC layers, 20 mm thick iron absorber plates are inserted to further suppress hadronic punch-through. Additional spacing between layers improves pattern recognition and track matching with the upstream tracking detectors and calorimeter system. The resulting multilayer configuration provides robust muon identification while maintaining a compact detector footprint.

\begin{table}[!htbp]
\centering
\caption{Main parameters of the forward RPC-based muon identification system.}
\label{tab:rpc}
\begin{tabular}{lc}
\hline
Parameter & Value \\
\hline
Detector technology & RPC \\
Number of RPC layers & 8 \\
Gas gap thickness & 2 mm \\
Electrode thickness & 2 mm \\
Gas mixture & Ar-CO$_2$ \\
Absorber material & Iron \\
Absorber thickness & 20 mm \\
Inter-layer spacing & 30 mm \\
Readout area (Phase-I) & $100\times100$ cm$^2$ \\
Readout area (Phase-II) & TBD  \\
Location & Downstream of HCAL \\
\hline
\end{tabular}
\end{table}

\subsubsection{Data Acquisition System}
The data acquisition (DAQ) system could provides high-throughput readout, system synchronization, and slow-control management for all detector subsystems. It is designed to operate in a continuous, triggerless readout mode, reflecting the moderate event rates expected in muon-nucleus scattering and enabling maximal efficiency for both inclusive and exclusive physics channels.

The DAQ architecture follows the modern PCIe-based high-bandwidth paradigm adopted in several contemporary high-energy physics experiments, including ATLAS (FELIX)~\cite{Anderson:2015zsi}, ALICE (CRU)~\cite{Boeschoten:2017aff}, and sPHENIX. In this framework, front-end electronics are directly interfaced with high-speed optical links and PCIe-based readout boards, allowing for scalable, modular, and detector-agnostic data transport. The system naturally supports integration of multiple sub-detectors through a unified hardware and software architecture.

\begin{figure}[!htbp]
    \centering
    \includegraphics[width=0.9\linewidth]{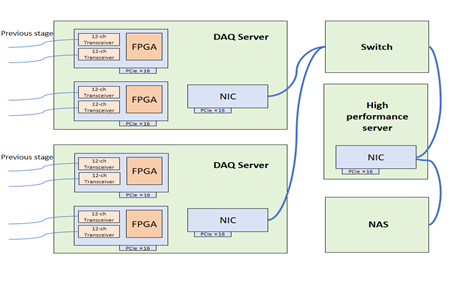}
    \caption{Schematic layout of the \lune DAQ architecture.}
    \label{fig:daq}
\end{figure}

As shown in Fig.~\ref{fig:daq}, in the \lune implementation, detector data are transmitted from front-end electronics via high-speed optical links to FPGA-based PCIe readout units, which perform initial data formatting and buffering. Data are then transferred through high-throughput DMA engines to commodity servers equipped with high-performance network interfaces. Event building, data reduction, and real-time monitoring are carried out on a computing farm, with data ultimately stored in a distributed storage system.

A dedicated timing and synchronization system distributes a global clock and control signals to all detector subsystems, ensuring coherent time stamping across the experiment. The design targets a synchronization precision at the level of $\mathcal{O}(10~\mathrm{ps})$, which is sufficient for the timing requirements of the silicon tracking system, calorimeters, and RPC-based muon detectors.
The triggerless architecture significantly simplifies the overall DAQ complexity by removing the need for a dedicated hardware trigger system. 

The DAQ system benefits from existing expertise within the Chinese nuclear and particle physics community. In particular, the electronics and readout system is supported by the technical capabilities developed at CCNU, where high-speed FPGA-based optical readout boards, PCIe/Gen3-Gen4 DMA firmware, and full DAQ software stacks have been successfully developed for large-scale experiments. These developments provide a solid foundation for scaling to the \lune requirements, while maintaining compatibility with future upgrades in data throughput and detector complexity.

Overall, the \lune DAQ system is a scalable, modular, and high-throughput architecture that aligns with modern detector readout paradigms and provides a robust infrastructure for precision muon-scattering measurements at \hiaf.

\subsection{Technical Readiness and Cost Strategy}
The \lune detector is designed following a staged implementation strategy that combines high-resolution tracking, precision calorimetry, and PID capabilities while maintaining a strong emphasis on technical readiness, cost effectiveness, and future scalability.

A design principle of the project is the extensive use of mature detector technologies that have already been successfully demonstrated in existing high-energy and nuclear physics experiments. This approach minimizes technical risk and enables a realistic construction schedule under the anticipated beam conditions at \hiaf. Major detector subsystems are based on established technologies with substantial domestic expertise, including MAPS silicon tracking developed within the H-NS program, Shashlyk electromagnetic calorimetry from the MPD@NICA project, BES-III/STCF style multi-wire drift chambers, and RPC-based muon detectors derived from the CMS endcap program.

Cost optimization is incorporated into the detector design from the outset. Several subsystems are being evaluated with multiple technology options in order to achieve the best balance between performance and construction cost. Examples include the possible replacement of peripheral silicon tracking area by GEM or scintillating-fiber detectors~\cite{Yang:2026uzl}, alternative muon-identification systems based on plastic scintillator strips and iron absorbers, and optimized hadronic calorimeter designs. In addition, a common software framework based on the {\sc Key4hep} ecosystem will be shared across both phases, reducing long-term development and maintenance costs.

Table~\ref{tab:lune_detector_summary} summarizes the principal detector subsystems, technology choices, and their technical heritage within the Chinese nuclear and particle physics community. The detector concept therefore benefits from substantial existing R\&D efforts and industrial experience, providing a solid foundation for future construction and upgrades.

\begin{table*}[!htbp]
\centering
\caption{Summary of the \lune detector subsystems, focusing on technology choices and domestic technical heritage. SDU and PKU stand for Shandong University and Peking University, respectively.}
\label{tab:lune_detector_summary}
\small
\begin{tabular}{p{3.2cm} p{4.0cm} p{6.8cm}}
\hline
Subsystem &Technology &Technical Heritage in China \\ \hline
Beam Monitor &
MAPS silicon pixel detector (3 layers, $10\times10~\mathrm{cm}^2$) &
H-NS MAPS technology developed by \textbf{CCNU} and \textbf{IMP} \\
\hline

Forward Scattering Tracker (Phase-I) &
MAPS silicon pixel tracking system (5 stations, low-material design) &
H-NS MAPS technology developed by \textbf{CCNU} and \textbf{IMP} \\
\hline

Central Tracking System (Phase-II) &
MAPS silicon tracker + cylindrical tracking integration in solenoid &
H-NS MAPS technology developed by \textbf{CCNU} and \textbf{IMP} \\
\hline

Central Drift Chamber (Phase-II) &
Multi-wire drift chamber (CDC), 29-layer cylindrical structure &
BES-III/STCF MDC technology (\textbf{IHEP, USTC}) \\
\hline

Target System (Phase-I) &
Solid target station (CH$_2$, CD$_2$, C, Al, Cu, Fe, Pb) &
Rotating target system developed at \textbf{IMP} \\
\hline

Cryogenic Target System (Phase-II) &
Liquid hydrogen target (5 cm / 50 cm) &
Cryogenic target technology at \textbf{IMP and other institutions} \\
\hline

Dipole Magnet &
Warm iron-core dipole magnet with high-permeability DT4 steel &
Magnet design and construction experience at \textbf{IMP} \\
\hline

Superconducting Solenoid (Phase-II) &
2 T large-aperture superconducting solenoid (H-NS based design) &
H-NS superconducting magnet program at \textbf{IMP} \\
\hline

Electromagnetic Calorimeter &
Shashlyk sampling ECAL (Pb + scintillator, SiPM readout) &
MPD@NICA Shashlyk ECAL technology at \textbf{SDU} \\
\hline

Hadronic Calorimeter &
Sampling HCAL (scintillator + iron, SiPM readout) &
Sampling calorimeter development at \textbf{SDU} and collaborators \\
\hline

Barrel Hadronic Calorimeter (Phase-II) &
Sampling calorimeter for full azimuthal coverage &
Extension of calorimeter expertise at \textbf{SDU} \\
\hline

Forward Muon ID System &
RPC-based muon detector (8 layers, Fe absorber stack) &
CMS RPC endcap technology developed by \textbf{PKU} \\
\hline

DAQ \& Electronics &
FPGA-based trigger and data acquisition system &
\hiaf detector electronics infrastructure \\
\hline
\end{tabular}
\end{table*}

Preliminary cost estimates for the Phase-I detector subsystems are summarized in Table~\ref{tab:phase1_cost}. The total cost of the proposed Phase-I detector is estimated to be approximately 20 million Chinese Yuan (CNY). The overall construction cost remains modest compared with other modern fixed-target facilities, demonstrating the cost-effectiveness of the proposed detector while maintaining the performance required for the precision proton charge-radius measurement program. The cost estimates presented here include the beamline upgrade necessary for transporting the muon beam from HIRIBL to the \lune experimental area, as detailed in Appendix~\ref{app:beamline}. A reduced-cost (`descoping') option for Phase-I is also under consideration, in which the calorimeter coverage would be reduced. This would lower the total budget to approximately 14.1 M CNY, albeit with a more limited geometrical acceptance. Such a configuration would degrade the PID performance and is expected to increase the systematic uncertainty of the proton charge radius measurement. Preliminary estimates for the Phase-II detector remain under study, since the final cost will depend on the detector scale, technology selections, and the outcome of ongoing optimization efforts.

\begin{table}[!htbp]
\centering
\caption{Preliminary cost estimate for the \lune Phase-I detector. The estimates include detector construction, front-end electronics, mechanical support structures, and installation where applicable. The final cost will depend on the selected detector technologies and the scope of international contributions.}
\label{tab:phase1_cost}
\begin{tabular}{lcc}
\hline
Subsystem & Baseline & Descoping \\
& (million CNY)  & (million CNY) \\
\hline
Beam monitoring + Tracking & 3.0 & -\\
Target + Dipole + Beamline& 2.3 & -\\
Calorimeters + RPC  & 8.0 & 4.0\\
DAQ and trigger system & 4.3 & 3.3 \\
Infrastructure and contingency & 2.0 & 1.5\\
\hline
Total &  19.6 & 14.1\\
\hline
\end{tabular}
\end{table}

\subsection{Collaboration and Project Organization}
The \lune project is currently being developed by a core team of researchers participating in the \hiaf scientific program and is expected to evolve into a broad international collaboration during the design, construction, and operation phases.

The scientific goals of \lune are closely connected to ongoing international efforts in lepton-nucleon scattering and hadron physics, including programs at PSI, JLab, CERN (AMBER and COMPASS), J-PARC, the future Electron-Ion Collider (EIC)~\cite{Willeke:2021ymc} and EicC. These shared scientific interests naturally provide opportunities for collaboration in detector development, software infrastructure, and physics analysis.

Several detector subsystems have been identified as promising areas for international contributions. These include alternative tracking technologies, such as GEM and scintillating-fiber detectors for large-area outer tracking regions; cost-effective muon-identification systems based on scintillator technologies; advanced hadronic calorimeter concepts; Cherenkov detectors for particle PID in the 2–7\gevc range; and TOF detectors for particle PID below 2\gevc. Additionally, target technologies include liquid-hydrogen, liquid-deuterium, and polarized targets. 

International expertise in detector simulation, reconstruction algorithms, electronics, and computing infrastructure will also play an important role in optimizing detector performance.
As introduced in Sec.~\ref{sec:simulation}, the software framework of \lune is built upon widely adopted international tools, including {\sc Geant4}, {\sc Key4hep}, {\sc GenFit}, {\sc Pandora} SDK, and {\sc ROOT}, facilitating collaborative development and long-term maintainability.

The collaboration welcomes participation from institutions worldwide and will establish dedicated working groups for detector systems, software and computing, physics analysis, and project management. Future detector workshops and collaboration meetings will be organized to define technical contributions and strengthen international partnerships.

\clearpage

%% file: physicsprojection.tex
\section{Simulation and Projection}
\label{sec:phys}

\subsection{Simulation and reconstruction framework}
\label{sec:simulation}

To validate the feasibility of the proposed detector design and to support physics performance studies, a full simulation and reconstruction framework has been developed within the {\sc key4hep} software ecosystem~\cite{Key4hep:2021zms}, deployed via CVMFS~\cite{Buncic:2010zz}. The framework builds upon a modified version of the {\sc Geant4}-based simulation~\cite{GEANT4:2002zbu} developed for the PRad experiment, which has been extended to accommodate the geometry, magnetic configuration, and readout characteristics of the present detector design.

The detector geometry, material composition, and magnetic field configuration are implemented in detail, including a 0.5~T dipole field (Phase-I and Phase-II) and a 2~T solenoidal field (Phase-II). The simulation supports flexible integration of different physics event generators, allowing for a comprehensive coverage of relevant reaction channels. In particular, elastic and inelastic lepton-proton interactions are simulated using generators such as {\sc ESEPP}~\cite{Gramolin:2014pva} (for elastic scattering and radius measurements), {\sc DJANGOH}~\cite{Charchula:1994kf} (for DIS), {\sc HEPGEN++}~\cite{Sandacz:2012at} (for exclusive DVCS processes), and {\sc EpIC}~\cite{Aschenauer:2022aeb} (for DVCS and deeply virtual meson production (DVMP)). For interactions involving nuclear or solid targets, {\sc GiBUU}~\cite{Buss:2011mx} is also employed to model the corresponding nuclear effects and final-state interactions. Complementing this, a dedicated event generator for elastic scattering, {\sc Lumen} (LUNE Muon Elastic scattering Nuclear Generator), has been developed based on first-principle treatments of the measured form factors of the proton, deuteron, and carbon.

Generated events are propagated through the detector using {\sc Geant4}, where secondary particles are tracked and their interactions with detector materials are fully simulated. The resulting detector responses, including tracking hits and calorimetric energy deposits, are 
stored for subsequent reconstruction.
Track reconstruction is performed using {\sc GenFit2}~\cite{Rauch:2014wta}, a generic Kalman-filter-based fitting toolkit that combines hit information from tracking detectors, such as silicon pixel layers and the drift chamber, with the magnetic field map to reconstruct charged particle trajectories in three dimensions. Primary vertex reconstruction is carried out using {\sc RAVE}~\cite{Waltenberger:2008zza}, while calorimeter cluster reconstruction and particle-level object building are handled by the {\sc Pandora SDK}~\cite{Marshall:2015rfa}. The entire reconstruction workflow is managed using standard {\sc ROOT}-based data structures and analysis tools~\cite{Antcheva:2009zz}, ensuring compatibility with downstream physics analysis.

\subsection{Detector Performance}
\label{sec:detector_performance}
The detector performance has been evaluated using the full simulation and reconstruction framework described in Sec.~\ref{sec:simulation}. Unless otherwise specified, the simulated muon beam is assumed to have a beam spot of 2 cm, and an internal divergence of 10 mrad, consistent with the beam parameters summarized in Table~\ref{tab:HIAFmuon}. Samples of muons ($\mun$) and other particles ($\gamma$, $e$, $\pim$, $\piz$, $\Km$, $\proton$) covering the relevant momentum and angular ranges are generated and propagated through the detector geometry. The resulting detector responses are reconstructed using the standard tracking, vertexing, PID, and calorimeter reconstruction chain.

The principal performance metrics considered in this study include the tracking efficiency, momentum resolution, angular resolution, vertex resolution, PID capability, and calorimeter energy resolution. These quantities provide the essential inputs for evaluating the physics reach of the \lune experiment.

\subsubsection{Phase-I Detector Performance}

The Phase-I detector is optimized for precision measurements of elastic muon-proton scattering at low-\qsq. The detector consists primarily of silicon pixel tracking stations, a solid target, and a forward calorimeter.

\begin{figure}[!htbp]
    \centering
    \includegraphics[width=0.75\linewidth]{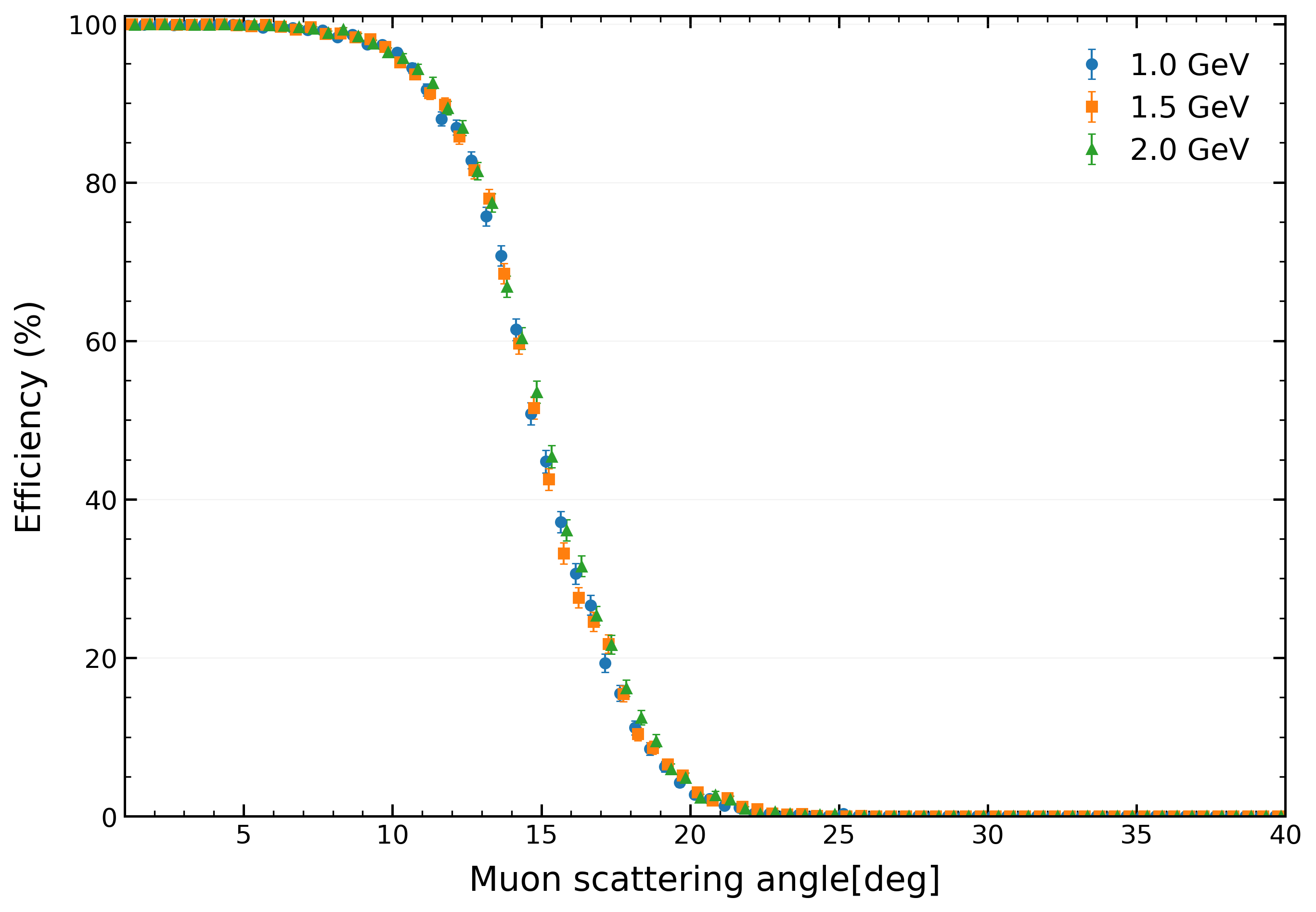}
    \caption{Reconstruction efficiency of the Phase-I detector as a function of particle (\mun) scattering angle.}
    \label{fig:eff_phase1}
\end{figure}

\paragraph{Tracking Efficiency} The reconstruction efficiency is defined as
\begin{equation}
\epsilon_{\rm reco}=\frac{N_{\rm reconstructed}}{N_{\rm generated}},
\end{equation}
where $N_{\rm reconstructed}$ and $N_{\rm generated}$ denote the numbers of reconstructed and generated events, respectively. Figure~\ref{fig:eff_phase1} presents the reconstruction efficiency of \mun obtained from the full simulation chain. High tracking efficiency is achieved throughout the nominal detector acceptance of $1^\circ$-$10^\circ$. The non-zero statistics observed at scattering angles above $10^\circ$ originate primarily from the finite beam spot size and the intrinsic divergence of the incident muon beam, which broaden the effective angular distribution.

\begin{figure}[!htbp]
    \centering
        \includegraphics[width=0.45\linewidth]{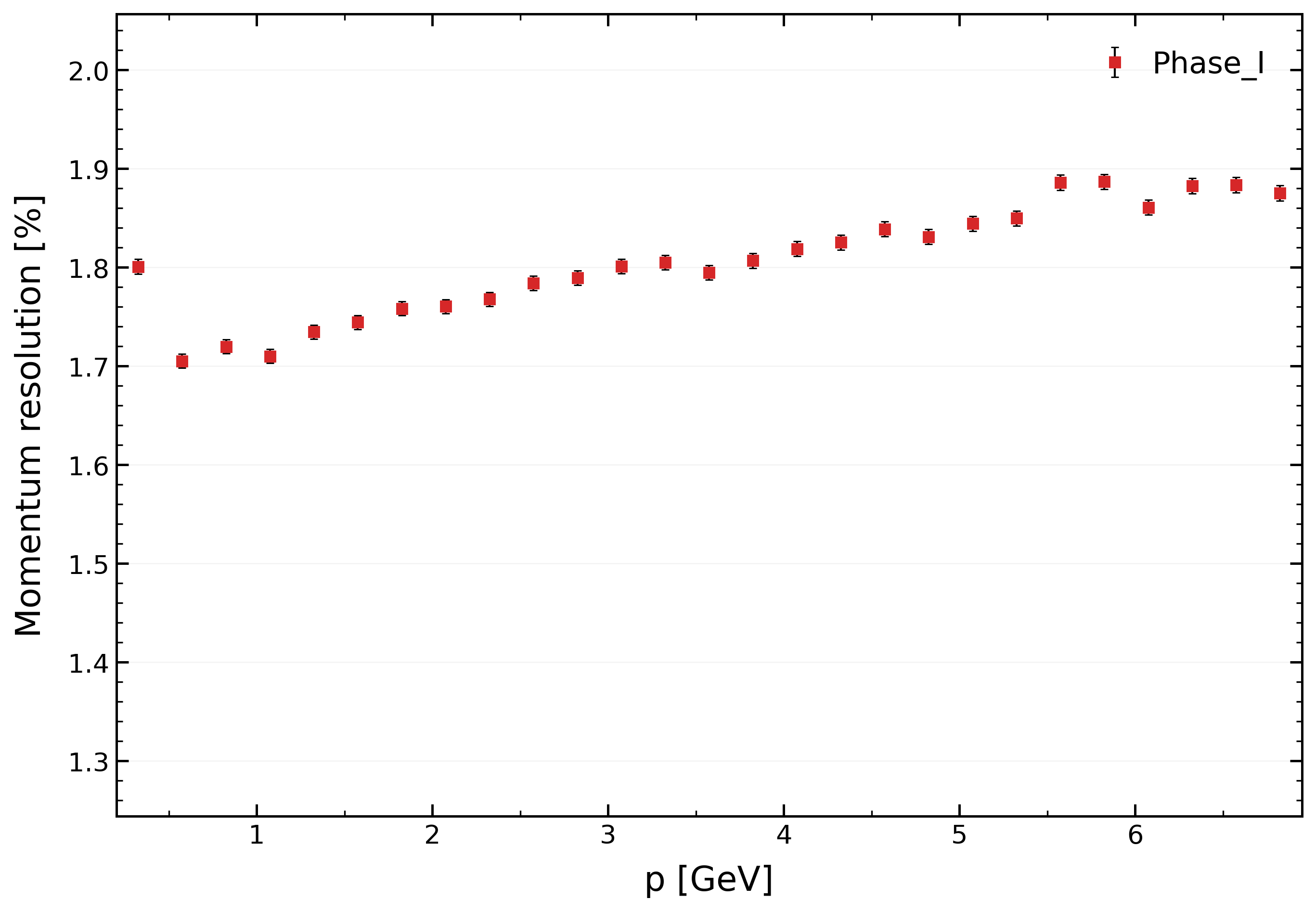}
    \includegraphics[width=0.45\linewidth]{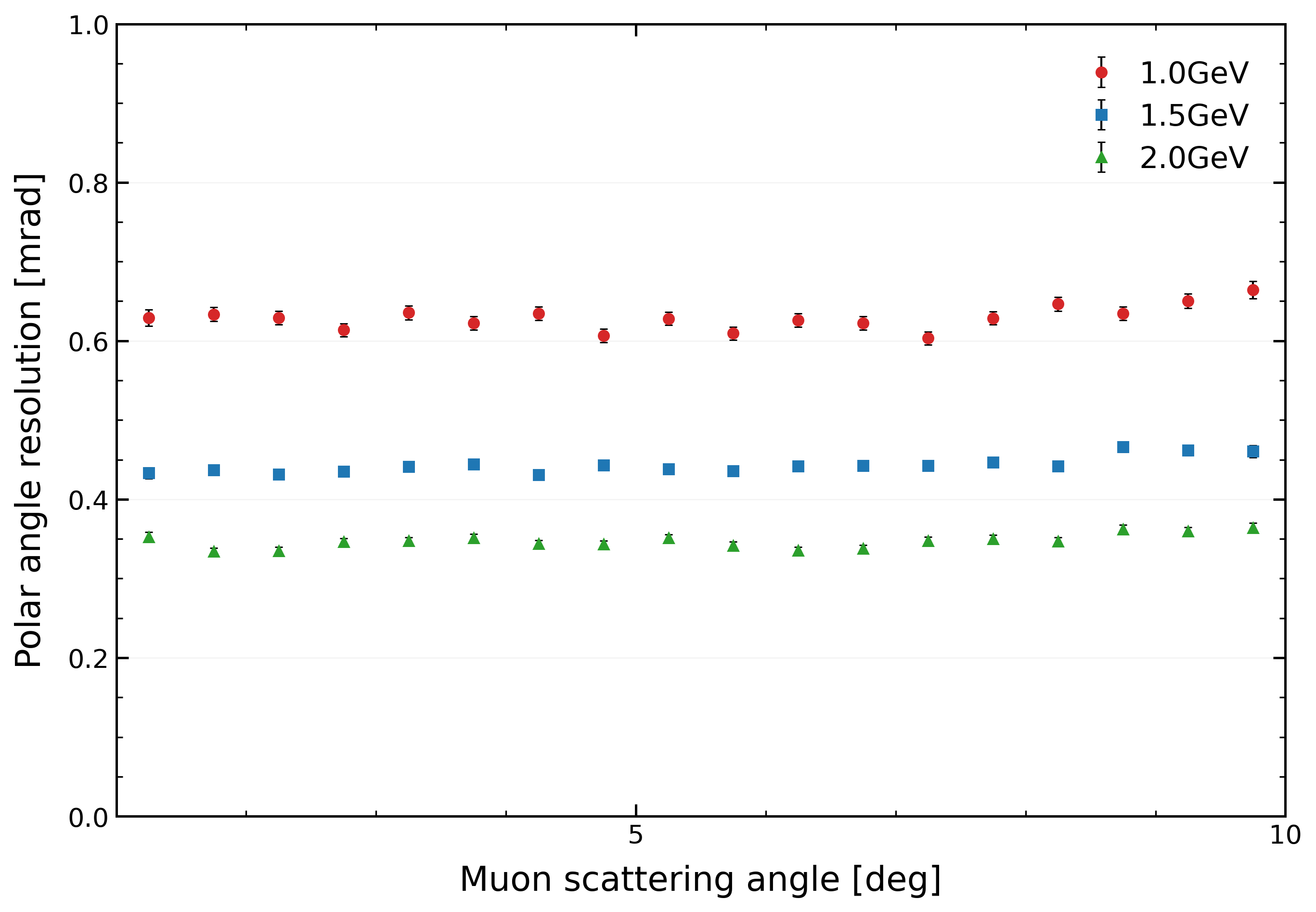}
    \caption{Momentum (left) and angular (right) resolutions of the \lune Phase-I detector.}
    \label{fig:areso_phase1}
\end{figure}

\paragraph{Momentum Resolution} The momentum resolution is evaluated using the relative residual
\begin{equation}
 \frac{p_{\rm reco}-p_{\rm true}}{p_{\rm true}},
\end{equation}
where $p_{\rm reco}$ and $p_{\rm true}$ denote the reconstructed and generated momenta. The resulting momentum resolution is shown in Fig.~\ref{fig:areso_phase1} (left figure). The excellent resolution ($<2.0\%$) provide by the silicon tracking system ensures accurate momentum reconstruction across the relevant kinematic region.

\paragraph{Angular Resolution}
The angular resolution is determined from the distribution of
\begin{equation}
 \theta_{\rm reco}-\theta_{\rm true},
\end{equation}
where $\theta_{\rm reco}$ and $\theta_{\rm true}$ are the reconstructed and generated scattering angles, respectively. Figure~\ref{fig:areso_phase1} (right figure) shows the achieved angular resolution. The obtained precision ($<0.8$ mrad) satisfies the requirements for proton-radius and low-\qsq form-factor measurements.

\begin{figure}[!htbp]
    \centering
    \includegraphics[width=0.7\linewidth]{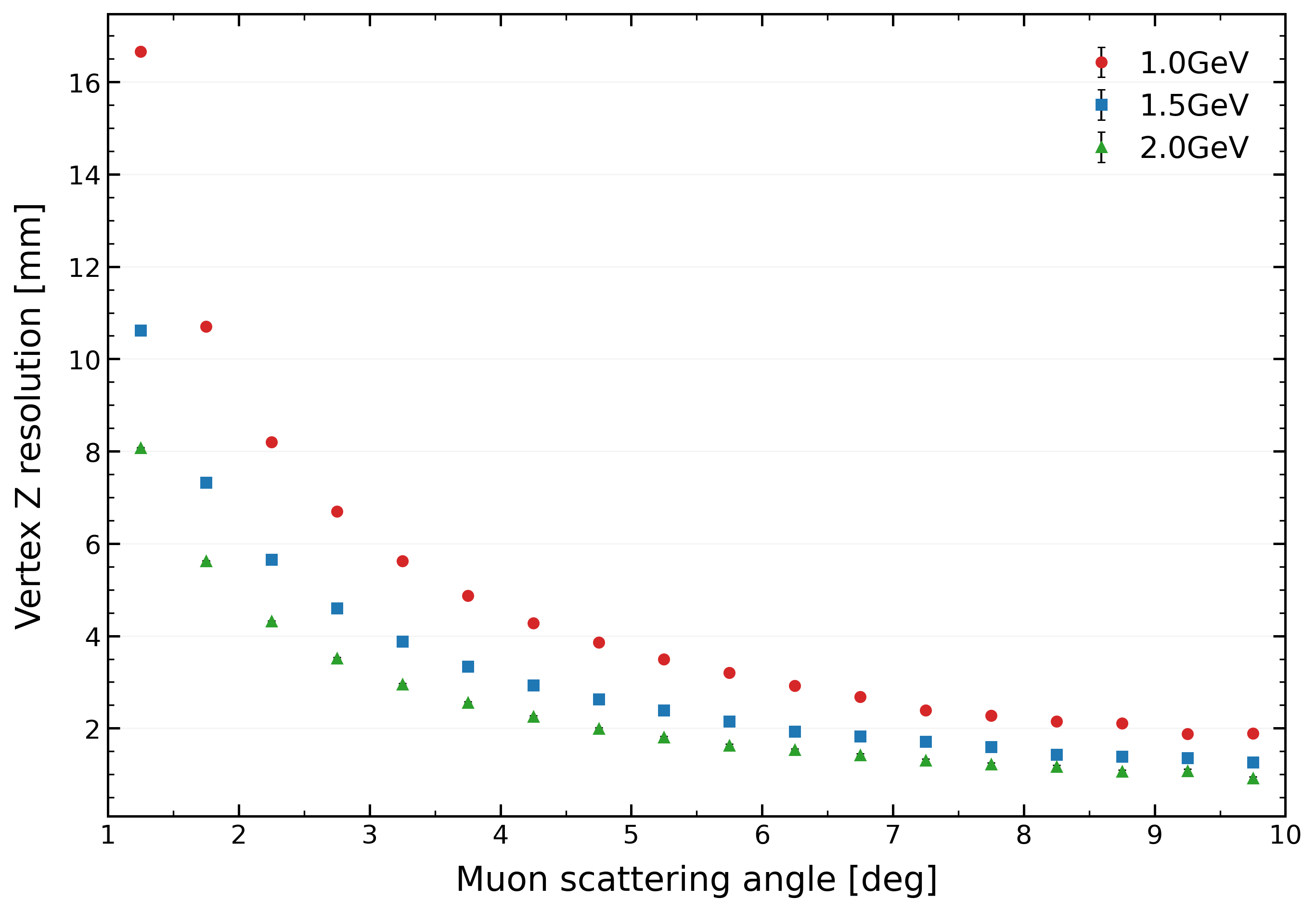}
    \caption{Primary vertex resolution of the Phase-I detector.}
    \label{fig:vtx_phase1}
\end{figure}

\paragraph{Vertex Resolution}
Primary interaction vertices are reconstructed using the {\sc RAVE} vertex fitting package. The vertex resolution is evaluated from the difference between reconstructed and generated vertex positions in the transverse and longitudinal directions, using the {\sc ESEPP} elastic scattering events. Figure~\ref{fig:vtx_phase1} shows the corresponding performance. The achieved vertex precision enables efficient separation of target interactions from beam-related backgrounds and provides a robust foundation for future measurements involving solid nuclear targets and exclusive reaction channels.

\paragraph{Particle Identification Performance}
PID is achieved through the combined use of tracking information and calorimeter responses. The performance is quantified in terms of identification efficiency and misidentification probability for electrons, muons, pions, kaons, and protons.

\begin{figure}[!htbp]
    \centering
    \includegraphics[width=0.7\linewidth]{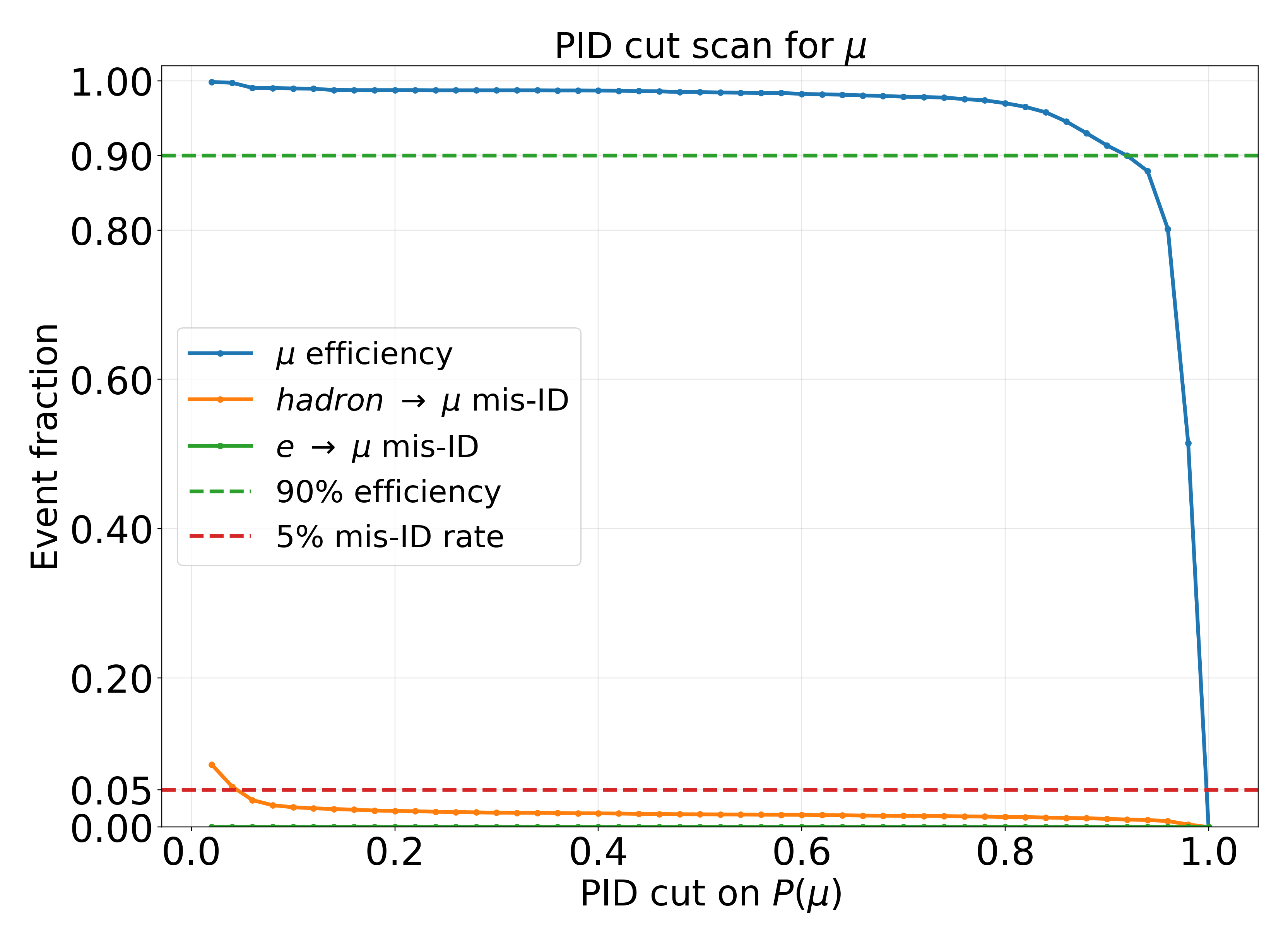}
    \caption{Particle-identification performance of the Phase-I detector at 1.5\gevc.}
    \label{fig:pid_phase1}
\end{figure}

Figure~\ref{fig:pid_phase1} and Table~\ref{tab:pid_efficiency_scan} summarizes the expected PID performance under 1.5\gevc. The detector provides effective separation among charged-particle species over the momentum range relevant to the \lune physics program, thereby enabling the rejection of background events.

\begin{table}[!htbp]
    \centering
    \caption{Muon identification efficiency and hadron misidentification rate at 1.5\gevc.}
    \begin{tabular}{c|cc}
        \hline
        $P(\mu)$ & $>0.92$ & $>0.04$ \\
        \hline

        $\varepsilon_{\mu}$ & $<90.0\%$ & $<99.7\%$ \\
        $\varepsilon_{h \rightarrow \mu}$ & $<1.0\%$ & $<5.0\%$ \\
        \hline
    \end{tabular}
    \label{tab:pid_efficiency_scan}
\end{table}

\begin{figure}[!htbp]
    \centering
    \includegraphics[width=0.55\linewidth]{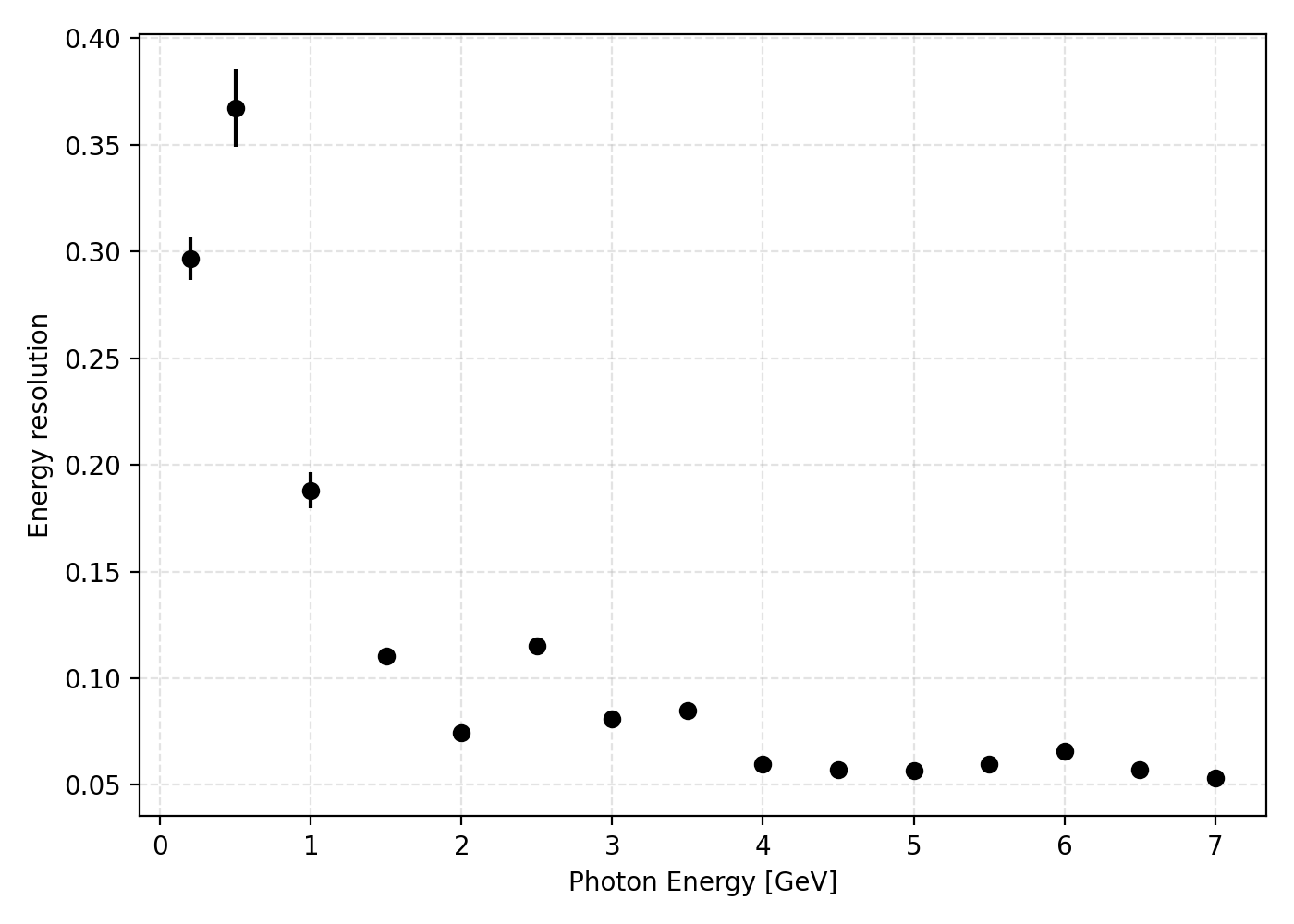}
    \caption{Electromagnetic calorimeter energy resolution of the Phase-I detector.}
    \label{fig:ecal_phase1}
\end{figure}

\paragraph{Calorimeter Performance}
The electromagnetic calorimeter performance is evaluated using single-particle simulations ($e$). The energy resolution is parameterized as
\begin{equation}
\frac{\sigma_E}{E}, 
\end{equation}
where $E$ denotes the deposited energy and $\sigma_E$ the corresponding resolution. Figure~\ref{fig:ecal_phase1} presents the reconstructed energy resolution as a function of particle energy.

\subsubsection{Phase-II Detector Performance}
The Phase-II detector extends the angular coverage and detector capabilities through the addition of a large-acceptance solenoidal spectrometer, multi-wire drift chambers, electromagnetic calorimeters, and PID detectors. This configuration is designed to support a broad physics program including elastic scattering, nucleon tomography, and DIS.

\begin{figure}[!htbp]
    \centering
    \includegraphics[width=0.75\linewidth]{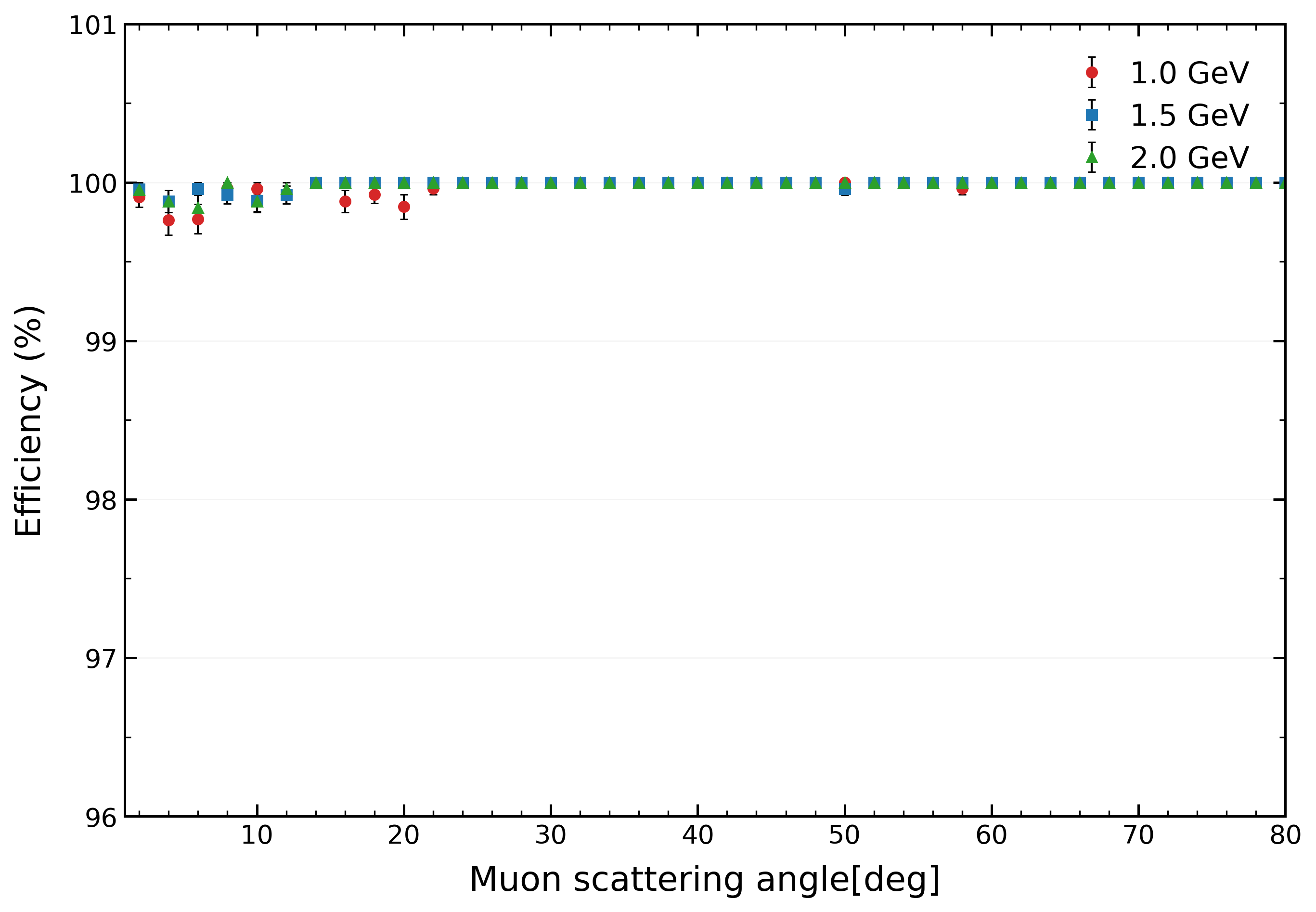}
    \caption{Reconstruction efficiency of the Phase-II detector.}
    \label{fig:eff_phase2}
\end{figure}

The reconstruction efficiency obtained for the Phase-II detector in the large scattering angle region is shown in Fig.~\ref{fig:eff_phase2}. The combination of silicon pixel detectors and large-volume drift chambers provides efficient charged-particle reconstruction over a broad kinematic range.

\begin{figure}[!htbp]
    \centering
    \includegraphics[width=0.45\linewidth]{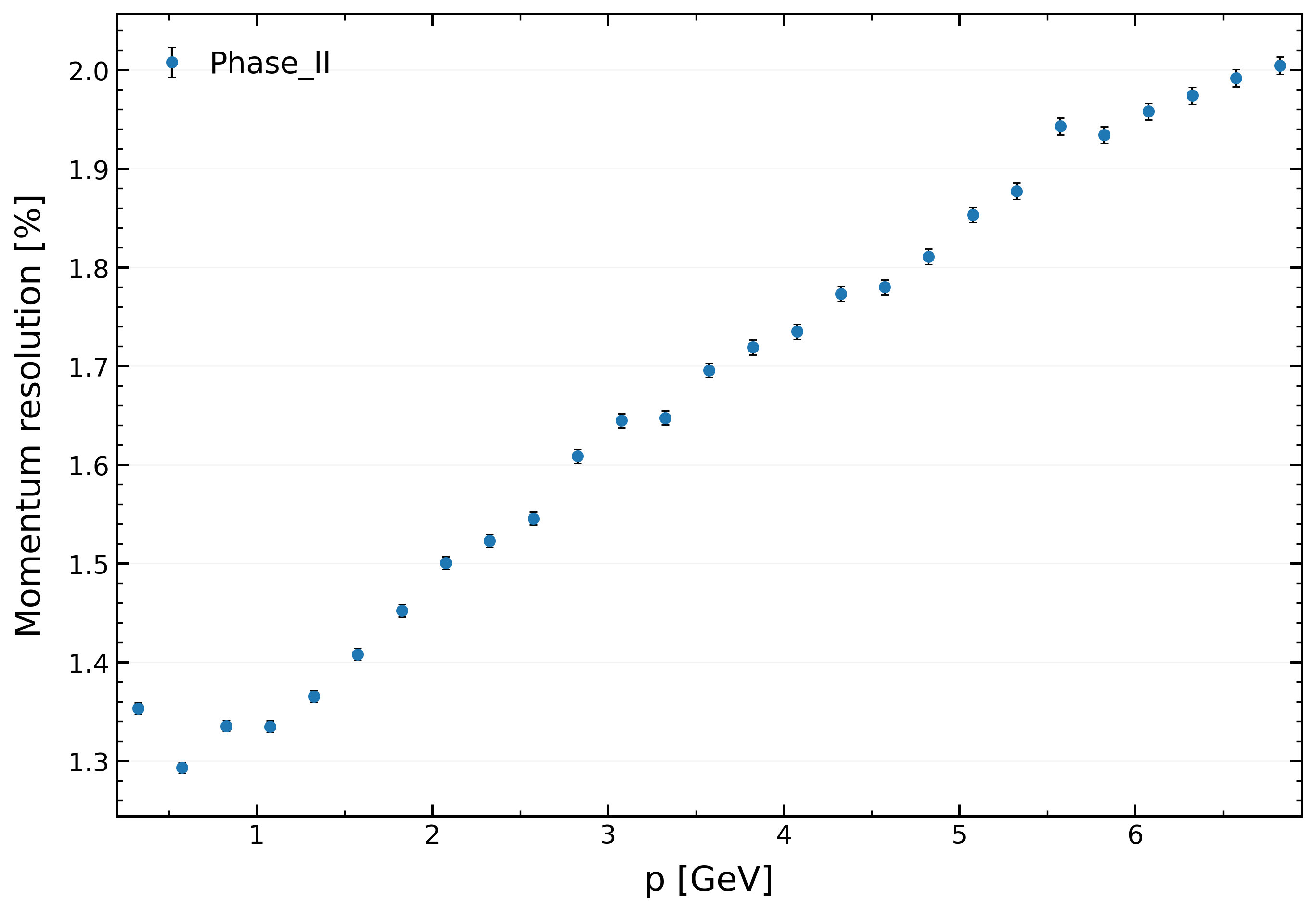}
    \includegraphics[width=0.45\linewidth]{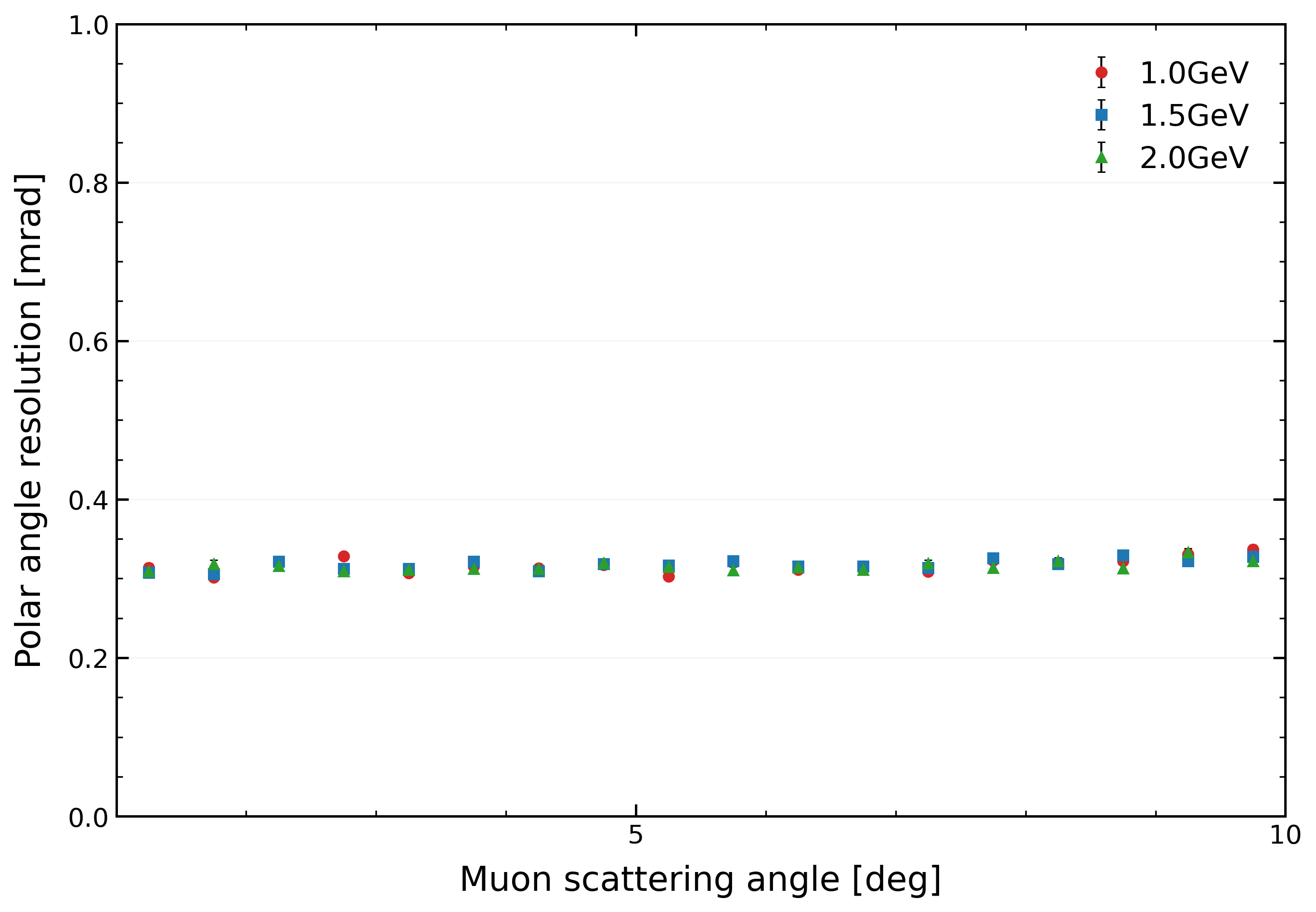}
    \caption{Momentum (left) and angular (right) resolutions of the \lune Phase-II detector.}
    \label{fig:areso_phase2}
\end{figure}

The momentum resolution is presented in Fig.~\ref{fig:areso_phase2} (left figure). Forward-going particles are primarily reconstructed by the silicon tracking system operating inside the dipole field, while particles emitted at larger angles are reconstructed using the drift chamber system within the solenoidal field. The resulting momentum resolution remains stable throughout the detector acceptance.
The angular resolution obtained from reconstructed tracks is shown in Fig.~\ref{fig:areso_phase2} (right figure). Across the full acceptance, the achieved precision fulfills the requirements for both exclusive and inclusive scattering measurements.

Overall, the simulation results demonstrate that the proposed detector concepts satisfy the technical requirements of the Phase-I and Phase-II physics programs. The achieved tracking, vertexing, particle-identification, and calorimetric performance provide a solid foundation for the precision measurements discussed in the following sections.

\subsection{Physics Projection}

The detector performance discussed in Sec.~\ref{sec:detector_performance} forms the basis for evaluating the expected physics reach of the \lune experiment. 
The projections presented in this section are based on the current detector configuration and beam parameters assumed for the \hiaf muon facility. Unless otherwise stated, the studies include realistic detector acceptance, reconstruction efficiencies, and detector resolutions obtained from the full simulation framework.

\subsubsection{Proton Charge Radius}

One of the primary goals of the Phase-I program is a precision determination of the proton charge radius using elastic muon-proton scattering. The long-standing discrepancy between measurements obtained from electronic probes and muonic hydrogen spectroscopy remains one of the most important unresolved issues in hadronic physics.

\begin{figure}[!htbp]
    \centering
    \includegraphics[width=0.45\linewidth]{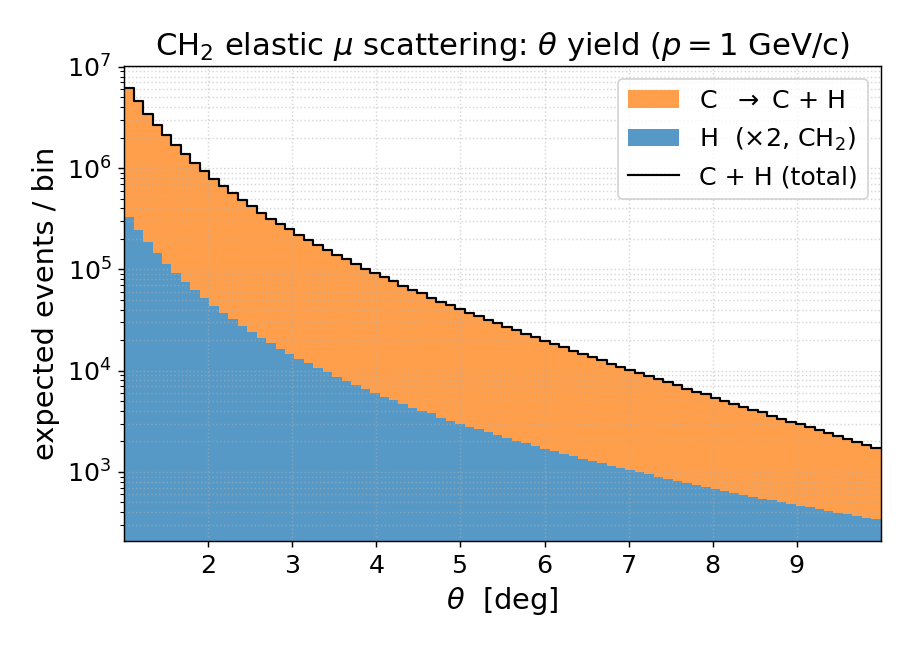}    \includegraphics[width=0.45\linewidth]{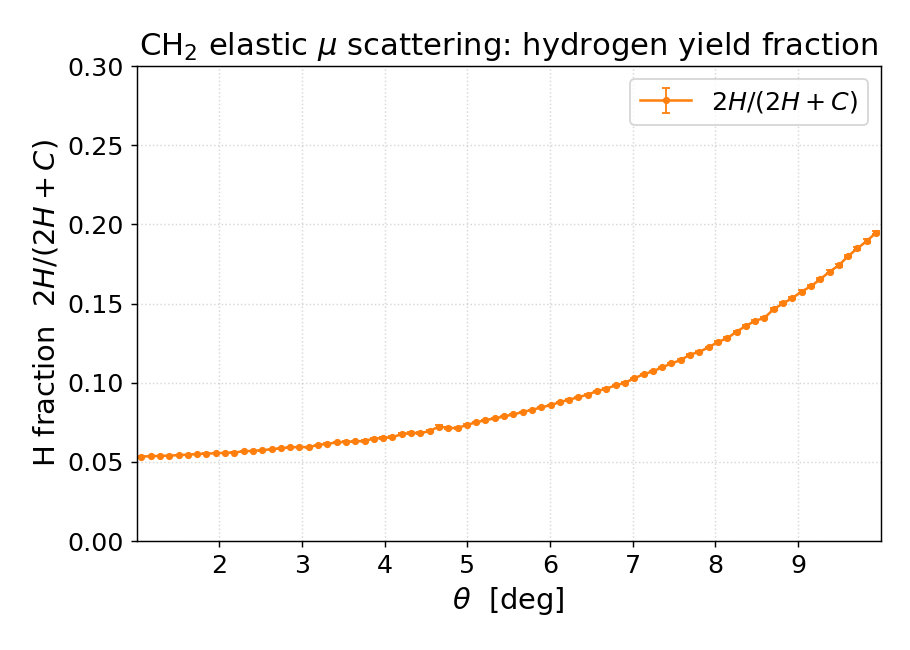}
    \caption{Comparison of the muon scattering angle distributions from CH$_2$ and carbon targets, with MC events generated using the {\sc Lumen} event generator. The different shapes arising from $\mu$-H and $\mu$-C interactions provide sensitivity for separating the carbon contribution through a shape-fitting procedure.}
    \label{fig:scattring_C_H}
\end{figure}

The proton charge radius is extracted from measurements of the electric form factor in the low-\qsq region (as shown in Eq.~\ref{eq:proton_radius}). 
The \lune experiment is designed to access a broad low-\qsq region using muon beams with momenta between 1 and 2\gevc. The high angular resolution provided by the silicon tracking system allows precise determination of the momentum transfer, while the high beam intensity enables the collection of large elastic scattering data samples. 
For the CH$_2$ and carbon targets, the muon scattering cross sections and angular distributions exhibit distinct differences, providing additional sensitivity to separate the $\mu$-C contribution through a shape-based fitting procedure, as demonstrated in Fig.~\ref{fig:scattring_C_H}. A dedicated evaluation of the systematic uncertainty associated with this subtraction procedure is required and is currently underway. We plan to employ the profile likelihood method~\cite{Cowan:2010js} for this analysis; based on the present detector configuration, the resulting systematic uncertainty is expected to be $<5\times 10^{-3}\fm$.

The projected precision of the proton charge-radius measurement is summarized in Table~\ref{tab:rp_error_budget}. The quoted values represent the current design goals adopted for the Phase-I sensitivity study rather than the final experimental performance. 

The statistical uncertainty is determined by the expected luminosity, detector acceptance, and data-taking period. Owing to the high muon flux provided by \hiaf, statistical uncertainties are expected to become subdominant after approximately one month of nominal beam operation. Consequently, the ultimate precision is expected to be limited primarily by systematic uncertainties rather than statistical fluctuations. In the present sensitivity study, a sample of $10^8$ elastic scattering events is assumed at each beam energies (1.0, 1.5, and 2.0\gevc), corresponding to approximately 1, 2, and 4 months of data taking, respectively. Under these assumptions, the statistical uncertainty is estimated to be only about 0.12\%. Even if the beam time were reduced by a factor of two, the statistical uncertainty would increase only modestly to approximately 0.17\%, indicating that the overall measurement remains predominantly limited by systematic effects.

Among the systematic uncertainties, the determination of the scattering angle constitutes one of the most critical experimental factors at small scattering angles. The detector design therefore emphasizes excellent angular resolution, precise detector alignment, and minimal multiple scattering in the tracking system. The carbon subtraction procedure introduces a dedicated source of systematic uncertainty. Additional contributions arise from QED radiative corrections, finite binning effects in the differential cross-section analysis, detector acceptance corrections, and the subtraction of backgrounds using the matched graphite target. The combined systematic uncertainty is expected to remain at the sub-percent level, leading to an overall projected precision on the proton charge radius of approximately $1\%$.

\begin{table}[!htbp]
\centering
\caption{Illustrative uncertainty budget for the projected proton charge-radius measurement with the baseline \lune Phase-I detector. The numerical values are representative estimates used for sensitivity studies and will be refined with full detector simulation and data-driven calibrations.}
\label{tab:rp_error_budget}
\renewcommand{\arraystretch}{1.2}
\begin{tabular}{lc}
\hline
Source & Estimated uncertainty \\
\hline
Statistical uncertainty & 0.12\% \\
Angular calibration and resolution & 0.10\% \\
QED radiative corrections & 0.16\% \\
Carbon subtraction & 0.50\% \\
Background, binning, acceptance \etc & 0.50\% \\
\hline
Total systematic uncertainty & 0.75\% \\
Total projected uncertainty & $<1.0\%$ \\
\hline
\end{tabular}
\end{table}

\begin{figure}[!htbp]
    \centering
    \includegraphics[width=0.95\linewidth]{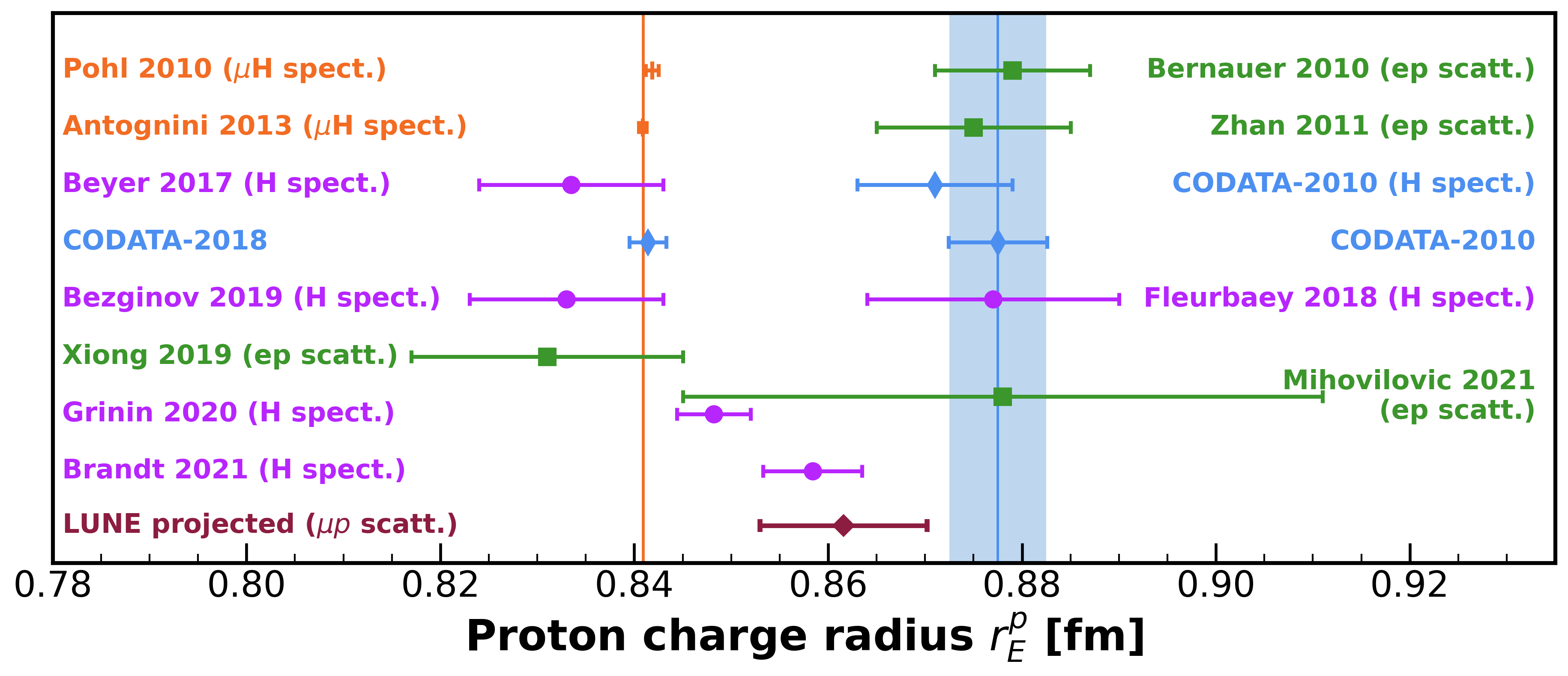}
    \caption{Projected precision of the proton charge-radius measurement from \lune, compared with existing experimental results. The central value of the \lune projection is arbitrary, while the corresponding uncertainty (1\%) includes both statistical and systematic components.}
    \label{fig:radius_projection}
\end{figure}

Figure~\ref{fig:radius_projection} shows the expected sensitivity to the proton charge radius based on the current detector design and projected integrated luminosity. The anticipated statistical precision is expected to be competitive with existing electron-scattering measurements while providing an independent determination using muon beam.

\subsubsection{Nuclear Form Factors}

Beyond proton structure measurements, the \lune facility offers unique opportunities for studying the electromagnetic structure of light nuclei and the dynamics of nucleons embedded in nuclear matter. The availability of high-intensity muon beams, combined with the flexibility to operate with both liquid and solid targets, enables a broad nuclear-physics program extending from precision form-factor measurements to investigations of spin-dependent nuclear structure.

\begin{figure}[!htbp]
    \centering
    \includegraphics[width=0.60\linewidth]{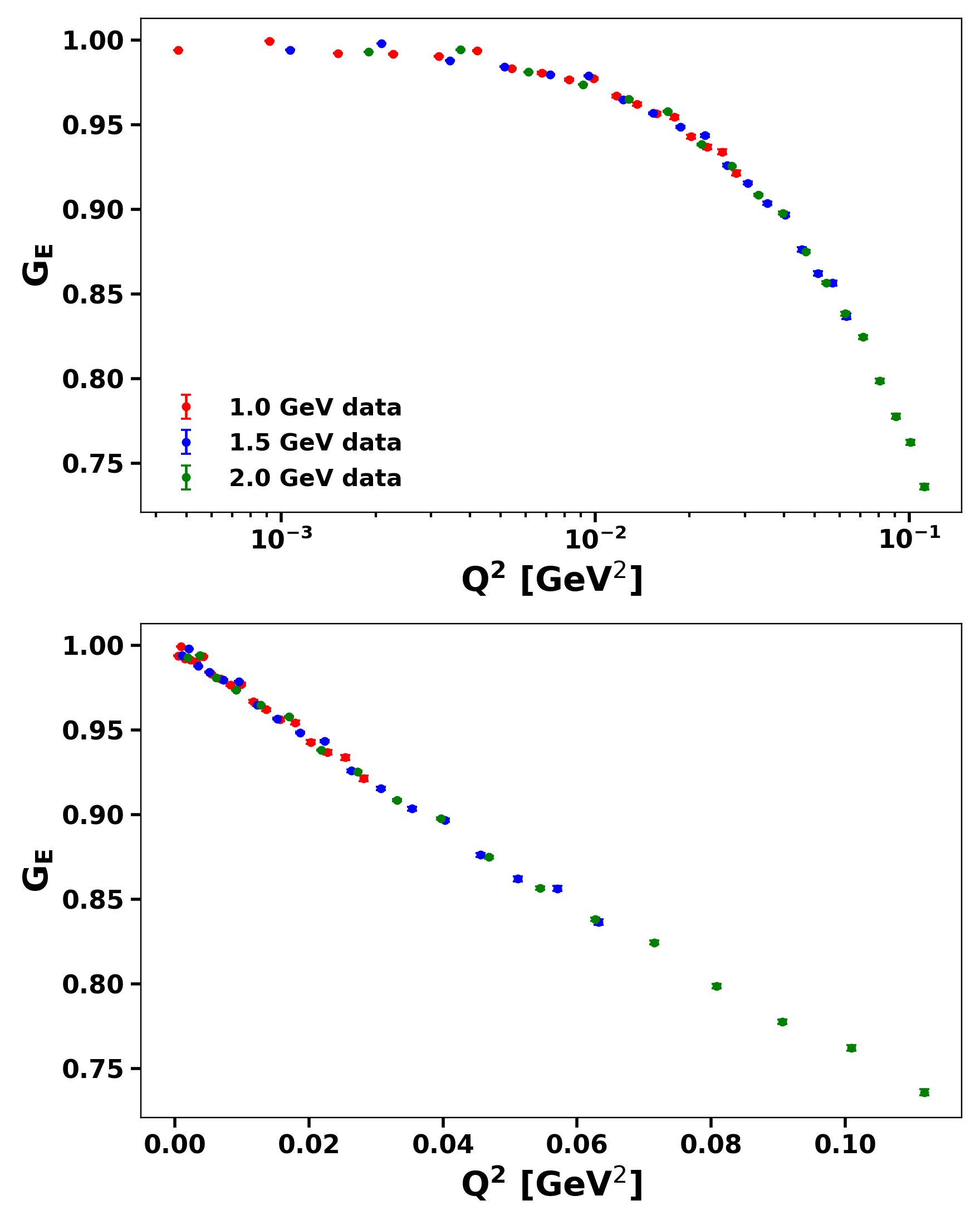}
    \caption{Projected precision for representative measurements of proton electromagnetic form factors.}
    \label{fig:nuclear_ff_projection}
\end{figure}

For light nuclear targets, such as deuterium and helium, elastic scattering measurements can be used to determine charge radii and electromagnetic form factors with unprecedented precision. 
Figure~\ref{fig:nuclear_ff_projection} illustrates the projected precision for representative proton form-factor measurements. The excellent tracking resolution of the detector allow measurements over a broad range of momentum transfer, enabling detailed studies of the spatial distributions of charge and magnetization in light nuclei (including proton).

\subsubsection{Nucleon Structure: DIS and Exclusive Processes}

The Phase-II detector extends the \lune physics program toward precision studies of nucleon structure through both inclusive and exclusive lepton-nucleon scattering.

\begin{figure}[!htbp]
    \centering
    \includegraphics[width=0.70\linewidth]{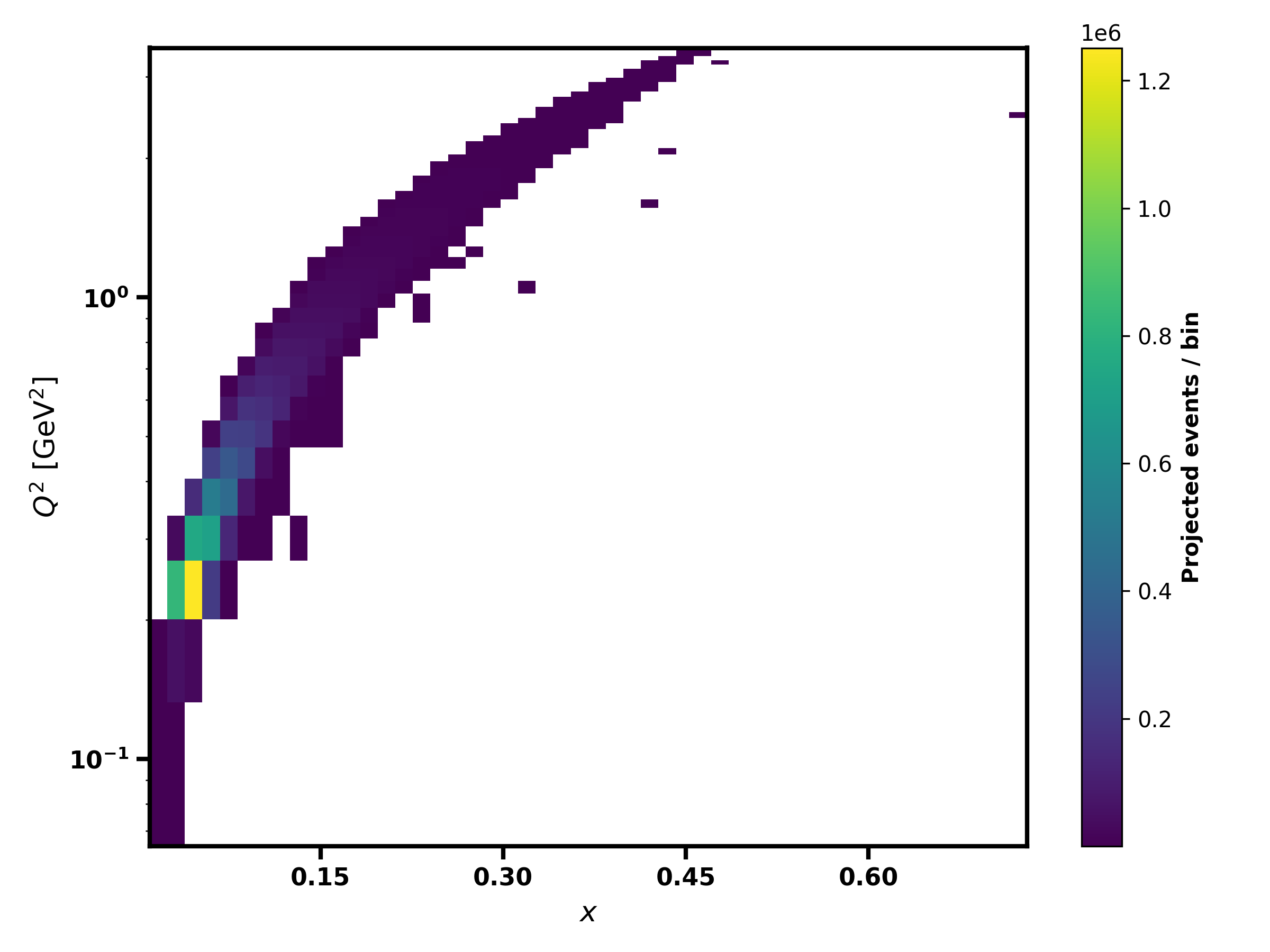}
    \caption{Projected DIS kinematic coverage in the $(x,\qsq)$ plane for the \lune Phase-II detector.}
    \label{fig:dis_coverage}
\end{figure}

For DIS, the large angular acceptance and high-resolution tracking system enable measurements of the double-differential cross section over a broad kinematic region in Bjorken-$x$ and four-momentum transfer \qsq. Figure~\ref{fig:dis_coverage} illustrates the expected kinematic coverage accessible to \lune.

\begin{figure}[!htbp]
    \centering
    \includegraphics[width=0.47\linewidth]{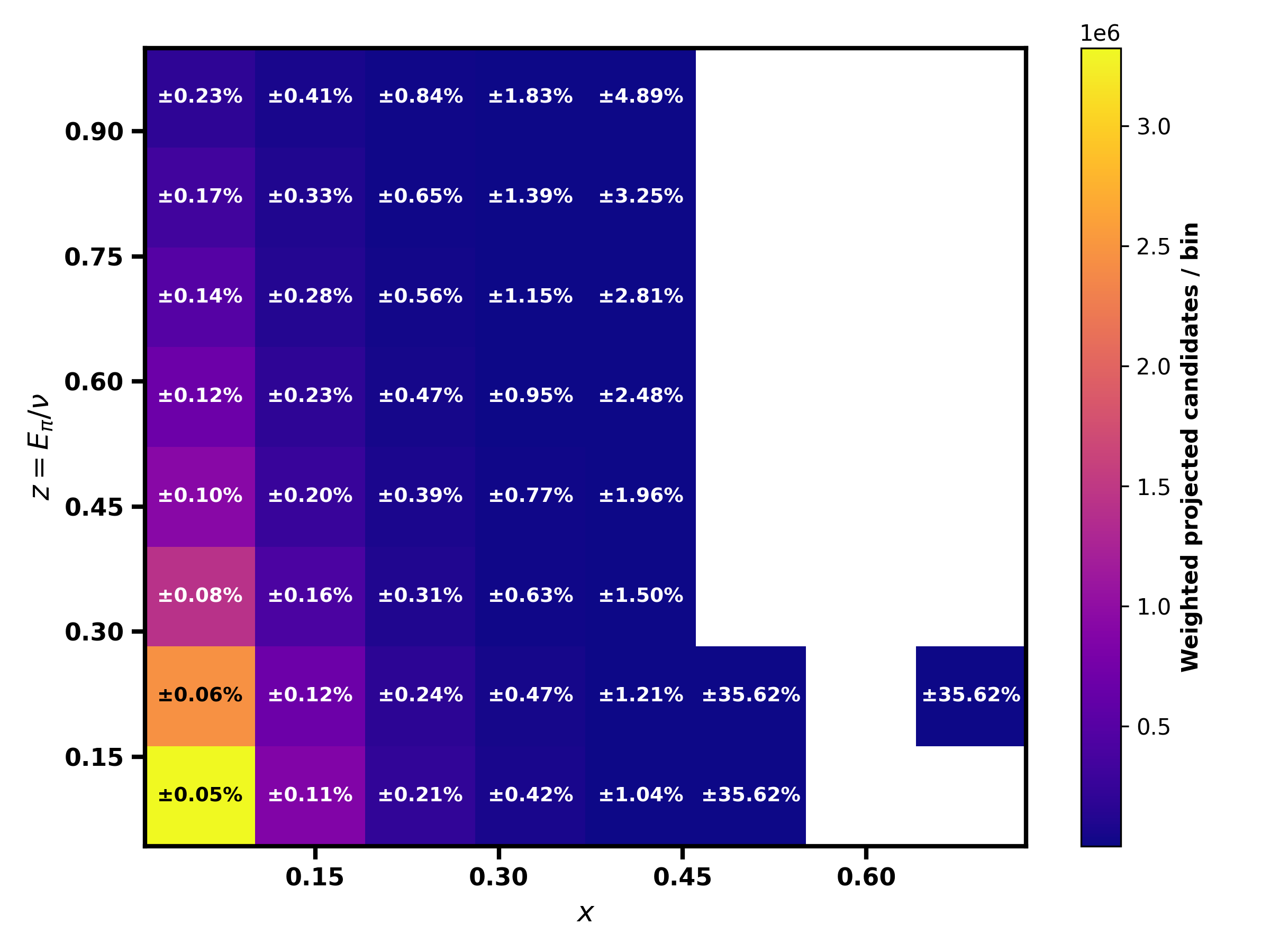}
    \includegraphics[width=0.47\linewidth]{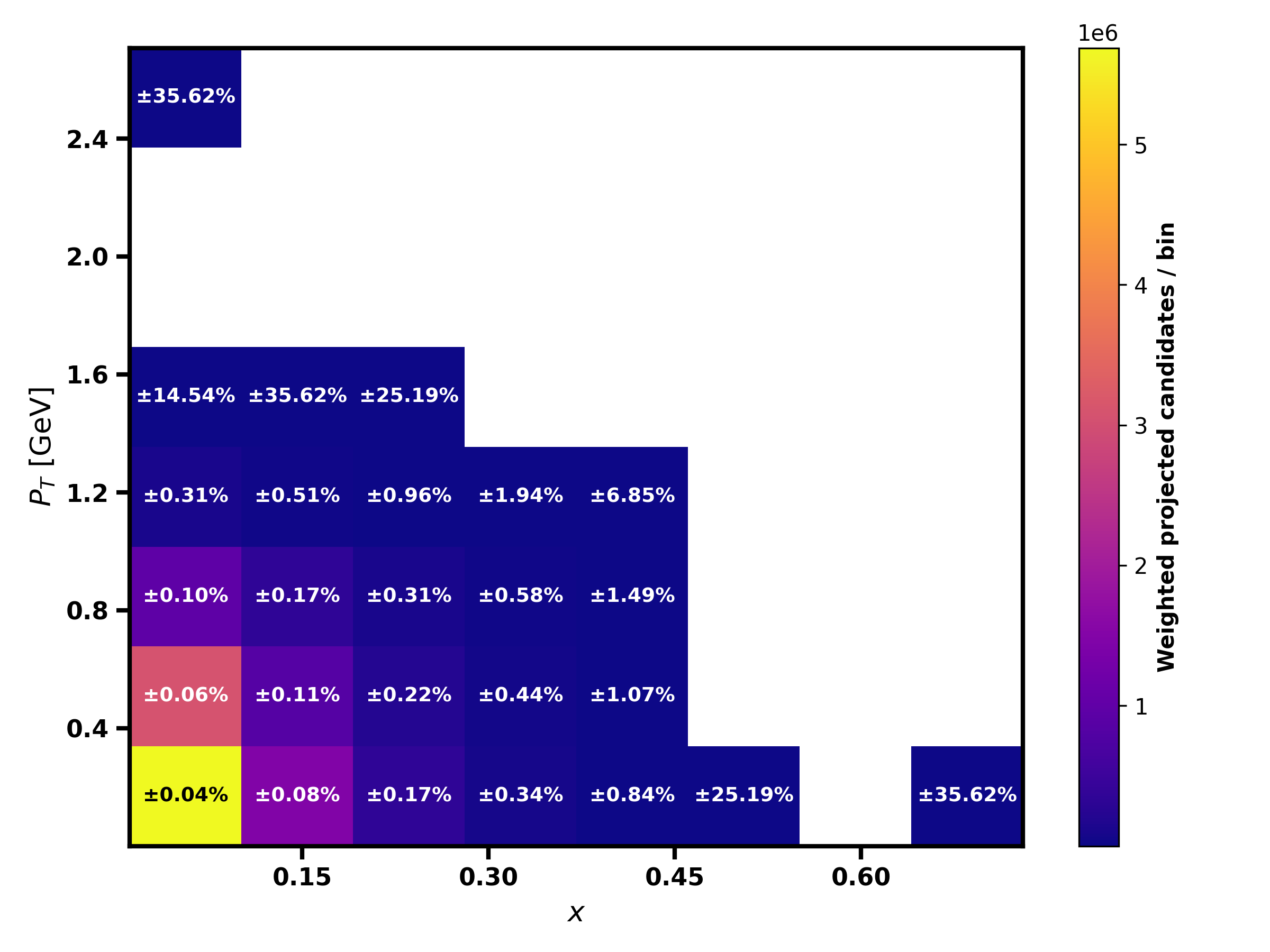}
    \caption{Projected statistical precision of SIDIS measurements in the $(x,z)$ (left) and $(x,P_{hT})$ (right) planes for the \lune Phase-II detector. The color scale represents the expected number of reconstructed SIDIS events in each kinematic bin, while the value quoted in each bin indicates the corresponding relative statistical uncertainty.}
    \label{fig:sidis_coverage}
\end{figure}

Beyond inclusive scattering, the detector is designed to enable comprehensive studies of SIDIS and DVCS, which provides direct access to TMD and GPDs. Through measurements of the differential cross sections and azimuthal asymmetries, \lune will probe the multidimensional structure of the nucleon and the orbital motion of quarks.
Figure~\ref{fig:sidis_coverage} shows the expected statistical precision for pion SIDIS measurements assuming one year of data taking ($10^7$ events). Here, $z$ denotes the fraction of the virtual-photon energy carried by the detected pion, and $P_{hT}$ represents the transverse momentum of the pion with respect to the virtual-photon direction.

\begin{figure}[!htbp]
    \centering
    \includegraphics[width=0.45\linewidth]{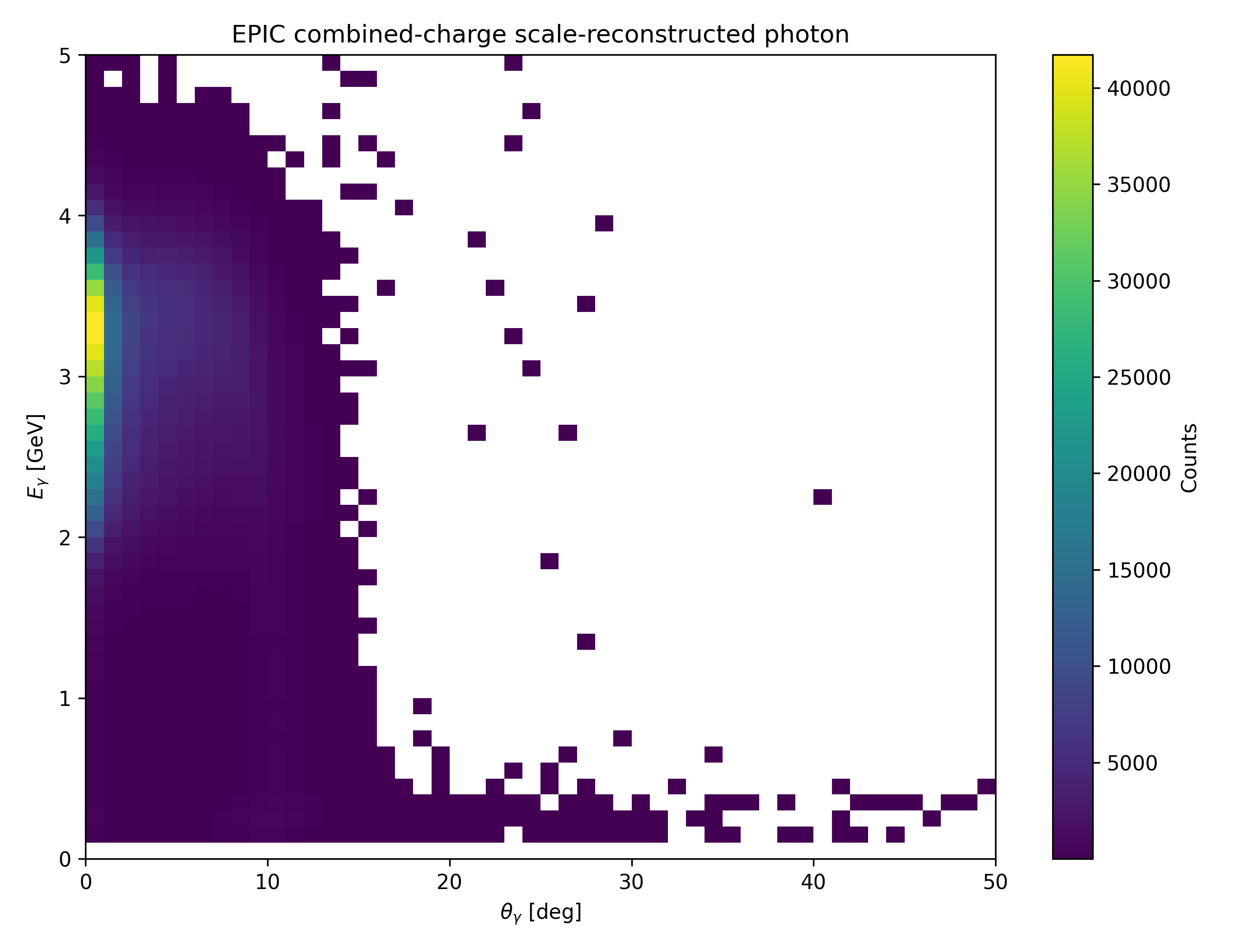}
    \includegraphics[width=0.45\linewidth]{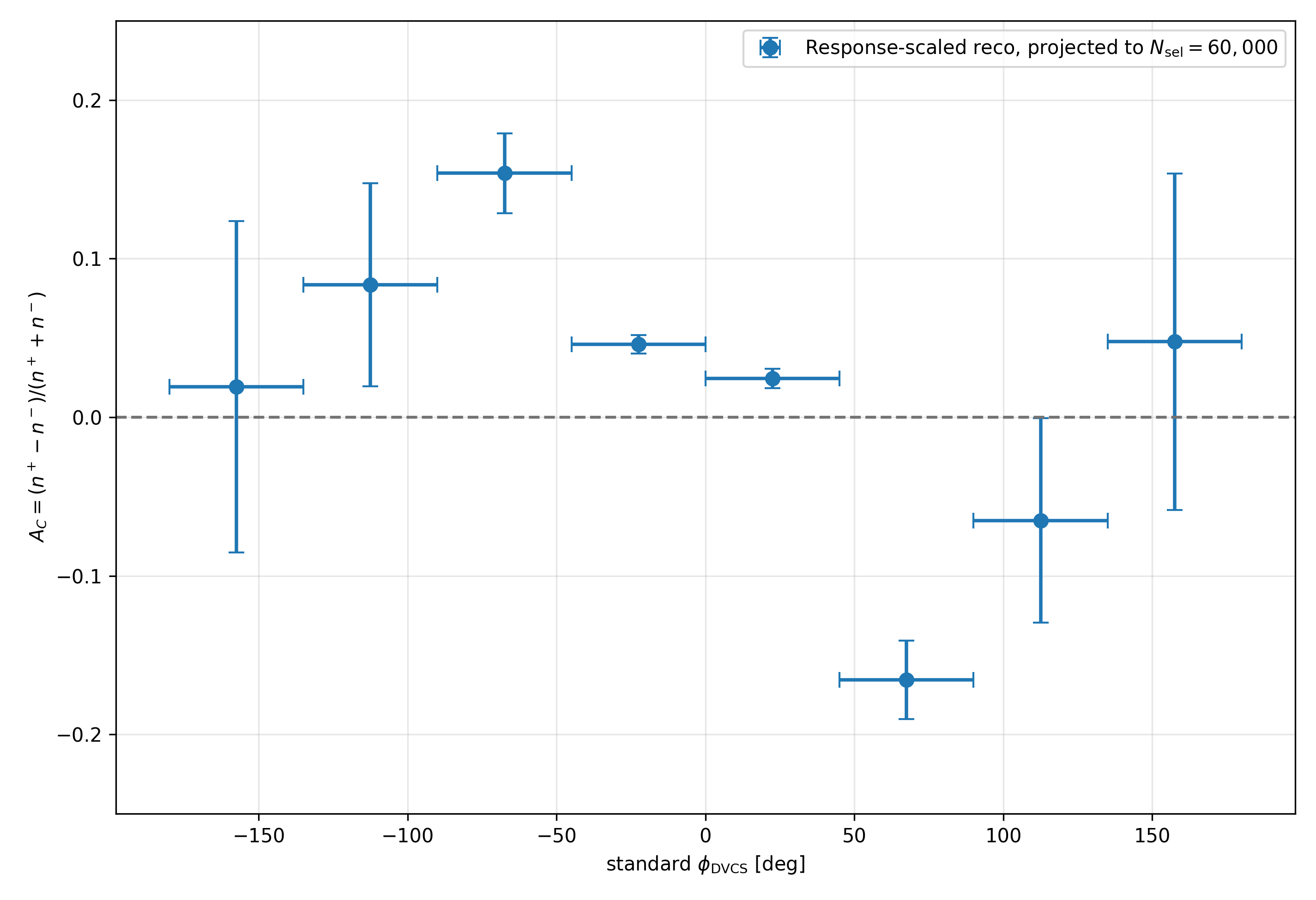}
    \caption{Projected DVCS performance for the \lune Phase-II detector. The left panel shows the expected distribution of DVCS photons in the energy-angle plane, with the color scale representing the number of events for one year of data taking ($6\times 10^4$ events). The right panel presents the projected statistical precision of the beam charge asymmetry $A_C(\phi)$ extracted from $\mu^+$ and $\mu^-$ beam measurements.}
    \label{fig:dvcs_projection}
\end{figure}

Figure~\ref{fig:dvcs_projection} shows the expected statistical precision for representative DVCS observables. The real photon produced in DVCS is predominantly emitted in the very forward direction; therefore, the current forward calorimeter design is well suited for its detection. The beam charge asymmetry is expected to be measured with a statistical precision better than 1\% using both $\mu^+$ and $\mu^-$ beams. The beam charge asymmetry provides direct sensitivity to the real part of the Compton form factors, and therefore offers an important constraint on the GPDs of the nucleon. The combination of tracking detectors and electromagnetic calorimetry enables efficient reconstruction of the scattered muon and the final-state photon, which are essential for exclusive-event identification. With future upgrades of \hiaf, the muon flux could be increased by approximately two orders of magnitude, enabling high-precision DVCS measurements with significantly reduced beam time.

The Phase-II program therefore provides a unique opportunity to connect traditional nucleon form-factor measurements with modern three-dimensional nucleon imaging studies within a single experimental facility.

Together, the precision nucleon-structure program and the capability for direct searches make \lune a versatile facility capable of addressing several of the most important open questions in contemporary nuclear and particle physics.

\clearpage

%% file: Summary.tex
\section{Summary and outlook}

The \lune experiment has been proposed as a dedicated muon-nucleon and muon-nucleus scattering facility at \hiaf, aiming to exploit the unique opportunities provided by high-intensity muon beams in the momentum range from 0.5 to 7.5\gevc. By combining precision lepton-scattering techniques with the capabilities of a modern fixed-target detector system, \lune will establish a new experimental platform for investigating the structure of nucleons and nuclei across multiple distance scales.

A broad physics program has been identified. In the Phase-I, \lune will focus on precision measurements of elastic muon-proton scattering at low momentum transfer, providing an independent determination of the proton charge radius and electromagnetic form factors. These measurements will directly address the proton-radius puzzle and offer important tests of lepton universality.
Beyond the proton, the experiment will enable precision studies of light nuclei, including measurements of charge radii and electromagnetic form factors for deuterium and other nuclear targets. The Phase-II detector configuration significantly broadens the experimental scope. Through measurements of DIS, DVCS, and related exclusive processes, \lune will provide access to parton distributions, generalized parton distributions, and multidimensional nucleon tomography. These studies will complement existing and planned programs at JLab, AMBER, and the future EicC/EIC, while offering unique opportunities associated with high-energy muon beams.

In addition to its core nuclear-physics program, the facility possesses significant potential for searches for physics beyond the SM. Precision comparisons of muon and electron scattering observables, together with dedicated searches for light weakly coupled particles, may provide sensitivity to new interactions connected to the muon sector.

The studies presented in this White Paper demonstrate the scientific potential and technical feasibility of a dedicated muon-scattering facility at \hiaf. By bridging precision low-energy measurements, nuclear structure physics, nucleon tomography, and searches for new phenomena, \lune has the potential to become a unique international facility serving both the nuclear-physics and particle-physics communities.

Future efforts will focus on further optimization of the detector design, detailed systematic studies, refinement of the projected physics sensitivities, and the preparation of a Conceptual Design Report. The successful implementation of \lune would constitute a significant new cornerstone of the \hiaf scientific portfolio, offering a uniquely capable experimental platform for probing fundamental aspects of hadron and nuclear structure.

%% file: appendix.tex
\section*{Appendix}
\section{Brief history of muon-scattering experiments}
\label{app:history}
The era of muon scattering experiments began in the early 1960s with the advent of high-energy muon beams at proton synchrotrons. Early measurements focused on elastic and inelastic scattering of muons off protons and light nuclei, primarily to investigate nucleon form factors and the onset of scaling behavior in DIS.
Notably, Muon scattering experiments performed at CERN during the 1960s and 1970s provided an important independent verification of the scaling behavior first observed in electron scattering at SLAC~\cite{Taylor:1991ew,Feynman:1969ej}. The muon scattering data complemented electron scattering results and were crucial in confirming the quark-parton model.

During 1970s, the European Muon Collaboration (EMC) at CERN conducted systematic measurements of muon DIS with improved energy and precision, using muon beams up to 280\gev~\cite{EuropeanMuon:1983wih}. The EMC experiment extended the kinematic coverage in Bjorken-$x$ and four-momentum transfer \qsq and provided key data that challenged prevailing parton distribution models.
One of the landmark discoveries from muon scattering in the early 1980s was the `EMC effect'. The observation that nucleon structure functions are modified inside nuclei compared to free nucleons~\cite{Geesaman:1995yd}. This phenomenon was revealed by comparing deep inelastic muon scattering cross sections on heavy nuclei to those on deuterium. The EMC effect sparked extensive theoretical and experimental efforts to understand nuclear medium modifications of quark distributions.

Beginning with the EMC polarized-scattering measurements in the late 1980s and continuing through the Spin Muon Collaboration (SMC) program in the 1990s, muon scattering experiments refined measurements of nucleon spin structure functions~\cite{SpinMuon:1998eqa}, quark distributions, and electroweak form factors. Their results contributed to the so-called `proton spin crisis', revealing that quarks carry only a fraction of the nucleon spin.

In addition to the European Muon Collaboration (EMC), two other major fixed-target muon scattering experiments significantly advanced our understanding of nucleon structure: the BCDMS and NMC experiments.
The BCDMS experiment, conducted at CERN during the 1980s, used high-energy muon beams (100–280\gev) to perform precise measurements of deep inelastic muon-proton and muon-deuteron scattering. BCDMS provided high-statistics data with excellent control of systematic uncertainties over a broad range of Bjorken-$x$ and $Q^2$, enabling detailed studies of the scaling violations predicted by perturbative QCD and precise extraction of the strong coupling constant $\alpha_s$~\cite{BCDMS:1989qop,BCDMS:1989ggw}. This experiment played a crucial role in confirming QCD as the theory of strong interactions.
The New Muon Collaboration (NMC) extended the scope of muon scattering measurements in the late 1980s and early 1990s. NMC utilized muon beams with energies up to 280\gev to measure structure functions of various nuclear targets with high precision. The NMC data provided key input to parton distribution function (PDF) global fits and delivered important insights into nuclear effects in DIS, complementing and expanding upon the earlier EMC findings~\cite{NewMuon:1996yuf,NewMuon:1996fwh}.

Together, BCDMS and NMC, along with EMC, form the foundation of our modern understanding of nucleon structure through muon scattering experiments, paving the way for the precision measurements carried out later by COMPASS and upcoming experiments such as AMBER and MUSE.
With the development of high-luminosity muon beams and advanced detectors, the 1990s and early 2000s saw precision tests of QCD predictions through muon scattering experiments. The COMPASS experiment at CERN, starting in 2002, represented a major leap forward in muon scattering studies~\cite{COMPASS:2007rjf}. COMPASS employed polarized muon beams at energies of 160 GeV and diverse polarized targets to measure spin-dependent structure functions with unprecedented accuracy.
COMPASS also investigated SIDIS, allowing the extraction of TMDs and GPDs. These measurements enhanced our understanding of the three-dimensional momentum and spatial structure of nucleons.

The E665 experiment at Fermilab extended muon DIS measurements to significantly lower values of Bjorken-$x$ using 470 GeV muon beams, providing important information on nuclear shadowing and the low-$x$ behavior of structure functions.

Muon scattering was also instrumental in precise determinations of electroweak parameters and searches for physics beyond the Standard Model, through measurements of weak neutral current interactions and lepton flavor universality tests.
In recent years, muon scattering experiments continue to contribute critically to nuclear and particle physics. Advances in detector technology and beam intensity have enabled high-statistics measurements of elastic and inelastic muon-proton scattering, with improved control over systematic uncertainties~\cite{COMPASS:2010wkz}.

\begin{sidewaysfigure}
    \centering
    \includegraphics[width=0.95\textheight]{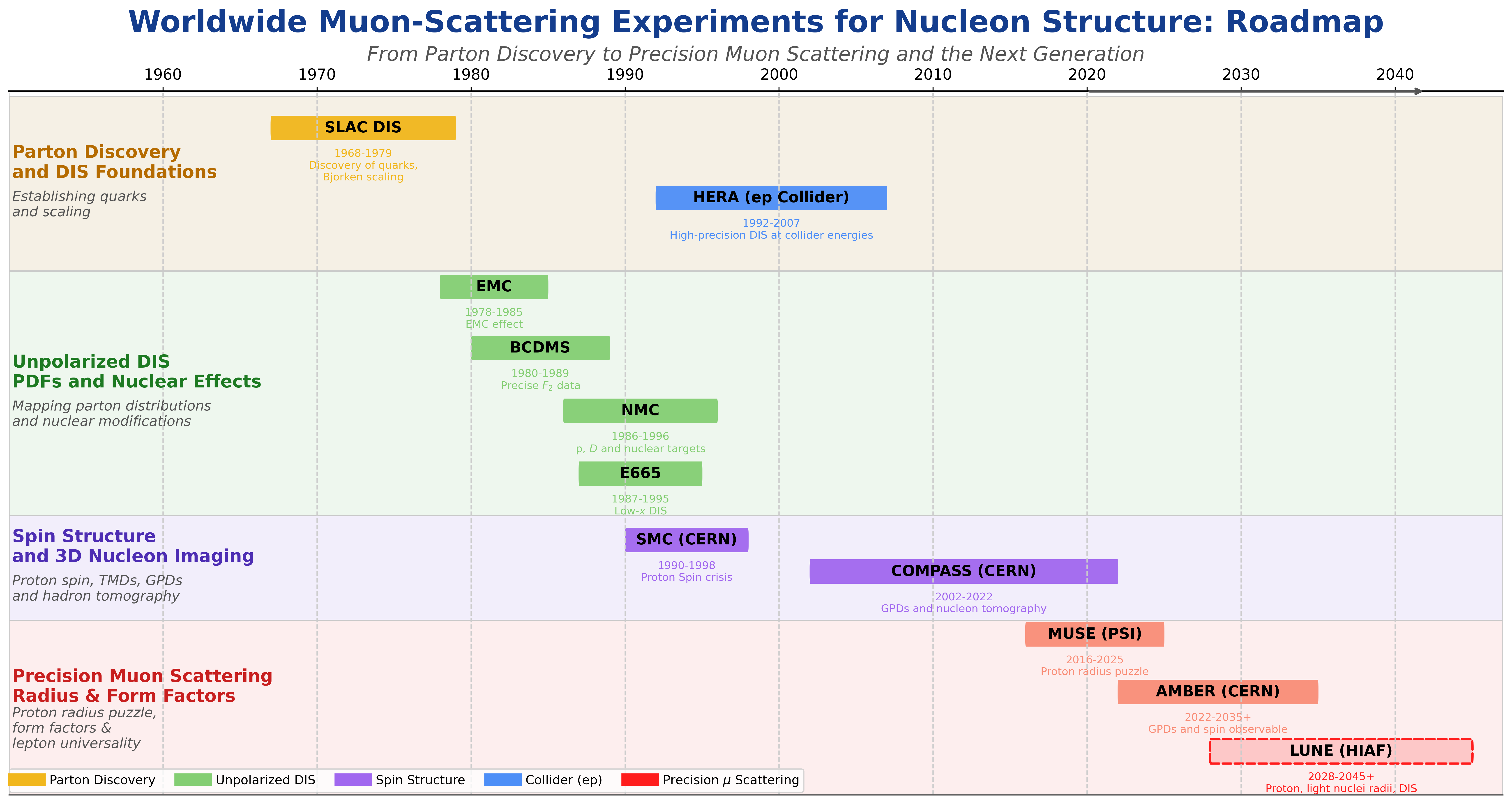}
    \caption{Historical and projected timeline of muon scattering experiments.}
    \label{fig:roadmap}
\end{sidewaysfigure}

Two prominent ongoing efforts are the MUSE (MUon proton Scattering Experiment) and AMBER (Apparatus for Meson and Baryon Experimental Research) experiments, both aiming to deepen our understanding of nucleon structure and fundamental interactions through precision muon scattering measurements.
The MUSE experiment at the PSI is designed to resolve the `proton radius puzzle' by comparing elastic scattering of electrons and muons (both positive and negative) on protons at low momentum transfer. By performing simultaneous measurements with muon and electron beams, MUSE aims to test lepton universality and improve the precision of proton charge radius extraction, thus shedding light on discrepancies between electronic and muonic determinations of proton size~\cite{MUSE:2013uhu}.
The AMBER experiment at CERN is a next-generation fixed-target facility under development that plans to utilize high-intensity muon and hadron beams to perform precision studies of hadron structure. Among its goals are measurements of elastic and inelastic muon-proton scattering over a wide kinematic range, investigations of GPDs, and studies of spin-dependent observables. AMBER will build upon the experience of COMPASS, extending the capabilities for high-precision muon scattering with modern detectors and advanced data acquisition systems~\cite{Wallner:2022scd}.
Together, these experiments represent a new era of muon scattering studies with the potential to address outstanding questions in hadronic physics and test the SM with unprecedented accuracy. As shown in Fig.~\ref{fig:roadmap}.

\section{International Landscape and Scientific Opportunities}
\label{app:international}

High-energy muon facilities are currently concentrated in Europe and North America and have already made important contributions to nuclear and particle physics. The PiM1 beamline at PSI provides muon beams with energies of approximately 100-500\mevc and has supported a wide range of precision nuclear physics experiments~\cite{MUSE:2013uhu}. At Fermilab, a 3.1\gevc muon beam serves the Muon \textit{g}-2 experiment~\cite{Muong-2:2025xyk} and will provide the basis for future charged-lepton-flavor-violation searches such as Mu2e~\cite{Mu2e:2014fns}. At CERN, the AMBER experiment~\cite{Adams:2018pwt} employs muon beams with energies above 100\gevc to investigate hadron structure through DIS and related processes.

Several new muon facilities are currently under construction or under development, including the \hiaf muon source~\cite{Xu:2025spd}, the CiADS muon source~\cite{Cai:2023caf}, the China Spallation Neutron Source (CSNS) muon facility~\cite{Chen:2023uvp}, the SHINE muon source~\cite{Lv:2023wsc}, and the muon program at J-PARC~\cite{Miyake:2010zz}. However, most existing and planned muon facilities focus on beam energies below 1\gevc, while facilities operating at much higher energies are primarily designed for DIS studies. As a result, a significant gap exists in the intermediate-energy region between approximately 1 and 7\gevc.

The importance of this energy regime has recently been recognized by the international community. In particular, JLab, which has achieved numerous landmark results using high-intensity electron beams, has initiated research and development efforts toward a GeV-scale muon beam facility~\cite{Achenbach:2025ynn}. According to current plans, JLab aims to deliver high-quality muon beams with energies up to approximately 5\gevc and intensities of order $10^7$ muons per second within the next decade. This development highlights the growing interest in muon scattering as a next-generation tool for precision studies of nucleon structure and QCD.

Compared with existing and planned facilities, the \hiaf muon source offers a unique combination of broad momentum coverage, flexible beam tuning, high beam quality, and excellent particle purity. These capabilities provide an unprecedented opportunity to establish a comprehensive muon-proton and muon-nucleus scattering program in the largely unexplored 0.5-7.5\gevc energy region, enabling China to play a leading role in this emerging field.

\section{HIRIBL Beamline Upgrade for the LUNE Experiment}
\label{app:beamline}

\subsection{Beamline Adaptation and Optics Design}

The HIAF Radioactive Ion Beam Line (HIRIBL) was not originally designed to accommodate a dedicated GeV-scale muon beamline. To support the \lune physics program, a dedicated experimental area is planned approximately 14.5~m downstream of the present HIRIBL terminal. The available space, measuring approximately $6~\mathrm{m}$ in the horizontal direction and $10~\mathrm{m}$ along the beam direction, is sufficient for the installation and operation of the \lune spectrometer.

The existing downstream section of HIRIBL is equipped with five room-temperature quadrupole magnets of approximately 120~mm aperture together with the associated vacuum system. These components were originally optimized for the transport of primary and secondary ion beams with relatively small transverse emittance. Owing to the limited aperture and geometrical acceptance, the current configuration is not suitable for the transport of the large-emittance muon beams required by \lune.

To address this limitation, a dedicated beamline upgrade is proposed as part of the \lune project. The primary objective is to establish an efficient transport channel between the muon production target and the experimental area while maintaining optical compatibility with the existing HIRIBL lattice. The upgraded system will provide stable and low-loss transport of muon beams over the momentum range required by the experiment.

The beam transport section from the present HIRIBL terminal to the \lune experimental area has a total length of approximately 14.5~m. The design requirements include a maximum magnetic rigidity of 15~Tm, horizontal and vertical acceptances of $30\pi~\mathrm{mm,mrad}$ and $37.5\pi~\mathrm{mm,mrad}$, respectively, and a momentum acceptance of $\pm2\%$.

Beam-optics calculations indicate that the required transport performance can be achieved using a system of four quadrupole magnets. The first quadrupole can be realized using existing HIRIBL hardware, while the remaining three magnets require newly developed large-aperture designs. The principal design challenge is to provide sufficient good-field region and field gradient while maintaining compatibility with the existing power-supply infrastructure.

The new quadrupole magnets will employ high-permeability DT4 electrical iron cores and water-cooled copper coils operating in DC mode. Detailed optimization of the pole geometry will be carried out to minimize higher-order field distortions and maximize field uniformity within the required aperture. The magnet excitation parameters will be designed to remain compatible with the existing HIRIBL quadrupole power supplies, thereby reducing construction cost and facilitating integration into the current accelerator infrastructure.

The proposed upgrade will significantly increase the transverse acceptance of the transport line and provide the beam quality required for precision muon-scattering measurements at LUNE.

\subsection{Vacuum and Beam Diagnostic Systems}

A dedicated vacuum and beam-diagnostic system will be constructed together with the upgraded transport line. The layout 
includes target chambers, beam pipes, ion pumps, vacuum gauges, gate valves, fluorescent beam-profile monitors, and a vacuum-isolation window located immediately upstream of the \lune detector.

Since the muon beam traverses the transport line only once and beam losses due to residual-gas interactions are negligible, the vacuum specification follows the existing HIRIBL standard of approximately $1\times10^{-6}$~Pa. This vacuum level is sufficient to ensure reliable beam transport and stable detector operation.

The vacuum system consists of five four-way vacuum chambers distributed along the beamline. These chambers provide installation ports for pumping, diagnostics, and instrumentation. The beamline itself is based on large-aperture vacuum pipes with an inner diameter of approximately 260~mm in order to accommodate the required beam acceptance.

Vacuum pumping is provided by five ion pumps, each with a nominal pumping speed of 600~L/s. Pressure monitoring is achieved using a full-range vacuum gauge system covering the range from $5\times10^{-8}$~Pa to $10^{5}$~Pa. Two pneumatic gate valves are incorporated to allow isolation of different beamline sections during maintenance and commissioning.

Beam tuning and commissioning will be supported by two fluorescent beam-profile monitors installed at strategic locations along the transport line. These devices provide direct observation of the beam spot and facilitate optics verification and alignment procedures. A large-aperture vacuum-isolation window will be installed at the entrance of the LUNE experimental area, providing vacuum separation between the beamline and detector systems.

The proposed vacuum and diagnostic infrastructure follows established HIRIBL engineering standards and can be implemented using mature technologies with low technical risk.